%% file: article.tex
\documentclass[12pt,a4paper,french,english,PSfrag,BCOR8mm, most, usenames, dvipsnames]{article}
\usepackage[T1]{fontenc}
\usepackage[latin9]{inputenc}
\usepackage{xcolor}
\usepackage{babel}
\makeatletter
\addto\extrasfrench{%
   \providecommand{\fg}{\ifdim\lastskip>\z@\unskip\fi~\frqq}%
}

\makeatother
\usepackage{calc}
\usepackage{textcomp}
\usepackage{tcolorbox}
\usepackage{amsmath}
\usepackage{amsthm}
\usepackage{amssymb}
\usepackage{graphicx}
\usepackage{wasysym}
\usepackage[unicode=true,pdfusetitle,
 bookmarks=true,bookmarksnumbered=false,bookmarksopen=false,
 breaklinks=false,pdfborder={0 0 0},pdfborderstyle={},backref=false,colorlinks=true]
 {hyperref}

\makeatletter

\newcommand*\LyXZeroWidthSpace{\hspace{0pt}}
\pdfpageheight\paperheight
\pdfpagewidth\paperwidth

\newcommand{\noun}[1]{\textsc{#1}}
\providecommand{\tabularnewline}{\\}
\numberwithin{equation}{section}
\numberwithin{figure}{section}
\theoremstyle{plain}
\newtheorem{thm}{\protect\theoremname}[section]
\theoremstyle{remark}
\newtheorem{rem}[thm]{\protect\remarkname}
\theoremstyle{remark}
\newtheorem{acknowledgement}[thm]{\protect\acknowledgementname}
\theoremstyle{plain}
\newtheorem{assumption}[thm]{\protect\assumptionname}
\theoremstyle{definition}
\newtheorem{defn}[thm]{\protect\definitionname}
\theoremstyle{plain}
\newtheorem{question}[thm]{\protect\questionname}
\theoremstyle{plain}
\newtheorem{lem}[thm]{\protect\lemmaname}
\theoremstyle{plain}
\newtheorem{prop}[thm]{\protect\propositionname}
\theoremstyle{definition}
\newtheorem{example}[thm]{\protect\examplename}

\usepackage{a4wide}
\usepackage{color}
\usepackage{graphics}

\tcbuselibrary{breakable}

\makeatother

\addto\captionsenglish{\renewcommand{\acknowledgementname}{Acknowledgement}}
\addto\captionsenglish{\renewcommand{\assumptionname}{Assumption}}
\addto\captionsenglish{\renewcommand{\definitionname}{Definition}}
\addto\captionsenglish{\renewcommand{\examplename}{Example}}
\addto\captionsenglish{\renewcommand{\lemmaname}{Lemma}}
\addto\captionsenglish{\renewcommand{\propositionname}{Proposition}}
\addto\captionsenglish{\renewcommand{\questionname}{Question}}
\addto\captionsenglish{\renewcommand{\remarkname}{Remark}}
\addto\captionsenglish{\renewcommand{\theoremname}{Theorem}}
\addto\captionsfrench{\renewcommand{\acknowledgementname}{Remerciements}}
\addto\captionsfrench{\renewcommand{\assumptionname}{Hypothèse}}
\addto\captionsfrench{\renewcommand{\definitionname}{Définition}}
\addto\captionsfrench{\renewcommand{\examplename}{Exemple}}
\addto\captionsfrench{\renewcommand{\lemmaname}{Lemme}}
\addto\captionsfrench{\renewcommand{\propositionname}{Proposition}}
\addto\captionsfrench{\renewcommand{\questionname}{Question}}
\addto\captionsfrench{\renewcommand{\remarkname}{Remarque}}
\addto\captionsfrench{\renewcommand{\theoremname}{Théorème}}
\providecommand{\acknowledgementname}{Acknowledgement}
\providecommand{\assumptionname}{Assumption}
\providecommand{\definitionname}{Definition}
\providecommand{\examplename}{Example}
\providecommand{\lemmaname}{Lemma}
\providecommand{\propositionname}{Proposition}
\providecommand{\questionname}{Question}
\providecommand{\remarkname}{Remark}
\providecommand{\theoremname}{Theorem}

\begin{document}
\title{Manifestation of the topological index formula in quantum waves and
geophysical waves}
\author{\href{https://www-fourier.ujf-grenoble.fr/~faure/}{Frédéric Faure}\\
{\small{}Univ. Grenoble Alpes, CNRS, Institut Fourier, F-38000 Grenoble,
France}\\
{\small{}\href{mailto:frederic.faure@univ-grenoble-alpes.fr}{frederic.faure@univ-grenoble-alpes.fr}}}
\date{2022 April 06}
\maketitle
\begin{abstract}
Using semi-classical analysis in $\mathbb{R}^{n}$ we present a quite
general model for which the topological index formula of Atiyah-Singer
predicts a spectral flow with the transition of a finite number of
eigenvalues between clusters (energy bands). This model corresponds
to physical phenomena that are well observed for quantum energy levels
of small molecules \cite{fred-boris,fred-boris01}, also in geophysics
for the oceanic or atmospheric equatorial waves \cite{matsuno1966quasi,Delplace_Venaille_2018}
and expected to be observed in plasma physics \cite{2022_qin_plasma_physics}.
\end{abstract}
\newtcolorbox{cBoxA}[1][]{enhanced, frame style={purple!80}, interior style={red!0}, #1}

\newtcolorbox{cBoxB}[2][]{enhanced, frame style={teal!80}, interior style={cyan!0}, #2}

\footnote{2010 Mathematics Subject Classification:\\
35Q86 PDEs in connection with geophysics,

81V55 Molecular physics,

55R50 Stable classes of vector space bundles, K-theory,

47A53 (Semi-) Fredholm operators; index theories,

19K56 Index theory,

81Q20 Semiclassical techniques, including WKB and Maslov methods

81Q05 Closed and approximate solutions to the Schrödinger, Dirac,
Klein-Gordon and other equations of quantum mechanics}

\newpage{}

\global\long\def\eq#1{\underset{(#1)}{=}}%

\tableofcontents{}

\newpage{}
\begin{rem}
On this pdf file, you can click on the colored words, they contain
an hyper-link to \href{https://en.wikipedia.org/wiki/Main_Page}{wikipedia}
or other multimedia contents.
\end{rem}

\section{Introduction}

The famous \href{https://en.wikipedia.org/wiki/Atiyah\%E2\%80\%93Singer_index_theorem}{index theorem of Atiyah Singer}
obtained in the 60' relates two different domains of mathematics:
spectral theory of pseudo-differential operators and differential
topology \cite{booss_85}\cite{fedosov96}. This theorem has a strong
importance in mathematics with many applications (e.g. the Riemann-Roch-Hirzebruch
index formula that is used in geometric quantization \cite{hawkins00})
but also in physics: in quantum field theory with anomalies \cite[chap.19]{peskin_TQFT_95},
in molecular physics with energy spectrum \cite{fred-boris,fred-boris01,fred-boris02}.
Recently P. Delplace, J.B. Marston and A. Venaille \cite{Delplace_Venaille_2018}
have discovered that a famous model of oceanic equatorial waves established
by Matsuno in 1966 \cite{matsuno1966quasi} has remarkable topological
properties, namely that the existence of $\mathcal{N}=+2$ equatorial
modes in the Matsuno's model is related to the fact that the dispersion
equation of this model defines a vector bundle over\footnote{Here $S^{2}$ is not related to the surface of the earth but is a
surface in $\mathbb{R}^{3}$ that enclosed a singularity at the origin.} $S^{2}$ whose topology is characterized by a Chern index with value
$\mathcal{C}=+2$. In the similar context of waves but in \href{https://en.wikipedia.org/wiki/Plasma_(physics)}{plasma physics},
Hong Qin, Yichen Fu \cite{2022_qin_plasma_physics} have recently
predicted a manifestation of the index formula.

In this paper we propose a general mathematical model that contains
as particular cases the normal form used for molecular physics in
\cite{fred-boris,fred-boris01} and the model of Matsuno \cite{matsuno1966quasi}\cite{Delplace_Venaille_2018}
of equatorial waves. For this general model we have on one side a
spectral index $\mathcal{N}\in\mathbb{Z}$ that counts the number
of eigenvalues that move upwards as a parameter $\mu$ increases and
on the other side a topological Chern index $\mathcal{C}\in\mathbb{Z}$
associated to a vector bundle that characterizes the equivalence class
of the model. We establish the index formula $\mathcal{N}=\mathcal{C}$.

There are many studies about topological phenomena in condensed matter
physics. Closely related to this paper are the works related to bulk-interface
correspondence by Guillaume Bal \cite{bal2019continuous}, Chris Bourne,
Johannes Kellendonk, and Adam Rennie \cite{bourne2017k}\cite{bourne2018chern},
Alexis Drouot \cite{drouot2021microlocal}, A Elgart, GM Graf, and
JH Schenker \cite{elgart2005equality}, Gian Michele Graf and Marcello
Porta \cite{graf2013bulk}, Yosuke Kubota \cite{kubota2017controlled},
Emil Prodan and Hermann Schulz-Baldes \cite{prodan2016bulk}, and
Julio Cesar Avila, Hermann Schulz-Baldes, and Carlos Villegas-Blas\cite{avila2013topological}.
In particular the work of Alexis Drouot \cite{drouot2021microlocal}
uses microlocal analysis as here. There are also the works by C Dembowski,
H-D Gräf, HL Harney, A Heine, WD Heiss, H Rehfeld, and A Richter \cite{dembowski2001experimental},
Jacob Shapiro and Clément Tauber \cite{shapiro2019strongly}, Alex
Bols, Jeffrey Schenker, and Jacob Shapiro\cite{bols2021fredholm}.

The paper is organized as follows. In Section \ref{sec:The-model}
we present the general model and the main result of this paper, Theorem
\ref{thm:Formule-de-l'indice.-1}. In Section \ref{sec:Proof-of-Theorem}
we give the proof of Theorem \ref{thm:Formule-de-l'indice.-1}. The
proof relies on the index theorem on Euclidean space of Fedosov-Hörmander
given in \cite[thm 7.3 p. 422]{hormander1979weyl},\cite[Thm 1, page 252]{booss_85}
and explained in the appendix.

Sections \ref{subsec:Spectral-flow-and} and \ref{subsec:Spectral-flow-and-1}
are applications of this general model in physics. In Section \ref{subsec:Spectral-flow-and}
we present a simple model used in \cite{fred-boris,fred-boris01}
to show the manifestation of the index formula in experimental molecular
spectra of quantum waves. In Section \ref{subsec:Spectral-flow-and-1}
we present the model of equatorial geophysics waves of Matsuno \cite{matsuno1966quasi}
and the topological interpretation from \cite{Delplace_Venaille_2018}.

The reader may prefer to read first Section \ref{subsec:Spectral-flow-and}
and \ref{subsec:Spectral-flow-and-1} that present the examples with
detailed computations before Section \ref{sec:The-model} that presents
the general but more abstract model.

Appendix \ref{sec:Quantification,-op=0000E9rateurs-pseud} gives a
short overview of symbols and pseudo-differential operators. Appendix
\ref{sec:Espaces-fibr=0000E9s-vectoriels} gives a short overview
of vector bundles over spheres.

This article is made from the lecture notes in French \cite{faure_cours_thm_adiabatique_2018}.
\begin{acknowledgement}
The author thanks P. Delplace and A. Venaille for interesting discussions
about models of geophysical waves.
\end{acknowledgement}

\section{\label{sec:The-model}A general model on $\mathbb{R}^{n}$ and index
formula}

In this Section we propose a general framework that will contains
the particular models of molecular physics of Section \ref{subsec:Spectral-flow-and}
and of geophysics of Section \ref{subsec:Spectral-flow-and-1}. For
this general model we define a spectral index $\mathcal{N}$ that
corresponds to the number of eigenvalues that move upwards with respect
to an external parameter $\mu$ and we define a topological index
(Chern index) $\mathcal{C}$ of a vector bundle that characterizes
the (stable) isomorphism class of the model. We establish the index
formula $\mathcal{N}=\mathcal{C}$.

\subsection{Admissible family of symbols $\left(H_{\mu}\right)_{\mu}$}

Let $\mu\in]-2,2[$ be a parameter. Let $n\in\mathbb{N}\backslash\left\{ 0\right\} $
and $\left(x,\xi\right)\in T^{*}\mathbb{R}^{n}=\mathbb{R}^{n}\times\mathbb{R}^{n}$
a point on the cotangent space $T^{*}\mathbb{R}^{n}$ called ``\textbf{slow
phase space}''. Let $d\geq2$ be an integer and $\mathrm{Herm}\left(\mathbb{C}^{d}\right)$
denotes \href{https://en.wikipedia.org/wiki/Hermitian_adjoint\#Hermitian_operators}{Hermitian operators}
on $\mathbb{C}^{d}$. We consider a function $\left(\mu,x,\xi\right)\rightarrow H_{\mu}\left(x,\xi\right)$
smooth with respect to $\mu,x,\xi$ and valued in $\mathrm{Herm}\left(\mathbb{C}^{d}\right)$
:
\begin{equation}
H_{\mu}:\begin{cases}
T^{*}\mathbb{R}^{n} & \rightarrow\mathrm{Herm}\left(\mathbb{C}^{d}\right)\\
\left(x,\xi\right) & \mapsto H_{\mu}\left(x,\xi\right)
\end{cases}\label{eq:symbole_H_mu-1}
\end{equation}
called \textbf{symbol} (we suppose that $H_{\mu}\in S_{\rho,\delta}^{m}\left(T^{*}\mathbb{R}^{n}\right)$
belongs to the class of Hörmander \href{https://en.wikipedia.org/wiki/Pseudo-differential_operator}{symbols}.
This corresponds to suitable hypothesis of regularity at infinity,
see Section \ref{sec:Quantification,-op=0000E9rateurs-pseud}).

For fixed values of $\mu,x,\xi$, the eigenvalues of the matrix $H_{\mu}\left(x,\xi\right)$
are real and are denoted
\begin{equation}
\omega_{1}\left(\mu,x,\xi\right)\leq\ldots\leq\omega_{d}\left(\mu,x,\xi\right).\label{eq:eigenvalues}
\end{equation}

\begin{figure}
\begin{centering}
\scalebox{0.9}[0.9]{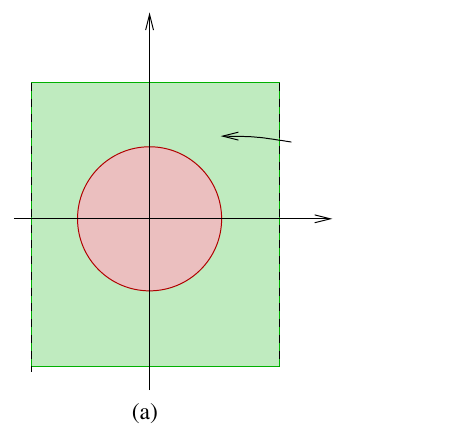}~~~~~\scalebox{0.9}[0.9]{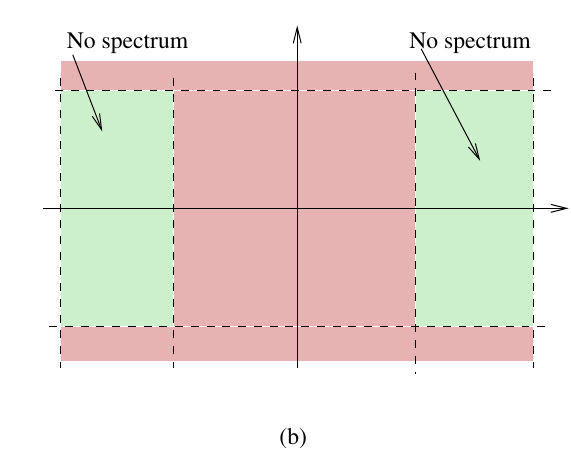}
\par\end{centering}
\caption{\label{fig:Hypoth=0000E8se}Illustration of Assumption \ref{assu:Gap.-On-suppose}.
On figure (a), for parameters $\left(\mu,x,\xi\right)\in\mathbb{R}\times\mathbb{R}^{n}\times\mathbb{R}^{n}$
in the green domain, we assume that the spectrum of the hermitian
matrix $H_{\mu}\left(x,\xi\right)$, has $r$ eigenvalues smaller
than $-C$ and that the others are greater than $C>0$. Equivalently,
on figure (b), the spectrum of $H_{\mu}\left(x,\xi\right)$ for any
$\left(x,\xi\right)$ is contained in the red domain.}
\end{figure}

We will assume the following hypothesis\footnote{\label{fn:Here--is}Here $\left\Vert \left(\mu,x,\xi\right)\right\Vert :=\sqrt{\mu^{2}+\sum_{j=1}^{n}\left(x_{j}^{2}+\xi_{j}^{2}\right)}$
is the Euclidean distance from $\left(\mu,x,\xi\right)$ to the origin
in $\mathbb{R}^{2n+1}$.} for the family of symbols $\left(H_{\mu}\right)_{\mu}$. This hypothesis
is illustrated on Figure \ref{fig:Hypoth=0000E8se}.

\begin{cBoxA}{}
\begin{assumption}
\label{assu:Gap.-On-suppose}<<\textbf{Spectral gap assumption}>>.
For the family of symbols $\left(H_{\mu}\right)_{\mu}$, (\ref{eq:symbole_H_mu-1}),
we suppose that there exists an index $r\in\left\{ 1,\ldots d-1\right\} $
and $C>0$ such that for every $\left(\mu,x,\xi\right)\in\mathbb{R}^{3}$
such that $\left\Vert \left(\mu,x,\xi\right)\right\Vert \geq1$ and
$\left|\mu\right|\leq2$, we have
\[
\omega_{r}\left(\mu,x,\xi\right)<-C\text{ and }\omega_{r+1}\left(\mu,x,\xi\right)>+C.
\]
\end{assumption}

\end{cBoxA}

\subsection{Spectral index $\mathcal{N}$ for the family of symbols $\left(H_{\mu}\right)_{\mu}$}

The reader may read first the appendix \ref{sec:Quantification,-op=0000E9rateurs-pseud}
that gives an introduction with examples to \textbf{pseudo-differential
operators} (PDO) and \textbf{pseudo-differential calculus}.

Let us introduce a new parameter $\epsilon>0$ called \textbf{adiabatic
parameter} or \textbf{semi-classical parameter}. We define the \textbf{pseudo-differential
operator} \footnote{The operator $\hat{H}_{\mu,\epsilon}$ belongs to $\mathrm{Herm}\left(L^{2}\left(\mathbb{R}^{n}\right)\otimes\mathbb{C}^{d}\right)\equiv\mathrm{Herm}\left(L^{2}\left(\mathbb{R}^{n};\mathbb{C}^{d}\right)\right)$,
i.e. is a self-adjoint operator in the space of functions on $\mathbb{R}^{n}$
with $d$ complex components.} (PDO)
\begin{equation}
\hat{H}_{\mu,\epsilon}:=\mathrm{Op}_{\epsilon}\left(H_{\mu}\right)\quad\in\mathrm{Herm}\left(L^{2}\left(\mathbb{R}^{n}\right)\otimes\mathbb{C}^{d}\right),\label{eq:def_OpH}
\end{equation}
obtained by \href{https://en.wikipedia.org/wiki/Wigner\%E2\%80\%93Weyl_transform}{Weyl quantization}
of the symbol $H_{\mu}$.

\begin{cBoxB}{}
\begin{thm}
\label{prop:Avec-l'hypoth=0000E8se-,}We do the assumption \ref{assu:Gap.-On-suppose}.
Then for every $\alpha>0$ there exists $\epsilon_{0}>0$ such that
for every $0<\epsilon<\epsilon_{0}$,
\begin{itemize}
\item for any $\mu$ such that $1+\alpha<\left|\mu\right|<2$, the operator
$\hat{H}_{\mu,\epsilon}$ has \textbf{no spectrum} in the interval
$]-C+\alpha,+C-\alpha[$.
\item for any $\mu$ such that $\left|\mu\right|\leq1+\alpha$, the operator
$\hat{H}_{\mu,\epsilon}$ has \textbf{discrete spectrum} in the interval
$]-C+\alpha,C-\alpha[$ that depends continuously on $\mu,\epsilon$.
\end{itemize}
\end{thm}

\end{cBoxB}

\begin{figure}
\begin{centering}
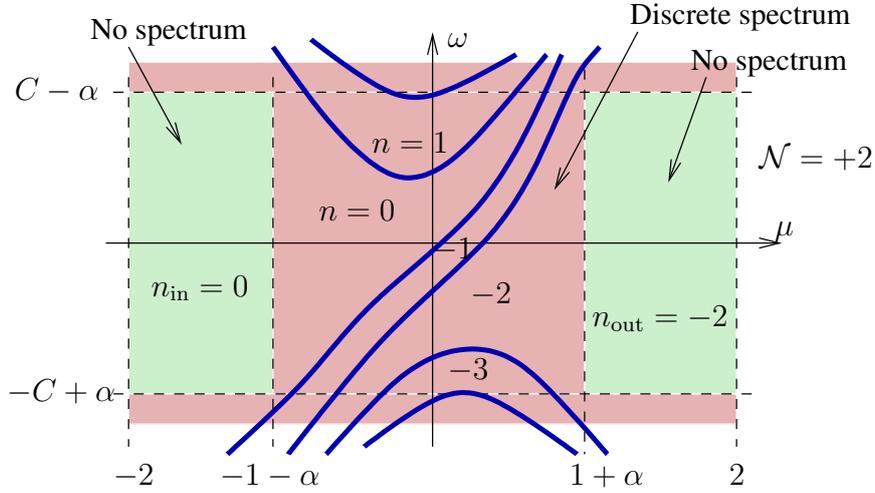
\par\end{centering}
\caption{\label{fig:def_N}For $\epsilon>0$ fixed, this is a schematic picture
of the spectrum of the operator $\hat{H}_{\mu,\epsilon}$. From Theorem
\ref{prop:Avec-l'hypoth=0000E8se-,}, in the green domain, there is
no spectrum. In the red domain, the spectrum is discrete: \textbf{discrete
eigenvalues are shown in blue} and depend continuously on $\mu,\epsilon$.
Consequently we can label the eigenvalues by some increasing number
$n$ and \textbf{label each spectral gap }by the index $n$ of the
first eigenvalue below it. We define then the spectral index of the
family of symbols $\left(H_{\mu}\right)_{\mu}$ by $\mathcal{N}=n_{\mathrm{in}}-n_{\mathrm{out}}$.
In this example, $\mathcal{N}=n_{\mathrm{in}}-n_{\mathrm{out}}=0-\left(-2\right)=+2$
corresponding to the fact that $\mathcal{N}=+2$ eigenvalues are moving
upward as $\mu$ increases.}
\end{figure}

See figure \ref{fig:def_N}.

\begin{proof}
We follow quite standard techniques from micro-local analysis. From
assumption \ref{assu:Gap.-On-suppose}, if $\left|\mu\right|>1+\alpha$
then the symbol has no spectrum in the interval $\left[-C,C\right]$.
Hence for any $z\in I:=\left[-C+\alpha,C-\alpha\right]$, the operator
$\left(z\mathrm{Id}-\hat{H}_{\mu,\epsilon}\right)$ is invertible
with approximate inverse given by $\mathrm{Op}_{\epsilon}\left(\left(z-H_{\mu}\left(x,\xi\right)\right)^{-1}\right)$.
This means that there is no spectrum for $\hat{H}_{\mu}$ in this
interval $I$.

If $\left|\mu\right|<1+\alpha$, the symbol $H_{\mu}\left(x,\xi\right)$
may have some spectrum in this spectral range $I$. However, from
assumption \ref{assu:Gap.-On-suppose}, the points $\left(\mu,x,\xi\right)$
for which $\mathrm{Ran}\left(H_{\mu}\left(x,\xi\right)\right)\cap I$
is non empty are included in the compact ball $\left\Vert \left(\mu,x,\xi\right)\right\Vert \leq1$
(here $\mathrm{Ran}\left(H_{\mu}\left(x,\xi\right)\right)$ stands
for the \href{https://en.wikipedia.org/wiki/Numerical_range}{numerical range}
of the matrix $H_{\mu}\left(x,\xi\right)$). Let us take $\chi_{\mu}:\left(x,\xi\right)\rightarrow\chi_{\mu}\left(x,\xi\right)\geq0$
that is a smooth regularization of the characteristic function of
the set $\left\Vert \left(\mu,x,\xi\right)\right\Vert \leq1$. We
take $A>0$ large enough such that the perturbed symbol $H'_{\mu}\left(x,\xi\right)=H_{\mu}\left(x,\xi\right)+A\chi_{\mu}\left(x,\xi\right)$
has no spectrum in $\left[-C,C\right]$ for every $\mu$ such that
$\left|\mu\right|<1+\alpha$. Since $\chi_{\mu}$ has compact support,
then $\mathrm{Op}_{\epsilon}\left(\chi_{\mu}\right)$ is trace class
hence compact, see (\ref{eq:Trace_OPD}). Then as before, $\mathrm{Op}_{\epsilon}\left(H'_{\mu}\right)$
has no spectrum in $\left[-C+\alpha,C-\alpha\right]$ for $\mu$ s.t.
$\left|\mu\right|<1+\alpha$ (i.e. the perturbation $A\chi_{\mu}$
has pushed the spectrum above). For $z\in\left[-C+\alpha,C-\alpha\right]$,
we write
\[
\left(z-H_{\mu}\right)^{-1}=\left(z-H'_{\mu}+A\chi_{\mu}\right)^{-1}=\left(z-H'_{\mu}\right)^{-1}\left(1+\left(z-H'_{\mu}\right)^{-1}A\chi_{\mu}\right)^{-1}
\]
By quantization of this relation, we have that $\mathrm{Op}_{\epsilon}\left(\left(z-H_{\mu}'\right)^{-1}\right)$
is bounded and analytic in $z$, $\mathrm{Op}_{\epsilon}\left(A\chi_{\mu}\right)$
is compact, hence from analytic Fredholm theorem \cite[p.201]{reed-simon1},
$\mathrm{Op}_{\epsilon}\left(1+\left(z-H'_{\mu}\right)^{-1}A\chi_{\mu}\right)^{-1}$
and therefore $\left(z-\hat{H}_{\mu,\epsilon}\right)^{-1}$ are meromorphic
in $z$ with residues that are operators of finite rank, i.e. the
spectrum is discrete.
\end{proof}
As a consequence of Theorem \ref{prop:Avec-l'hypoth=0000E8se-,} we
can define the spectral index $\mathcal{N}$ as follows, as shown
on figure \ref{fig:def_N}.

\begin{cBoxA}{}
\begin{defn}
\textbf{\label{def:Spectral-index-}<<Spectral index $\mathcal{N}$
of the family of symbols $\left(H_{\mu}\right)_{\mu}$>>. }With assumption
\ref{assu:Gap.-On-suppose} and from Theorem \ref{prop:Avec-l'hypoth=0000E8se-,},
for fixed $\epsilon$, \textbf{each spectral gap can be labeled }as
follows. Let $\left(\omega_{n}\left(\mu,\epsilon\right)\right)_{n\in\mathbb{Z}}$
be the eigenvalues of the operator $\hat{H}_{\mu,\epsilon}$ that
belongs to the interval $I_{\alpha}:=]-C+\alpha,+C-\alpha[$, labeled
by $n\in\mathbb{Z}$ and sorted by increasing values (this is well
defined up to a constant). For a given $n$, the eigenvalue $\omega_{n}\left(\mu,\epsilon\right)\in\mathbb{R}$
is continuous w.r.t. $\mu,\epsilon$. For a point $\left(\mu,\omega\right)\in\left(-1-\alpha,1+\alpha\right)\times I_{\alpha}$
different from an eigenvalue, we associate the index $n\left(\mu,\omega\right)\in\mathbb{Z}$
of the eigenvalue just below it, i.e. such that $\omega_{n}\left(\mu,\epsilon\right)<\omega<\omega_{n+1}\left(\mu,\epsilon\right)$.
We denote $n_{\mathrm{in}}:=n\left(-1,0\right)$ the index of the
first gap and $n_{\mathrm{out}}:=n\left(1,0\right)$ the index of
the last gap. This defines an integer
\begin{equation}
\mathcal{N}:=n_{\mathrm{in}}-n_{\mathrm{out}}\in\mathbb{Z}\label{eq:def_N}
\end{equation}
called the \textbf{spectral index }of the family of symbols $\left(H_{\mu}\right)_{\mu}$.
This integer $\mathcal{N}$ counts the number of eigenvalues that
go upwards as $\mu$ increases. $\mathcal{N}$ is independent on $\epsilon$
and more generally invariant under any continuous variation of the
symbol $\left(H_{\mu}\right)_{\mu}$ family satisfying the assumption
\ref{assu:Gap.-On-suppose}. Hence, $\mathcal{N}$ is a \textbf{topological
index.}
\end{defn}

\end{cBoxA}

\begin{rem}
The last remark that $\mathcal{N}$ is invariant under continuous
variation of the symbol comes from the fact that the map $H_{\mu}\rightarrow\mathcal{N}\in\mathbb{Z}$
is continuous hence locally constant
\end{rem}

~
\begin{rem}
We have used Weyl quantization in (\ref{eq:def_OpH}) to define the
operator $\hat{H}_{\mu,\epsilon}$. We could have choose any other
quantization procedure. The index $\mathcal{N}$ does not depend on
the choice of quantization.
\end{rem}

\begin{question}
How to compute the spectral index $\mathcal{N}\in\mathbb{Z}$ directly
from the symbol $\left(H_{\mu}\right)_{\mu}$?
\end{question}

Answer: in the next section, with Theorem \ref{thm:Formule-de-l'indice.-1},
we will see that $\mathcal{N}$ is simply related to the \href{https://en.wikipedia.org/wiki/Degree_of_a_continuous_mapping}{degree}
of a certain map $f:S^{2n-1}\rightarrow S^{2n-1}$ that is obtained
from the symbol $\left(H_{\mu}\right)_{\mu}$.

\subsection{Chern topological index $\mathcal{C}$ and index formula}

The reader may read first the appendix \ref{subsec:Vector-bundles-in}
that gives an introduction and general informations about topology
of vector bundles over spheres.

Let
\[
S^{2n}:=\left\{ \left(\mu,x,\xi\right)\in\mathbb{R}^{1+2n},\quad\left\Vert \left(\mu,x,\xi\right)\right\Vert =1\right\} ,
\]
be the unit sphere in the space of parameters. From assumption \ref{assu:Gap.-On-suppose},
for every parameter $\left(\mu,x,\xi\right)\in S^{2n}$, we have a
spectral gap between eigenvalues $\omega_{r}\left(\mu,x,\xi\right)$
and $\omega_{r+1}\left(\mu,x,\xi\right)$. Then we can define the
spectral projector associated to the first $r$ eigenvalues by \href{https://en.wikipedia.org/wiki/Holomorphic_functional_calculus}{Cauchy formula}
\[
\Pi_{1;r}\left(\mu,x,\xi\right):=\frac{i}{2\pi}\varoint_{\gamma}\left(z-H_{\mu}\left(x,\xi\right)\right)^{-1}dz
\]
where the integration path $\gamma\subset\mathbb{C}$ enclosed the
segment $\left[\omega_{1}\left(\mu,x,\xi\right),\omega_{r}\left(\mu,x,\xi\right)\right]$
and crosses the spectral gaps. The spectral space associated to the
first $r$ eigenvalues $\omega_{1}\ldots\omega_{r}$ is then the image
of this projector
\begin{equation}
F\left(\mu,x,\xi\right):=\mathrm{Ran}\Pi_{1;r}\left(\mu,x,\xi\right).\label{eq:F}
\end{equation}
The linear space $F\left(\mu,x,\xi\right)\subset\mathbb{C}^{d}$ has
complex dimension $r$ and defines a smooth complex vector bundle
of rank $r$ over the sphere $S^{2n}$, that we denote $F\rightarrow S^{2n}$.
From remark \ref{rem:ajout_fibre_trivial} below, we can suppose that
$r\geq n$.

From Bott's theorem \ref{thm:def_indice_topologique}, the topology
of $F\rightarrow S^{2n}$ is characterized by an integer $\mathcal{C}\in\mathbb{Z}$
called \textbf{Chern index} defined in (\ref{eq:def_D-1}) from the
degree $\mathrm{deg}\left(f\right)$ of a map $f:S^{2n-1}\rightarrow S^{2n-1}$
in (\ref{eq:def_f-1}), by $\mathcal{C}=\frac{\mathrm{deg}\left(f\right)}{\left(n-1\right)!}$,
and $f$ is directly obtained from the clutching function 
\begin{equation}
g:S^{2n-1}\rightarrow U\left(r\right)\label{eq:g}
\end{equation}
of the bundle $F\rightarrow S^{2n}$ on the equator $S^{2n-1}$ with
respect to some local trivialization. In dimension $n=1$ this is
more simple because $\mathcal{C}$ is just the winding number of the
clutching function $g:S^{1}\rightarrow U\left(1\right)\equiv S^{1}$
on the equator $S^{1}$. The physical applications considered later
in this paper correspond to dimension $n=1$.

\begin{cBoxB}{}
\begin{thm}
\textbf{\label{thm:Formule-de-l'indice.-1}<<Index formula>>}. Let
$\left(H_{\mu}\right)_{\mu}$ be a family of symbols that satisfies
assumption \ref{assu:Gap.-On-suppose}. Let $\mathcal{N}\in\mathbb{Z}$
be the spectral index defined in (\ref{eq:def_N}) and let $\mathcal{C}\in\mathbb{Z}$
be the Chern topological index defined from the vector bundle $F\rightarrow S^{2n}$
by (\ref{eq:def_D-1}). We have
\begin{equation}
\mathcal{N}=\mathcal{C}.\label{eq:formule_indice}
\end{equation}
\end{thm}

\end{cBoxB}

The proof of Theorem \ref{thm:Formule-de-l'indice.-1} is given in
Section \ref{sec:Proof-of-Theorem}. It is based on the index theorem
on Euclidean space of Fedosov-Hörmander given in \cite[thm 7.3 p. 422]{hormander1979weyl},\cite[Thm 1, page 252]{booss_85}.
\begin{rem}
\label{rem:ajout_fibre_trivial}If one replaces the symbol $H_{\mu}\left(x,\xi\right)\in\mathrm{Herm}\left(\mathbb{C}^{d}\right)$
in (\ref{eq:symbole_H_mu-1}) by the symbol $\tilde{H}_{\mu}\left(x,\xi\right)\in\mathrm{Herm}\left(\mathbb{C}^{d+m}\right)$
obtained by adding a constant diagonal term
\[
\tilde{H}_{\mu}\left(x,\xi\right)=\left(\begin{array}{cc}
H_{\mu}\left(x,\xi\right) & 0\\
0 & \omega_{0}\mathrm{Id}_{\mathbb{C}^{m}}
\end{array}\right),
\]
with $\omega_{0}<-C$ then one observes that
\begin{itemize}
\item $\tilde{H}_{\mu}$ satisfies assumption \ref{assu:Gap.-On-suppose}.
\item The spectral index of $\tilde{H}_{\mu}$ and $H_{\mu}$ are equal,
i.e. $\tilde{\mathcal{N}}=\mathcal{N}$. This is because $\mathrm{Op}_{\epsilon}\left(\tilde{H}_{\mu}\right)\equiv\mathrm{Op}_{\epsilon}\left(H_{\mu}\right)\oplus\mathrm{Op}_{\epsilon}\left(\omega_{0}\mathrm{Id}_{\mathbb{C}^{m}}\right)$
and the spectrum of $\mathrm{Op}_{\epsilon}\left(\omega_{0}\mathrm{Id}_{\mathbb{C}^{m}}\right)=\omega_{0}\mathrm{Op}_{\epsilon}\left(\mathrm{Id}_{\mathbb{C}^{m}}\right)$
is on the constant horizontal line $\omega=\omega_{0}<-C$, so does
not give moving eigenvalues.
\item The associated vector bundle $\tilde{F}\rightarrow S^{2n}$ is $\tilde{F}=F\oplus T_{m}$
where $T_{m}=S^{2n}\times\mathbb{C}^{m}$ is the trivial bundle and
$\mathrm{rank}\left(\tilde{F}\right)=\mathrm{rank}\left(F\right)+m$.
\end{itemize}
This remark shows that the spectral index does not change if one adds
a trivial bundle $T_{m}$ to the bundle $F$. It means that $\mathcal{N}$
depends only on the equivalence class of $F$ (or $H$) in the \href{https://en.wikipedia.org/wiki/K-theory}{K-theory}
group $\tilde{K}\left(S^{2n}\right)$, cf \cite{hatcher_ktheory}.
\end{rem}

\subsection{\label{subsec:Cas-particulier-de}Special case of matrix symbols
that are linear in $\left(\mu,x,\xi\right)$}

In this section, we give a simple but important remark to understand
why the model of Matsuno presented in Section \ref{subsec:Spectral-flow-and-1}
does not depend on a small parameter $\epsilon$ but nevertheless
belongs to the general model presented here. This is the same for
the normal form model presented in Section \ref{subsec:Spectral-flow-and}.

Suppose that 
\[
\tilde{H}:\left(\tilde{\mu},\tilde{x},\tilde{\xi}\right)\in\mathbb{R}^{1+2n}\rightarrow H\left(\tilde{\mu},\tilde{x},\tilde{\xi}\right)\in\mathrm{Herm}\left(\mathbb{C}^{d}\right)
\]
is a \textbf{linear map }with respect to $\left(\tilde{\mu},\tilde{x},\tilde{\xi}\right)$
and consider the quantization rule $\mathrm{Op}_{1}\left(\tilde{\xi}\right)=-i\partial_{\tilde{x}}$
(i.e. with $\epsilon=1$). For example, see the normal form symbol
(\ref{eq:symbole_H_mu}) or the Matsuno's symbol (\ref{eq:symbol_matsuno}).

For any $\epsilon>0$, we do the change of variables
\[
\mu=\sqrt{\epsilon}\tilde{\mu},\quad x=\sqrt{\epsilon}\tilde{x},
\]
that gives
\[
\mathrm{Op}_{\epsilon}\left(\xi\right)=-i\epsilon\partial_{x}=-i\sqrt{\epsilon}\partial_{\tilde{x}}=\sqrt{\epsilon}\mathrm{Op}_{1}\left(\tilde{\xi}\right).
\]
Hence the symbol $H\left(\mu,x,\xi\right)=\sqrt{\epsilon}\tilde{H}\left(\mu,x,\xi\right)$
satisfies
\[
\mathrm{Op}_{\epsilon}\left(H_{\mu}\right)=\sqrt{\epsilon}\mathrm{Op}_{1}\left(\tilde{H}_{\tilde{\mu}}\right).
\]
In other words all these models with different $\epsilon$ are equivalent
up to a scaling of the parameters and the operator (and spectrum).
The benefit to consider an additional semi-classical (or adiabatic)
parameter $\epsilon\ll1$ is that \textbf{one can perturb the linear
symbol to a non linear symbol }and still get the index formula $\mathcal{N}=\mathcal{C}$
from Theorem \ref{thm:Formule-de-l'indice.-1}.

\subsection{\label{sec:Proof-of-Theorem}Proof of the index formula (\ref{eq:formule_indice})}

In this section we give a proof of Formula (\ref{eq:formule_indice}).
This proof relies on the index Theorem on Euclidean space of Fedosov-Hörmander
given in \cite[thm 7.3 p. 422]{hormander1979weyl},\cite[Thm 1, page 252]{booss_85}.

For a given family of symbols $H=\left(H_{\mu}\right)_{\mu\in\left(-2,2\right)}$
with assumption \ref{assu:Gap.-On-suppose}, we have defined two topological
indices $\mathcal{N}_{H}\in\mathbb{Z}$ and $\mathcal{C}_{H}\in\mathbb{Z}$.
These indices are topological, i.e. they depend only on the class
of equivalence of the symbols and we want to show that they are equal,
i.e. $\mathcal{N}_{H}=\mathcal{C}_{H}$.

Let us denote $F\rightarrow S^{2n}$ the smooth vector bundle of rank
$r$ defined from $H$ in (\ref{eq:F}). We will construct a new symbol
in the same equivalence class, so having the same indices $\mathcal{N}_{H},\mathcal{C}_{H}$,
but that will be easier to handle to show that $\mathcal{N}_{H}=\mathcal{C}_{H}$.
Let $g:S^{2n-1}\rightarrow U\left(r\right)$ be the clutching function
on the equator of the bundle $F$, as defined in (\ref{eq:g}) or
appendix \ref{subsec:Complex-Vector-bundles}. We extend $g$ outside
of $S^{2n-1}\subset\mathbb{R}_{x,\xi}^{2n}$ giving a 1-homogeneous
function $\tilde{g}:\mathbb{R}_{x,\xi}^{2n}\rightarrow\mathrm{Mat}\left(\mathbb{C}^{r}\right)$
by
\begin{equation}
\tilde{g}:\left(x,\xi\right)\in\mathbb{R}^{2n}\rightarrow\tilde{g}\left(x,\xi\right):=\left\Vert \left(x,\xi\right)\right\Vert g\left(\frac{\left(x,\xi\right)}{\left\Vert \left(x,\xi\right)\right\Vert }\right)\in\mathrm{Mat}_{r}\left(\mathbb{C}\right).\label{eq:g_tilde}
\end{equation}
Then we define the (new) symbol $H_{\mu}$ as follows. For $\mu\in\mathbb{R},\left(x,\xi\right)\in\mathbb{R}^{2n}$,
let
\begin{equation}
H_{\mu}\left(x,\xi\right):=\left(\begin{array}{cc}
-\mu\mathrm{Id}_{r} & -\tilde{g}\left(x,\xi\right)\\
-\tilde{g}^{\dagger}\left(x,\xi\right) & \mu\mathrm{Id}_{r}
\end{array}\right)\in\mathrm{Herm}\left(\mathbb{C}^{2r}\right).\label{eq:H_mu-1-1}
\end{equation}

\begin{cBoxB}{}
\begin{lem}
\label{lem:There-are-two}There are two eigenvalues of $H_{\mu}\left(x,\xi\right)$
defined in (\ref{eq:H_mu-1-1}), given by $\omega_{\pm}\left(\mu,x,\xi\right)=\pm\left\Vert \left(\mu,x,\xi\right)\right\Vert $,
each with multiplicity $r$. For $\left(\mu,x,\xi\right)\in S^{2n}$,
the eigenspace $F_{-}\left(\mu,x,\xi\right)$ associated to $\omega_{-}\left(\mu,x,\xi\right)=-1$
defines a vector bundle $F_{-}\rightarrow S^{2n}$ of rank $r$ isomorphic
to the initial given vector bundle $F\rightarrow S^{2n}$. 
\end{lem}

\end{cBoxB}

\begin{rem}
Eq.(\ref{eq:H_mu-1-1}) car be related to a more general construction
of a projector from a given vector bundle, see \cite[p.14]{fedosov96}.
\end{rem}

\begin{proof}
For $\left(\mu,x,\xi\right)\in\mathbb{R}^{3}$, we denote $R=\left\Vert \left(\mu,x,\xi\right)\right\Vert $.
Since $g$ is unitary on $S^{2n-1}$, we get that $\tilde{g}^{\dagger}\tilde{g}=\tilde{g}\tilde{g}^{\dagger}=\left\Vert \left(x,\xi\right)\right\Vert ^{2}=R^{2}-\mu^{2}$
and easily check that eigenvalues $\omega^{\pm}$ and eigenvectors
$U_{j}^{\pm}$ defined by $H_{\mu}\left(x,\xi\right)U_{j}^{\pm}=\omega^{\pm}U_{j}^{\pm}$
are given for $j=1,\ldots r$ by 
\[
\omega_{\pm}\left(\mu,x,\xi\right)=\pm R,\qquad U_{j}^{\pm}\left(\mu,x,\xi\right)=\left(\begin{array}{c}
\left(-\mu\pm R\right)\delta_{j}\\
-\tilde{g}^{\dagger}\delta_{j}
\end{array}\right),
\]
where $\delta_{j}:=\left(0,\ldots,\underbrace{1}_{j},0\ldots\right)\in\mathbb{C}^{r}$
denotes the canonical basis vector of $\mathbb{C}^{r}$. So there
are two eigenvalues $\omega_{\pm}\left(\mu,x,\xi\right)$ each with
multiplicity $r$. We denote $F_{\pm}\left(\mu,x,\xi\right):=\mathrm{Vect}\left(U_{j}^{\pm},j\in\left\{ 1\ldots r\right\} \right)\subset\mathbb{C}^{2}$
the associated eigenspaces. We compute that
\[
\left\Vert U_{j}^{\pm}\right\Vert ^{2}=\left(-\mu\pm R\right)^{2}+\sum_{j'}\left|\langle\delta_{j'}|\tilde{g}^{\dagger}\delta_{j}\rangle\right|^{2}=\left(-\mu\pm R\right)^{2}+R^{2}-\mu^{2}=2R\left(R\mp\mu\right).
\]
Since the vectors $\left(U_{j}^{\pm}\right)_{j}$ are orthogonal,
the spectral projector $\pi_{-}$ on $F_{-}$ is given by
\begin{equation}
\pi^{-}=\sum_{j=1}^{r}\frac{1}{\left\Vert U_{j}^{-}\right\Vert ^{2}}U_{j}^{-}\langle U_{j}^{-}|.\rangle\quad:\mathbb{C}^{2r}\rightarrow F_{-}\left(\mu,x,\xi\right).\label{eq:pi_--1}
\end{equation}
Consider $S^{2n}=\left\{ \left(\mu,x,\xi\right)\in\mathbb{R}^{2n+1},R=1\right\} $
the unit sphere in the parameter space,  the northern hemisphere $H_{1}:=\left\{ \left(\mu,x,\xi\right)\in S^{2n},\mu\geq0\right\} $
and southern hemisphere 

\noindent$H_{2}:=\left\{ \left(\mu,x,\xi\right)\in S^{2n},\mu\leq0\right\} $.
For a given $j\in\left\{ 1,\ldots,r\right\} $, the orthogonal projection
of the fixed vector $\left(\begin{array}{c}
\delta_{j}\\
0
\end{array}\right)\in\mathbb{C}^{2r}$ onto $F_{-}\left(\mu,x,\xi\right)$ gives the global section:
\begin{equation}
s_{1}^{\left(j\right)}\left(\mu,x,\xi\right):=\pi_{-}\left(\begin{array}{c}
\delta_{j}\\
0
\end{array}\right)\eq{\ref{eq:pi_--1}}-\frac{1}{2}U_{j}^{-}\label{eq:s1-3}
\end{equation}
We compute $\left\Vert s_{1}^{\left(j\right)}\right\Vert ^{2}=\frac{1}{2}\left(1+\mu\right)$
hence $\left\Vert s_{1}^{\left(j\right)}\right\Vert ^{2}\neq0$ does
not vanish on $H_{1}$. Hence $\left(s_{1}^{\left(j\right)}\right)_{j\in\left\{ 1,\ldots r\right\} }$
is a trivialization of $F_{-}\rightarrow H_{1}$. We consider also
the following trivialization of $F_{-}\rightarrow H_{2}$:
\begin{equation}
s_{2}^{\left(j\right)}\left(\mu,x,\xi\right):=\pi_{-}\left(\begin{array}{c}
0\\
\delta_{j}
\end{array}\right)=\frac{-1}{2\left(1-\mu\right)}\sum_{j'=1}^{r}U_{j'}^{-}\langle\tilde{g}^{\dagger}\delta_{j'}|\delta_{j}\rangle,\label{eq:s2-1}
\end{equation}
We have $\left\Vert s_{2}^{\left(j\right)}\right\Vert ^{2}=\frac{1}{2}\left(1-\mu\right)$
hence $\left\Vert s_{2}^{\left(j\right)}\right\Vert ^{2}\neq0$ on
$H_{2}$ and $\left(s_{2}^{\left(j\right)}\right)_{j\in\left\{ 1,\ldots r\right\} }$
is a trivialization of $F_{-}\rightarrow H_{2}$. We observe that
\begin{align*}
s_{2}^{\left(j\right)} & =\frac{-1}{2\left(1-\mu\right)}\sum_{j'=1}^{r}U_{j'}^{-}\tilde{g}_{j',j}\eq{\ref{eq:s1-3}}\frac{1}{\left(1-\mu\right)}\sum_{j'=1}^{r}\tilde{g}_{j',j}s_{1}^{\left(j\right)}
\end{align*}
Hence on the equator $S^{2n-1}=\left\{ \mu=0,\left(x,\xi\right)\in S^{2n-1}\right\} $
the clutching function $f_{21}:S^{2n-1}\rightarrow U\left(r\right)$
of $F_{-}$ defined by $s_{2}^{\left(j\right)}\left(0,x,\xi\right)=\sum_{k=1}^{r}f_{21}^{\left(j,k\right)}\left(x,\xi\right)s_{1}^{\left(k\right)}\left(0,x,\xi\right)$
is given by $f_{21}\left(x,\xi\right)=\tilde{g}\left(x,\xi\right)=g\left(x,\xi\right)$,
that is the clutching function of $F$. Hence $F_{-}$ and $F$ are
isomorphic.
\end{proof}
From Lemma \ref{lem:There-are-two}, we see that the symbol $H_{\mu}$
in (\ref{eq:H_mu-1-1}) satisfies the assumption \ref{assu:Gap.-On-suppose}.
As in (\ref{eq:def_OpH}) we define the operator
\begin{equation}
\hat{H}_{\mu,\epsilon}\eq{\ref{eq:def_OpH}}\mathrm{Op}_{\epsilon}\left(H_{\mu}\right)\eq{\ref{eq:H_mu-1-1}}\left(\begin{array}{cc}
-\mu\mathrm{Id} & -\mathrm{Op}_{\epsilon}\left(\tilde{g}\right)\\
-\mathrm{Op}_{\epsilon}\left(\tilde{g}\right)^{\dagger} & \mu\mathrm{Id}
\end{array}\right)\quad\in\mathrm{Herm}\left(L^{2}\left(\mathbb{R}^{n}\right)\otimes\mathbb{C}^{2r}\right).\label{eq:H_hat}
\end{equation}
and from Theorem \ref{prop:Avec-l'hypoth=0000E8se-,} we can define
the spectral index $\mathcal{N}_{H}$ in (\ref{eq:def_N}).

\begin{cBoxB}{}
\begin{lem}
The operator $\mathrm{Op}_{\epsilon}\left(\tilde{g}\right)\in\mathrm{Herm}\left(L^{2}\left(\mathbb{R}^{n}\right)\otimes\mathbb{C}^{r}\right)$
is \href{https://en.wikipedia.org/wiki/Fredholm_operator}{Fredholm}
with index
\begin{equation}
\mathrm{Ind}\left(\mathrm{Op}_{\epsilon}\left(\tilde{g}\right)\right)=\mathcal{N}_{H}.\label{eq:ind1}
\end{equation}
\end{lem}

\end{cBoxB}

\begin{proof}
For simplicity of notation, we denote the operator $A:=\mathrm{Op}_{\epsilon}\left(\tilde{g}\right)$.
Since $\tilde{g}^{\dagger}\tilde{g}=\tilde{g}\tilde{g}^{\dagger}=\left\Vert \left(x,\xi\right)\right\Vert ^{2}$
we see that $A$ is elliptic hence Fredholm \cite[thm3 p.185]{booss_85},
with index \cite[thm2 p.16]{booss_85}
\begin{equation}
\mathrm{Ind}A=\mathrm{dim}\mathrm{Ker}A-\mathrm{dim}\mathrm{Ker}A^{\dagger}.\label{eq:indA}
\end{equation}

Since $\langle u|A^{\dagger}Au\rangle=\left\Vert Au\right\Vert ^{2}\geq0$,
we have that $A^{\dagger}A$ has discrete and positive spectrum denoted
$A^{\dagger}A=\sum_{k\in\mathbb{N}^{*}}\lambda_{k}\pi_{k}$, with
positive eigenvalues $0<\lambda_{1}\leq\lambda_{2}\leq\ldots$ and
$\pi_{k}$ being the spectral projector associated to $\lambda_{k}$.
We denote $\pi_{0}$ the projector on $\mathrm{Ker}A$. Similarly
we denote $AA^{\dagger}=\sum_{k\in\mathbb{N}^{*}}\lambda'_{k}\pi'_{k}$
and $\pi'_{0}$ the projector on $\mathrm{Ker}A^{\dagger}$. In fact
for a given $k>0$, we have $\lambda_{k}'=\lambda_{k}>0$ and $\pi'_{k}=\frac{1}{\lambda_{k}}A\pi_{k}A^{\dagger}$,
$\pi{}_{k}=\frac{1}{\lambda_{k}}A^{\dagger}\pi'_{k}A$, because $\mathrm{Tr}\left(A\pi_{k}A^{\dagger}\right)=\mathrm{Tr}\left(A^{\dagger}A\pi_{k}\right)=\lambda_{k}>0$.
For $k>0$, we have the isomorphism $A:\mathrm{Im}\pi_{k}\rightarrow\mathrm{Im}\pi'_{k}$
and $A^{\dagger}:\mathrm{Im}\pi'_{k}\rightarrow\mathrm{Im}\pi_{k}$.
If $\left(e_{l}\right)_{l=1\ldots\mathrm{dim}\mathrm{Im}\pi_{k}}$
is an orthonormal basis of $\mathrm{Im}\pi_{k}$ then $\frac{1}{\sqrt{\lambda_{k}}}\left(Ae_{l}\right)_{l},\left(e_{l}\right)_{l}$
is an orthonormal basis of $\mathrm{Im}\pi'_{k}\oplus\mathrm{Im}\pi{}_{k}$.
In this basis, the operator $\hat{H}_{\mu,\epsilon}$ is represented
by the matrix
\[
\hat{H}_{\mu,\epsilon}\equiv\left(\begin{array}{cc}
-\mu & -\sqrt{\lambda_{k}}\\
-\sqrt{\lambda_{k}} & \mu
\end{array}\right)
\]
The eigenvalues of this matrix are $\omega_{k}^{\pm}=\pm\left(\mu^{2}+\lambda_{k}\right)^{1/2}$
and never vanish for any $\mu\in\mathbb{R}$, since $\lambda_{k}>0$.
Additionally, for $k=0$, we have $\lambda_{0}=0$, hence $\hat{H}_{\mu,\epsilon}\equiv\left(\begin{array}{cc}
-\mu & 0\\
0 & \mu
\end{array}\right)$ has eigenvalue $-\mu$ with multiplicity $\mathrm{rank}\pi'_{0}$,
and eigenvalue $\mu$ with multiplicity $\mathrm{rank}\pi_{0}$. As
a function of $\mu\in\mathbb{R}$, these eigenvalues vanish transversely
for $\mu=0$, as on Figure \ref{fig:Spectre-de-().} and we get the
index $\mathcal{N}_{H}=\mathrm{rank}\pi{}_{0}-\mathrm{rank}\pi'{}_{0}$.
Consequently
\[
\mathcal{N}_{H}=\mathrm{rank}\pi{}_{0}-\mathrm{rank}\pi'{}_{0}=\mathrm{dim}\mathrm{Ker}A-\mathrm{dim}\mathrm{Ker}A^{\dagger}\eq{\ref{eq:indA}}\mathrm{Ind}A.
\]
\end{proof}
The index Theorem on Euclidean space of Fedosov-Hörmander given in
\cite[thm 7.3 p. 422]{hormander1979weyl} or Eq. (\ref{eq:index_formula_Euclidean})
gives
\begin{equation}
\mathrm{Ind}\left(\mathrm{Op}_{\epsilon}\left(\tilde{g}\right)\right)=\mathcal{C}_{H}.\label{eq:ind2}
\end{equation}
So we conclude that $\mathcal{N}_{H}\eq{\ref{eq:ind1},\ref{eq:ind2}}\mathcal{C}_{H}$.

\subsection{Some models with topological contact without exchange of states}

In Section \ref{sec:The-model}, we have seen a model constructed
from a symbol $H_{\mu}\left(x,\xi\right)$ on a phase space $\left(x,\xi\right)\in\mathbb{R}^{2n}$
(i.e. $n$ degrees of freedom) and parameter $\mu\in\left(-2,2\right)$,
with a spectral gap for $\mu<-1$ and $\mu>1$ and with a \textbf{spectral
index} $\mathcal{N}\in\mathbb{Z}$ that counts the exchange of discrete
energy eigenvalues (or states) between \textbf{two energy bands,}
as the parameter $\mu$ increases (energy bands are the spectrum below
the gap and the spectrum above the gap). We have seen that $\mathcal{N}$
is equal to the Chern index $\mathcal{C}$ of a vector bundle $F\rightarrow S^{2n}$
of rank $r$ that is defined from the symbol.
\begin{itemize}
\item If the \textbf{vector bundle $F$ is trivia}l, it means that the two
bands are not ``topologically coupled'' and we can perturb continuously
the symbol $\left(H_{\mu}\right)_{\mu}$ so that the gap may exist
for every values of $\mu\in\left(-2,2\right)$, i.e. we can ``\textbf{open
the gap}''.
\item If the \textbf{vector bundle $F$ is non trivial}, it means that the
two bands are ``topologically coupled'' with a ``\textbf{topological
contact}'' and we \textbf{can not ``open the gap''}, or remove
the contact between the two bands.
\end{itemize}
If $\mathcal{N}=\mathcal{C}\neq0$ then the bundle $F$ is not trivial
and we can not open the gap, since some energy levels pass through
it, and this situation cannot be changed by continuous perturbations.
From \textbf{Bott's theorem} \ref{thm:def_indice_topologique}, if
$r=\mathrm{rank}\left(F\right)\geq n$ then $\mathcal{C}\in\mathbb{Z}$
characterizes the topology of $F$. In other words, if $r\geq n$
then $\mathcal{C}=\mathcal{N}=0\Leftrightarrow F\text{ is trivial}$.

However for vector bundles $F$ of smaller ranks, $r<n$ this is not
always true (we only have the obvious fact $F\text{ is trivial}\Rightarrow\mathcal{C}=\mathcal{N}=0$
but not the converse). There exist some non trivial bundles $F\rightarrow S^{2n}$
with Chern index $\mathcal{C}\left(F\right)=0$. From table \ref{tab:Groupes-d'=0000E9quivalences-de},
the simplest example is for $F\rightarrow S^{6}$, i.e. $n=3$ degrees
of freedom, with rank $r=2$, because $\mathrm{Vect}^{2}\left(S^{6}\right)=\mathbb{Z}_{2}=\left\{ 0,1\right\} $.
Suppose for example that $F\rightarrow S^{6}$ is non trivial and
with topological class $\left[F\right]=1\in\mathrm{Vect}^{2}\left(S^{6}\right)=\mathbb{Z}_{2}$.
It means that the two bands have a ``topological contact'', i.e.
that we can not open the gap. Nevertheless $\mathcal{N}=\mathcal{C}=0$,
i.e. there is no exchange of states between the two bands at the contact
(since the spectrum is discrete, there is some small gap that goes
to zero as $\epsilon\rightarrow0$). See figure below.
\begin{center}
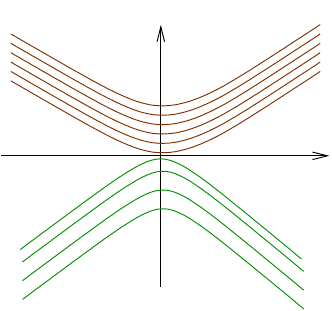
\par\end{center}

If one adds a second similar contact (at some other value of $\mu$),
then since $1+1=0$ in $\mathbb{Z}_{2}$, the result is that the two
contact annihilate themselves and one can finally ``open the gap''.
See figure below.
\begin{center}
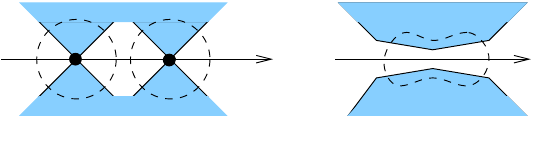
\par\end{center}

These kind of phenomena may occur with vector bundles $F\rightarrow S^{2n}$
that are in the ``\textbf{non stable range}'', where the homotopy
groups are very complicated, see the appendix \ref{subsec:Vector-bundles-in}.

For a different example of the role of topology in spectral phenomena,
in the paper \cite{fred-boris02} there is a simple model used molecular
physics, for which the energy bands are topological coupled and associated
to a rank $2$ vector bundle that can not be splitted into two rank
1 vector bundles. This involves Chern numbers $C_{1},C_{2}$ and shows
the manifestation of algebraic topology in quantum mechanics of molecules
or more generally quantum interacting systems.

\section{\label{subsec:Spectral-flow-and}Spectral flow and index formula
for quantum waves in molecules}

References for this Section are \cite{fred-boris,fred-boris-02bis,fred-boris01,fred_habilitation_06}.

\subsection{Introduction}

\begin{figure}
\begin{centering}
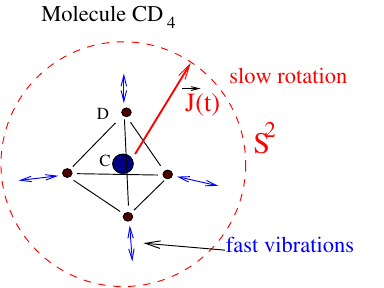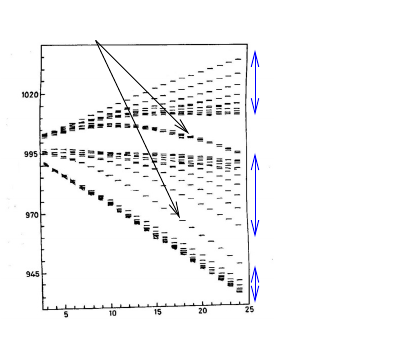
\par\end{centering}
\caption{\label{fig:Niveaux-d'=0000E9nergie-de}Energy levels (in $cm^{-1}$)
of the molecule $CD_{4}$ (carbon with 4 deuterium atoms) as a function
of the total angular momentum $J\in\mathbb{N}$ (rotation energy and
which is a preserved quantity). The fine structure of the spectrum
corresponds to the slow rotation motion and the broad structure to
the faster vibration motion. There are groups of levels and levels
that pass between these groups. The index formula gives the exact
values of number of levels $\mathcal{N}_{j}$ in each group \cite{fred-boris,fred-boris-02bis,fred-boris01,fred_habilitation_06}.}
\end{figure}

A small molecule is a set of atoms (electrons and nuclei) and can
be considered as an isolated but complex quantum system since many
degrees of freedom interact strongly on different time scales: the
electrons that are light evolve on very short scales of time $\tau_{e}\in\left[10^{-16}s,10^{-15}s\right]$,
which are small compared to the time scales of the vibration motion
of the atoms $\tau_{\mathrm{vib}}\in\left[10^{-15}s,10^{-14}s\right]$,
themselves small compared to the slower rotation of the molecule $\tau_{\mathrm{rot.}}\in\left[10^{-12}s,10^{-10}s\right]$.
In quantum mechanics the state of the molecule is described by a multivariate
``quantum wave function'' and a stationary state of the molecule
corresponds to an eigenfunction of the Hamiltonian operator. The corresponding
eigenvalue is the energy of this state. If the molecule is sufficiently
isolated from its environment, one can experimentally measure its
quantum energy levels (discrete spectrum) by \href{https://en.wikipedia.org/wiki/Spectroscopy}{spectroscopy}.
These quantum energy levels correspond to stationary collective states
of all the internal interactions between all these different degrees
of freedom. It seems to be (and it is) a very complicated problem,
but these different time scales allows to approximate the dynamics
by some ``fiber bundle description''. This is called the adiabatic
theory. In simple words the fast motion phase space is a fiber bundle
over the slow motion phase space. In quantum mechanics (or more generally
in wave mechanics, like optics, acoustics ...) one has to quantize
this fiber bundle description. Although this adiabatic approach does
not solve completely the problem it gives a geometric description
and some rough (and robust under perturbations) first description
of the spectrum can be obtained from topological properties of these
fiber bundles. This is the subject of this Section. See figure \ref{fig:Niveaux-d'=0000E9nergie-de}.

\subsection{\label{subsec:Mod=0000E8le-simple-(forme}Simple model (normal form)}

References for this section: \cite{fred-boris,fred-boris01}. The
following model not only is relevant in molecular physics to illustrate
the spectral behavior of rotational / vibrational (slow / fast) energy
levels of nuclei, but also plays an important role in the general
theory because it is an \textquotedbl elementary topological normal
form\textquotedbl .

Let $\mu\in\mathbb{R}$ be a parameter that is fixed. Let $\left(x,\xi\right)\in T^{*}\mathbb{R}\equiv\mathbb{R}\times\mathbb{R}$
\textquotedbl slow variables\textquotedbl{} on phase space $\mathbb{R}^{2}$.
We introduce the \textquotedbl symbol\textquotedbl
\begin{equation}
H_{\mu}\left(x,\xi\right):=\left(\begin{array}{cc}
-\mu & x+i\xi\\
x-i\xi & +\mu
\end{array}\right)\in\mathrm{Herm}\left(\mathbb{C}^{2}\right).\label{eq:symbole_H_mu}
\end{equation}
 We will call $\mathcal{H}=\mathbb{C}^{2}$ the fast Hilbert space.
The space of \textquotedbl slow Hilbert\textquotedbl{} is $L^{2}\left(\mathbb{R}\right)$
and corresponds to the quantification of the phase space $T^{*}\mathbb{R}$
of \textquotedbl slow variables\textquotedbl{} $x,\xi$ and replace
them by quantum operators. Let $\epsilon>0$, the \textquotedbl adiabatic
parameter\textquotedbl{} and set
\begin{equation}
\hat{H}_{\mu}:=\mathrm{Op}_{\epsilon}\left(H_{\mu}\right):=\left(\begin{array}{cc}
-\mu\,\mathrm{Id} & \hat{x}+i\hat{\xi}\\
\hat{x}-i\hat{\xi} & \mu\,\mathrm{Id}
\end{array}\right)\in\mathrm{Herm}\left(L^{2}\left(\mathbb{R}_{x}\right)\otimes\mathbb{C}^{2}\right)\label{eq:H_mu}
\end{equation}
where $\mathrm{Id}:L^{2}\left(\mathbb{R}\right)\rightarrow L^{2}\left(\mathbb{R}\right)$,
$\hat{\xi}:=\mathrm{Op}_{\epsilon}\left(\xi\right):=-i\epsilon\frac{d.}{dx}\in\mathrm{Herm}\left(L^{2}\left(\mathbb{R}\right)\right)$,
and $\hat{x}$ is the multiplication operator $x$ in $L^{2}\left(\mathbb{R}_{x}\right)$,
see Section \ref{sec:Quantification,-op=0000E9rateurs-pseud} for
more details.
\begin{rem}
In \cite{fred-boris,fred-boris01} it is shown how this normal form
gives a micro-local description of the interaction between the fast
vibration motion and the slow rotational motion of the molecule of
Figure \ref{fig:Niveaux-d'=0000E9nergie-de}. In few words, $\left(x,\xi\right)$
are local coordinates on the sphere $S^{2}$ of rotation in a vicinity
of a point where two spectral bands have a contact, and the $\mathbb{C}^{2}$
space describes the quantum dynamics of the fast vibrations by restricting
to an effective two level problem.
\end{rem}

\subsubsection{Spectral index $\mathcal{N}$}

In the following Theorem, $\left(\varphi_{n}\right)_{n\in\mathbb{N}}$
is the orthonormal basis of \href{https://en.wikipedia.org/wiki/Hermite_polynomials}{Hermite functions}
of $L^{2}\left(\mathbb{R}\right)$ defined by the Gaussian function
\begin{equation}
\varphi_{0}\left(x\right)=\frac{1}{\left(\pi\epsilon\right)^{1/4}}e^{-\frac{1}{2}\frac{x^{2}}{\epsilon}},\label{eq:phi_0}
\end{equation}
and
\begin{equation}
\varphi_{n+1}=\frac{1}{\sqrt{n+1}}a^{\dagger}\varphi_{n},\qquad a\varphi_{n}=\sqrt{n}\varphi_{n-1},\label{eq:phi_n}
\end{equation}
with the operators (so called \href{https://en.wikipedia.org/wiki/Creation_and_annihilation_operators}{annihilation and creation operators}
from \href{https://en.wikipedia.org/wiki/Quantization_of_the_electromagnetic_field}{quantum optics})
\begin{equation}
a:=\frac{1}{\sqrt{2\epsilon}}\left(\hat{x}+i\hat{\xi}\right),\quad a^{\dagger}:=\frac{1}{\sqrt{2\epsilon}}\left(\hat{x}-i\hat{\xi}\right).\label{eq:a_a+}
\end{equation}

\begin{figure}
\begin{centering}
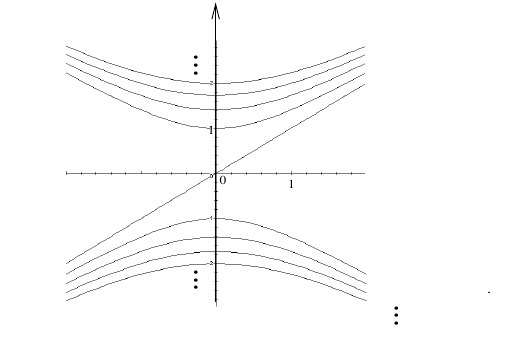
\par\end{centering}
\caption{\label{fig:Spectre-de-().}Spectrum of (\foreignlanguage{french}{\ref{eq:modele}}).\foreignlanguage{french}{}}
\end{figure}

\begin{cBoxB}{}
\begin{prop}
\textbf{\label{prop:Spectre-de-.}<<Spectrum of $\hat{H}_{\mu}$>>}.
For each parameter $\mu\in\mathbb{R}$, the operator $\hat{H}_{\mu}$,
(\ref{eq:H_mu}), has discrete spectrum in $L^{2}\left(\mathbb{R}_{x}\right)\otimes\mathbb{C}^{2}$
given by
\begin{equation}
\hat{H}_{\mu}\phi_{n}^{\pm}=\omega_{n}^{\pm}\phi_{n}^{\pm},\quad n\geq1,\label{eq:modele}
\end{equation}
with for any $n\in\mathbb{N}\backslash\left\{ 0\right\} $,
\begin{align}
\frac{\omega_{n}^{\pm}}{\sqrt{\epsilon}} & =\pm\sqrt{\left(\frac{\mu}{\sqrt{\epsilon}}\right)^{2}+2n}\label{eq:spectre}\\
\phi_{n}^{\pm} & =\left(\begin{array}{c}
\frac{\sqrt{2n\epsilon}}{\mu+\omega_{n}^{\pm}}\varphi_{n-1}\\
\varphi_{n}
\end{array}\right)\nonumber 
\end{align}
and for $n=0$,
\[
\hat{H}_{\mu}\phi_{0}=\omega_{0}\phi_{0},
\]
with 
\[
\omega_{0}=\mu
\]
\[
\phi_{0}=\left(\begin{array}{c}
0\\
\varphi_{0}
\end{array}\right)
\]
Observe that there is 
\begin{equation}
\mathcal{N}=+1\label{eq:N_transite}
\end{equation}
 eigenvalue transiting upwards, for $\mu$ increasing. See figure
\ref{fig:Spectre-de-().}.
\end{prop}

\end{cBoxB}

\begin{rem}
It appears in (\ref{eq:spectre}) that $\sqrt{\epsilon}$ is a natural
parameter of \textquotedbl scaling\textquotedbl . See Section \ref{subsec:Cas-particulier-de}
for a discussion.

For the moment we can not say that (\ref{eq:N_transite}) is a result
of topology. For $\mathcal{N}$ to be recognized as a \textquotedbl topological
index\textquotedbl , it would be necessary for this model to belong
to a set of models and to show that this number $\mathcal{N}=+1$
is model independent (robust by continuous perturbation within this
set). This is done in Section \ref{sec:The-model}.
\end{rem}

\begin{itemize}
\item 
\end{itemize}
\begin{proof}
We will see from (\ref{eq:val_p}) that the operator $\hat{H}_{\mu}$
is \href{https://en.wikipedia.org/wiki/Elliptic_operator}{elliptic}
hence\footnote{We can do without this argument by noticing at the end of the computation
that the found eigenvectors form a basis of the Hilbert space.} $\hat{H}_{\mu}$ has discrete spectrum that we will determine now
by different (but similar) methods.

\paragraph{Method 1:}

This first method is direct, simple and will be used again for the
proof of Theorem \ref{prop:Spectre-de-.-1}. Any vector $\psi\in L^{2}\left(\mathbb{R}\right)\otimes\mathbb{C}^{2}$
is written
\[
\psi=\sum_{n\geq0}\left(\begin{array}{c}
a_{n}\varphi_{n}\\
b_{n}\varphi_{n}
\end{array}\right),\qquad a_{n},b_{n}\in\mathbb{C}.
\]
Put $\omega=\sqrt{\epsilon}\tilde{\omega}$ and $\mu=\sqrt{\epsilon}\tilde{\mu}$.
We have
\[
\hat{H}_{\mu}\psi=\omega\psi\Leftrightarrow\sum_{n\geq0}\left(\begin{array}{cc}
-\mu-\omega & \sqrt{2\epsilon}a\\
\sqrt{2\epsilon}a^{\dagger} & \mu-\omega
\end{array}\right)\left(\begin{array}{c}
a_{n}\varphi_{n}\\
b_{n}\varphi_{n}
\end{array}\right)=0
\]
\[
\Leftrightarrow\begin{cases}
\sum_{n\geq0}-a_{n}\left(\tilde{\mu}+\tilde{\omega}\right)\varphi_{n}+b_{n}\sqrt{2}\sqrt{n}\varphi_{n-1}=0\\
\sum_{n\geq0}a_{n}\sqrt{2}\sqrt{n+1}\varphi_{n+1}+b_{n}\left(\tilde{\mu}-\tilde{\omega}\right)\varphi_{n}=0
\end{cases}
\]

\[
\Leftrightarrow\begin{cases}
-a_{n}\left(\tilde{\mu}+\tilde{\omega}\right)+b_{n+1}\sqrt{2}\sqrt{n+1}=0,\qquad\forall n\geq0\\
a_{n-1}\sqrt{2}\sqrt{n}+b_{n}\left(\tilde{\mu}-\tilde{\omega}\right)=0
\end{cases}
\]

If $\tilde{\omega}=-\tilde{\mu}$ then there is no non zero solution.

If $\tilde{\omega}\neq-\tilde{\mu}$ then
\[
\begin{cases}
a_{n'}=\frac{\sqrt{2\left(n'+1\right)}}{\tilde{\mu}+\tilde{\omega}}b_{n'+1},\qquad\forall n'\geq0\\
b_{n'}\left(-\frac{2n'}{\tilde{\mu}+\tilde{\omega}}+\left(\tilde{\omega}-\tilde{\mu}\right)\right)=0
\end{cases}
\]
Let $n\geq0$. If $b_{n}\neq0$ then
\[
\tilde{\mu}^{2}-\tilde{\omega}^{2}=-2n\Leftrightarrow\tilde{\omega}=\pm\sqrt{\tilde{\mu}^{2}+2n},\quad n\geq0.
\]

Hence
\begin{itemize}
\item If $n=0$, we have $\tilde{\omega}_{0}=\tilde{\mu}$ (because $\tilde{\omega}=-\tilde{\mu}$
is excluded) and $b_{0}=1$ gives $b_{n'}=0$ for $n'\geq1$ and $a_{n'}=0$
for $n'\geq0$.
\item For $n\geq1$, we have $\tilde{\omega}_{n}=\pm\sqrt{\tilde{\mu}^{2}+2n}$
and $b_{n}=1$ giving $b_{n'}=0$ for $n'\neq n$ and $a_{n-1}=\frac{\sqrt{2n}}{\tilde{\mu}+\tilde{\omega}}$
and $a_{n'}=0$ for $n'\neq n-1$.
\end{itemize}

\paragraph{Method 2:}

This second method explicitly uses a \textquotedbl symmetry\textquotedbl{}
of the problem. We first calculate the spectrum of $\hat{H}_{\mu}^{2}$
and then we diagonalize $\hat{H}_{\mu}$ in the eigenspace obtained.
Observe that
\begin{equation}
\left[\hat{x},\hat{\xi}\right]=\hat{x}\hat{\xi}-\hat{\xi}\hat{x}=i\epsilon\mathrm{Id}.\label{eq:commut}
\end{equation}
and
\begin{align}
a^{\dagger}a\eq{\ref{eq:a_a+}} & \frac{1}{2\epsilon}\left(\hat{x}-i\hat{\xi}\right)\left(\hat{x}+i\hat{\xi}\right)\nonumber \\
= & \frac{1}{2\epsilon}\left(\hat{x}^{2}+\hat{\xi}^{2}+i\left(\hat{x}\hat{\xi}-\hat{\xi}\hat{x}\right)\right)\nonumber \\
\eq{\ref{eq:commut}} & \frac{1}{2\epsilon}\left(\hat{x}^{2}+\hat{\xi}^{2}-\epsilon\mathrm{Id}\right)\label{eq:a+a}
\end{align}
\begin{equation}
a^{\dagger}a\varphi_{n}\eq{\ref{eq:phi_n}}n\varphi_{n}.\label{eq:spectre_a+a}
\end{equation}
We have
\begin{align*}
\hat{H}_{\mu}^{2}\eq{\ref{eq:H_mu}} & \left(\begin{array}{cc}
\mu^{2}+\hat{x}^{2}+\hat{\xi}^{2}-i\left(\hat{x}\hat{\xi}-\hat{\xi}\hat{x}\right) & 0\\
0 & \mu^{2}+\hat{x}^{2}+\hat{\xi}^{2}+i\left(\hat{x}\hat{\xi}-\hat{\xi}\hat{x}\right)
\end{array}\right)\\
\eq{\ref{eq:commut}} & \left(\begin{array}{cc}
\mu^{2}+\hat{x}^{2}+\hat{\xi}^{2}+\epsilon & 0\\
0 & \mu^{2}+\hat{x}^{2}+\hat{\xi}^{2}-\epsilon
\end{array}\right)\\
\eq{\ref{eq:a+a}} & \left(\begin{array}{cc}
\mu^{2}+2\epsilon a^{\dagger}a+2\epsilon & 0\\
0 & \mu^{2}+2\epsilon a^{\dagger}a
\end{array}\right)=\left(2\epsilon a^{\dagger}a+\mu^{2}\right)\otimes\mathrm{Id}_{\mathbb{C}^{2}}+\mathrm{Id}_{L^{2}}\otimes\left(\begin{array}{cc}
2\epsilon & 0\\
0 & 0
\end{array}\right)
\end{align*}
We deduce that the spectrum of $\hat{H}_{\mu}^{2}$ consists of eigenvalues
$\lambda_{n}=\omega_{n}^{2}=\left(2\epsilon n+\mu^{2}\right),n\geq0$,
and the associated eigenspace $E_{n}$ is

\[
E_{n}=\mathrm{Span}\left\{ \left(\begin{array}{c}
\varphi_{n-1}\\
0
\end{array}\right),\left(\begin{array}{c}
0\\
\varphi_{n}
\end{array}\right)\right\} \quad:\text{ if }n\geq1,
\]

\[
E_{0}=\mathrm{Span}\left\{ \left(\begin{array}{c}
0\\
\varphi_{0}
\end{array}\right)\right\} .
\]
It remains to diagonalize $\hat{H}_{\mu}$ in each space $E_{n}$.
For $n=0$, we observed that
\[
\hat{H}_{\mu}\left(\begin{array}{c}
0\\
\varphi_{0}
\end{array}\right)=\mu\left(\begin{array}{c}
0\\
\varphi_{0}
\end{array}\right),
\]
and for $n\geq1$,
\[
\hat{H}_{\mu}\left(\begin{array}{c}
\varphi_{n-1}\\
0
\end{array}\right)=\left(\begin{array}{c}
-\mu\varphi_{n-1}\\
\sqrt{2\epsilon n}\varphi_{n}
\end{array}\right),\qquad\hat{H}_{\mu}\left(\begin{array}{c}
0\\
\varphi_{n}
\end{array}\right)=\left(\begin{array}{c}
\sqrt{2\epsilon n}\varphi_{n-1}\\
\mu\varphi_{n}
\end{array}\right).
\]
So in the basis of $E_{n}$, $\hat{H}_{\mu}$ is represented by the
matrix
\[
\hat{H}_{\mu/E_{n}}\equiv\left(\begin{array}{cc}
-\mu & \sqrt{2\epsilon n}\\
\sqrt{2\epsilon n} & \mu
\end{array}\right),
\]
whose eigenvalues and eigenvectors are
\[
\omega_{n}^{\pm}=\pm\sqrt{\mu^{2}+2\epsilon n},
\]
and eigenvectors\footnote{With online \href{https://www-fourier.ujf-grenoble.fr/~parisse/xcasfr.html}{xcas},
write H:={[}{[}-mu,a{]},{[}a,mu{]}{]};eigenvals(H);eigenvects(H);}
\[
U_{n}^{\pm}=\left(\begin{array}{c}
\left(\omega_{n}^{\pm}-\mu\right)\varphi_{n-1}\\
\sqrt{2\epsilon n}\varphi_{n}
\end{array}\right)=\left(\begin{array}{c}
\frac{\sqrt{2n\epsilon}}{\mu+\omega_{n}^{\pm}}\varphi_{n-1}\\
\varphi_{n}
\end{array}\right).
\]
\end{proof}

\subsubsection{Topological Chern Index $\mathcal{C}$}

We can first consult the section \ref{sec:Espaces-fibr=0000E9s-vectoriels}
which introduces in simple terms the notion of topology of a complex
vector bundle of rank 1 on the sphere $S^{2}$.

\begin{cBoxB}{}
\begin{prop}
\textbf{\label{prop:Aspects-topologiques-du-1}<<Topological aspects
of the symbol>>}. The eigenvalues of the matrix $H_{\mu}\left(x,\xi\right)\in\mathrm{Herm}\left(\mathbb{C}^{2}\right)$
, Eq.(\ref{eq:symbole_H_mu}), are
\begin{equation}
\omega_{\pm}\left(\mu,x,\xi\right)=\pm\sqrt{\mu^{2}+x^{2}+\xi^{2}}\label{eq:val_p}
\end{equation}
There is therefore a degeneracy $\omega_{+}=\omega_{-}$ for $\left(\mu,x,\xi\right)=\left(0,0,0\right)$.
For $\left(\mu,x,\xi\right)\in S^{2}=\left\{ \left(\mu,x,\xi\right)\in\mathbb{R}^{3},\left|\left(\mu,x,\xi\right)\right|=1\right\} $,
i.e. on the unit sphere in the parameter space, the eigenspace $F_{-}\left(\mu,x,\xi\right)\subset\mathbb{C}^{2}$
associated with the eigenvalue $\omega_{-}$ defines a complex vector
bundle of rank 1, denoted $F_{-}$. Its isomorphism class is characterized
by the topological Chern index
\[
\mathcal{C}\left(F_{-}\right)=+1.
\]
Similarly for eigenvalue $\omega_{+}$, 
\[
\mathcal{C}\left(F_{+}\right)=-1.
\]
\end{prop}

\end{cBoxB}

\begin{figure}
\begin{centering}
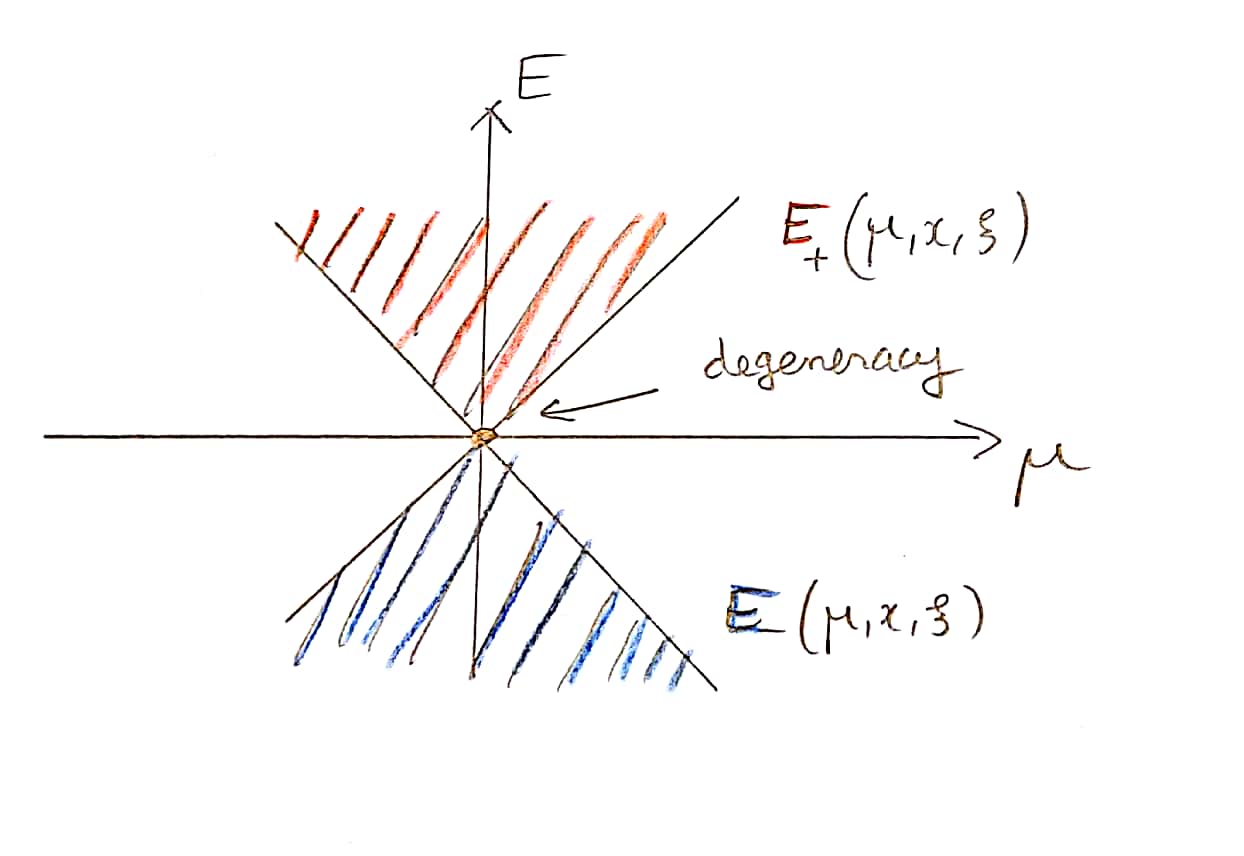
\par\end{centering}
\caption{We have $\omega^{-}\left(\mu,x,\xi\right)\protect\leq-\left|\mu\right|$,
$\omega^{+}\left(\mu,x,\xi\right)\protect\geq\left|\mu\right|$. The
red domain represents the possible values of $\omega_{-}\left(\mu,x,\xi\right)$
with $\mu$ fixed and $\left(x,\xi\right)\in\mathbb{R}^{2}$. Similarly,
the blue domain represent $\omega_{-}\left(\mu,x,\xi\right)$. The
degeneracy is at $\left(\mu,x,\xi\right)=\left(0,0,0\right)$.}
\end{figure}

\begin{itemize}
\item 
\end{itemize}
\begin{proof}
We will calculate the index $\mathcal{C}$ by two equivalent methods,
see section \ref{sec:Espaces-fibr=0000E9s-vectoriels}.

\paragraph{Method 1 (with a clutching function)}

We have 
\[
p\left(\omega\right):=\mathrm{det}\left(\omega\mathrm{Id}-H_{\mu}\left(x,\xi\right)\right)\eq{\ref{eq:symbole_H_mu}}\mathrm{det}\left(\begin{array}{cc}
\omega+\mu & -\left(x+i\xi\right)\\
-\left(x-i\xi\right) & \omega-\mu
\end{array}\right)=\omega-\left(\mu^{2}+x^{2}+\xi^{2}\right)
\]
hence $p\left(\omega\right)=0$ gives eigenvalues $\omega_{\pm}=\pm r$
with $r:=\sqrt{\mu^{2}+x^{2}+\xi^{2}}$, i.e. Eq. (\ref{eq:val_p}).
The eigenvectors of $H_{\mu}$ are respectively\footnote{In \href{https://www-fourier.ujf-grenoble.fr/~parisse/xcasfr.html}{xcas online},
write: \texttt{H:={[}{[}-mu,x+i{*}xi{]},{[}x-i{*}xi,mu{]}{]}; eigenvals(H);
eigenvects(H);}}
\begin{equation}
U_{+}=\left(\begin{array}{c}
-\mu+r\\
x-i\xi
\end{array}\right),\quad U_{-}=\left(\begin{array}{c}
-\mu-r\\
x-i\xi
\end{array}\right),\label{eq:U_+_U_-}
\end{equation}
i.e. $H_{\mu}\left(x,\xi\right)U_{\pm}=\omega_{\pm}U_{\pm}$. Write
$F_{\pm}\left(\mu,x,\xi\right):=\mathrm{Vect}\left(U_{\pm}\right)\subset\mathbb{C}^{2}$
the associated eigenspaces. The spectral projector $\pi_{-}$ on $F_{-}$
is
\begin{equation}
\pi_{-}=\frac{1}{\left\Vert U_{-}\right\Vert ^{2}}U_{-}\langle U_{-}|.\rangle\quad:\mathbb{C}^{2}\rightarrow F_{-}\left(\mu,x,\xi\right).\label{eq:pi_-}
\end{equation}
Consider $S^{2}=\left\{ \left(\mu,x,\xi\right)\in\mathbb{R}^{3},r=\left|\left(\mu,x,\xi\right)\right|=1\right\} $
the unit sphere in the parameter space and the northern and southern
hemispheres $H_{1}:=\left\{ \left(\mu,x,\xi\right)\in S^{2},\mu\geq0\right\} $,
$H_{2}:=\left\{ \left(\mu,x,\xi\right)\in S^{2},\mu\leq0\right\} $.
The projection of the fixed vector $\left(\begin{array}{c}
1\\
0
\end{array}\right)\in\mathbb{C}^{2}$ on $F_{-}$ gives the global section:
\begin{equation}
s_{1}\left(\mu,x,\xi\right):=\pi_{-}\left(\begin{array}{c}
1\\
0
\end{array}\right)\eq{\ref{eq:pi_-}}\frac{\left(-\mu-1\right)}{\left(\left(\mu+1\right)^{2}+x^{2}+\xi^{2}\right)}\left(\begin{array}{c}
-\mu-1\\
x-i\xi
\end{array}\right).\label{eq:s1}
\end{equation}
We have $\left\Vert s_{1}\right\Vert ^{2}=\frac{\left(\mu+1\right)^{2}}{\left(\left(\mu+1\right)^{2}+x^{2}+\xi^{2}\right)}=\frac{1+\mu}{2}$
hence $\left\Vert s_{1}\right\Vert ^{2}\neq0$ on $H_{1}$. Hence
$s_{1}$ is a trivialization of $F_{-}\rightarrow H_{1}$. We consider
also the following trivialization of $F_{-}\rightarrow H_{2}$:
\begin{equation}
s_{2}\left(\mu,x,\xi\right):=\pi_{-}\left(\begin{array}{c}
0\\
1
\end{array}\right)=\frac{\left(x+i\xi\right)}{\left(\left(\mu+1\right)^{2}+x^{2}+\xi^{2}\right)}\left(\begin{array}{c}
-\mu-1\\
x-i\xi
\end{array}\right),\label{eq:s2}
\end{equation}
We have $\left\Vert s_{2}\right\Vert ^{2}=\frac{\left(x^{2}+\xi^{2}\right)}{\left(\left(\mu+1\right)^{2}+x^{2}+\xi^{2}\right)}=\frac{1-\mu}{2}$
hence $\left\Vert s_{2}\right\Vert ^{2}\neq0$ on $H_{2}$. The clutching
function on the equator $S^{1}=\left\{ \mu=0,x+i\xi=e^{i\theta},\theta\in[0,2\pi[\right\} $
is defined by 
\begin{align*}
s_{2}\left(\theta\right) & =f_{21}\left(\theta\right)s_{1}\left(\theta\right)\\
\Leftrightarrow & \left(x+i\xi\right)\left(\begin{array}{c}
-1\\
x-i\xi
\end{array}\right)=-f_{21}\left(\theta\right)\left(\begin{array}{c}
-1\\
x-i\xi
\end{array}\right)\\
\Leftrightarrow & f_{21}\left(\theta\right)=-e^{i\theta}.
\end{align*}
The degree of the function $f_{21}:\theta\in S^{1}\rightarrow f_{21}\left(\theta\right)=-e^{i\theta}\in U\left(1\right)\equiv S^{1}$
is $\mathcal{C}=\mathrm{deg}\left(f_{21}\right)=+1$.

\paragraph{Method 2 (indices of zeroes of a global section)}

We consider the global section of $F_{-}\rightarrow S^{2}$, $s_{2}\left(\mu,x,\xi\right)$
given in (\ref{eq:s2}) that vanishes at $\left(\mu,x,\xi\right)=\left(1,0,0\right)$
from $\left\Vert s_{2}\right\Vert ^{2}=\frac{1-\mu}{2}$. In a neighborhood
of this point in first order of $\left(x,\xi\right)$ and writing
$x+i\xi=\epsilon e^{i\theta}\in\mathbb{C}$ with $\epsilon\ll1$,
we have
\begin{align}
s_{2}\left(\mu,x,\xi\right)\eq{\ref{eq:s2}} & \frac{\left(x+i\xi\right)}{4}\left(\begin{array}{c}
-2\\
0
\end{array}\right)+o\left(x,\xi\right)\label{eq:s1-1}\\
= & -\frac{\epsilon}{2}e^{i\theta}\left(\begin{array}{c}
1\\
0
\end{array}\right)+o\left(\epsilon\right)
\end{align}
It appears the function $e^{i\theta}\rightarrow e^{i\theta}$ whose
index (or <<winding number>>) is $\mathcal{C}=+1$.

\paragraph{Method 3 (with curvature integral)}

We will use spherical coordinates $\left(\theta,\varphi\right)\in]0,\pi[\times]0,2\pi[$
on the sphere $S^{2}$ defined by
\begin{align*}
x+i\xi & =\sin\theta e^{i\varphi}\\
\mu & =\cos\theta.
\end{align*}
A unit vector $v\left(\theta,\varphi\right)$ in the fiber $F_{-}$
over the unit sphere $S^{2}$ (except at points where $\left(1+\cos\theta\right)=0$)
is given by
\[
v\eq{\ref{eq:U_+_U_-}}\frac{U_{-}}{\left\Vert U_{-}\right\Vert }=\frac{1}{\sqrt{2\left(1+\mu\right)}}\left(\begin{array}{c}
-\mu-1\\
x-i\xi
\end{array}\right)=\frac{1}{\sqrt{2\left(1+\cos\theta\right)}}\left(\begin{array}{c}
-\cos\theta-1\\
\sin\theta e^{-i\varphi}
\end{array}\right),
\]
We will use the curvature integral formula (\ref{eq:curvature_integral_S2})
that gives
\[
\mathcal{C}=\frac{1}{2\pi}\int_{S^{2}}i\Omega
\]
with the curvature two form
\begin{align*}
\Omega\eq{\ref{eq:iOmega}}\langle dv\wedge|dv\rangle & =\left(\langle\frac{\partial v}{\partial\theta}|\frac{\partial v}{\partial\varphi}\rangle-\langle\frac{\partial v}{\partial\varphi}|\frac{\partial v}{\partial\theta}\rangle\right)d\theta\wedge d\varphi\\
 & =2i\mathrm{Im}\left(\langle\frac{\partial v}{\partial\theta}|\frac{\partial v}{\partial\varphi}\rangle\right)d\theta\wedge d\varphi
\end{align*}
We have
\[
\frac{\partial v}{\partial\theta}=\frac{\sin\theta}{\left(2\left(1+\cos\theta\right)\right)^{3/2}}\left(\begin{array}{c}
-\cos\theta-1\\
\sin\theta e^{-i\varphi}
\end{array}\right)+\frac{1}{\sqrt{2\left(1+\cos\theta\right)}}\left(\begin{array}{c}
\sin\theta\\
\cos\theta e^{-i\varphi}
\end{array}\right)
\]
\[
\frac{\partial v}{\partial\varphi}=\frac{1}{\sqrt{2\left(1+\cos\theta\right)}}\left(\begin{array}{c}
0\\
-i\sin\theta e^{-i\varphi}
\end{array}\right)
\]
\begin{align*}
\mathrm{Im}\left(\langle\frac{\partial v}{\partial\theta}|\frac{\partial v}{\partial\varphi}\rangle\right) & =-\frac{\sin^{3}\theta}{\left(2\left(1+\cos\theta\right)\right)^{2}}-\frac{\cos\theta\sin\theta}{\left(2\left(1+\cos\theta\right)\right)}\\
 & =-\frac{1}{4}\sin\theta
\end{align*}
We get
\begin{align*}
\mathcal{C} & =\frac{1}{2\pi}\int_{S^{2}}i\Omega=-\frac{1}{2\pi}\int_{\theta=0}^{\pi}\int_{\varphi=0}^{2\pi}2\mathrm{Im}\left(\langle\frac{\partial v}{\partial\theta}|\frac{\partial v}{\partial\varphi}\rangle\right)d\theta d\varphi\\
 & =\frac{1}{2}\int_{\theta=0}^{\pi}\sin\theta d\theta=1
\end{align*}
\end{proof}

\subsubsection{Conclusion on the model (\ref{eq:H_mu})}

In the model defined by (\ref{eq:H_mu}), we observe from the symbol,
a vector bundle $F_{-}$ whose index of Chern is $\mathcal{C}\left(F_{-}\right)=+1$
and we observe that there is $\mathcal{N}=+1$ level transiting (upwards)
in the spectrum of the operator. We see in Section \ref{sec:The-model},
Theorem \ref{thm:Formule-de-l'indice.-1}, that this equality
\[
\mathcal{N}=\mathcal{C}
\]
is a special case of a more general result, called the \textbf{index
formula}, valid for a continuous family of symbols and for spaces
and bundles of larger dimensions.

Another equivalent formulation given in \cite{fred-boris,fred-boris-02bis,fred-boris01}
in a more general context: for $\left|\mu\right|\gg1$, there are
two groups of levels $j=-,+$ in the spectrum of $\hat{H}_{\mu}$.
When changing $\mu=-\infty\rightarrow+\infty$ each group has a variation
$\Delta\mathcal{N}_{j}\in\mathbb{Z}$ of the number of levels. We
have the formula
\[
\Delta\mathcal{N}_{j}=-\mathcal{C}_{j}
\]
where $\mathcal{C}_{j}$ is the Chern index of the bundle $F_{j}\rightarrow S^{2}$.

\section{\label{subsec:Spectral-flow-and-1}Spectral flow and index formula
for oceanic equatorial waves}

In this Section we present the model of Matsuno (1966) \cite{matsuno1966quasi}
for equatorial waves and the topological interpretation given by P.
Delplace, J. B. Marston, and A. Venaille in \cite{Delplace_Venaille_2018}.

\subsection{Matsuno's model}

We first present the physical meaning of the Matsuno's model \cite{matsuno1966quasi}.
See also this \href{https://kiwi.atmos.colostate.edu/group/dave/pdf/Matsuno-Gill.pdf}{Document},
\cite{vallis2017atmospheric}.

\begin{figure}
\begin{centering}
\includegraphics[scale=0.2]{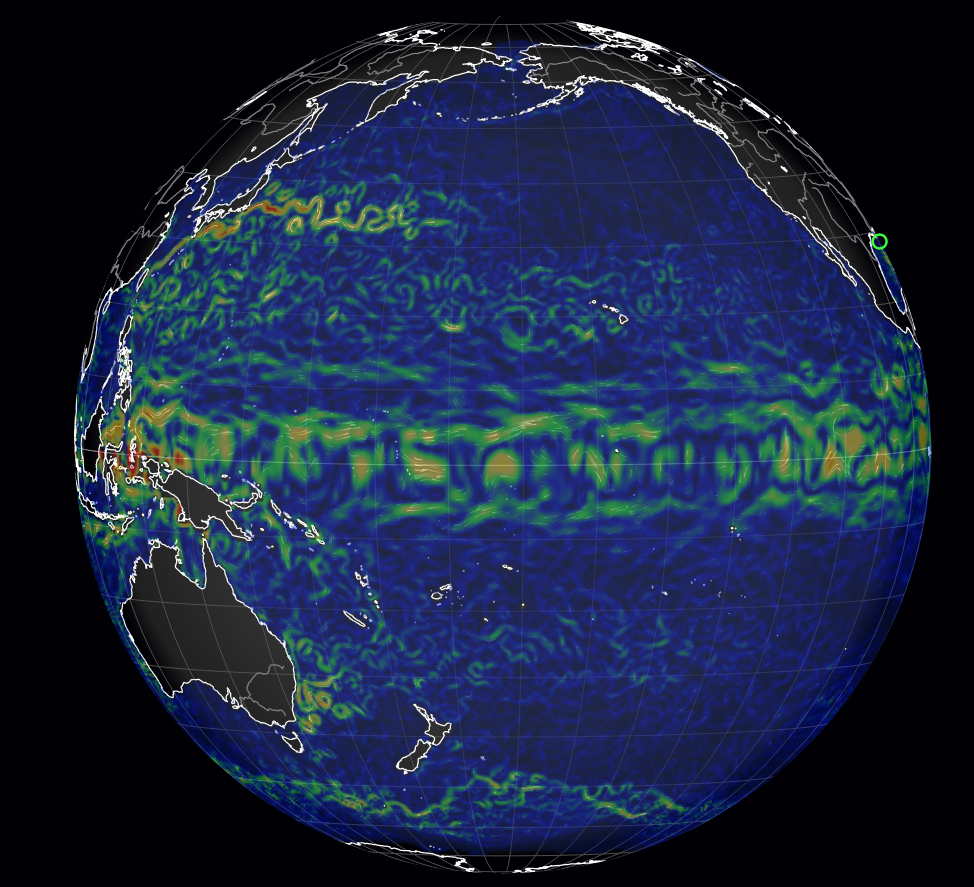}\includegraphics[scale=0.2]{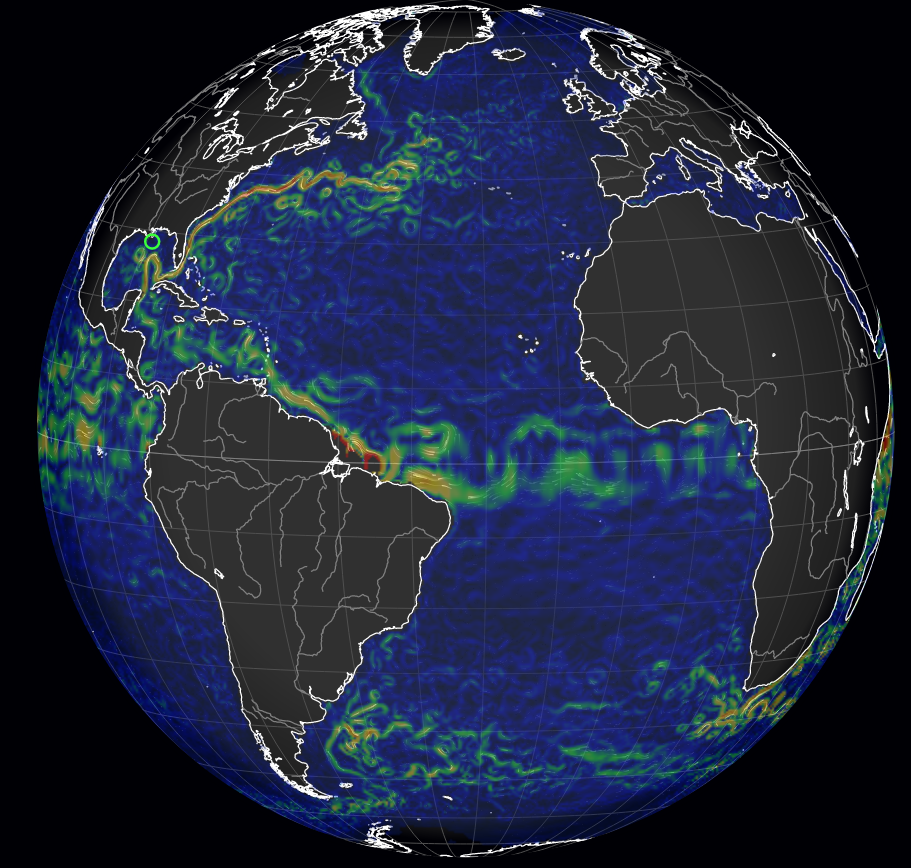}
\par\end{centering}
\caption{\label{fig:rossby-1}Marine currents May 27, 2018, according to the
website \protect\href{https://earth.nullschool.net}{nullschool}.
We observe the equatorial waves and the accumulation of energy on
the east coasts of continents.}
\end{figure}

\paragraph{The shallow water model:}

\begin{figure}
\begin{centering}
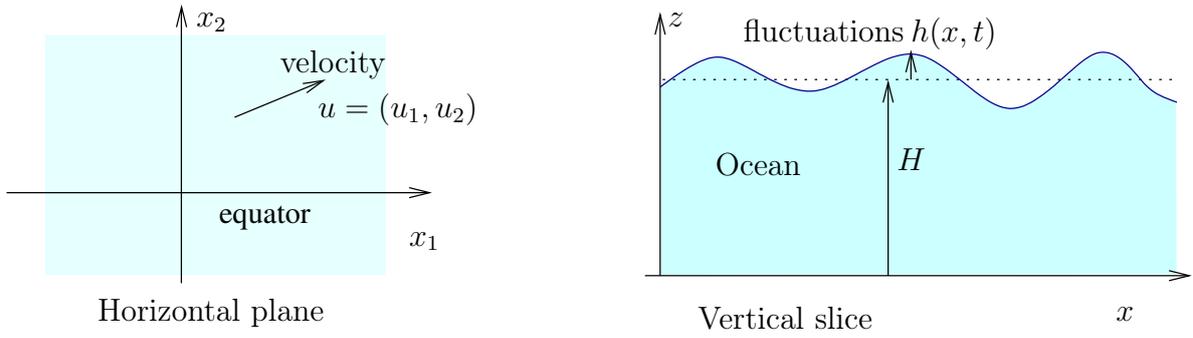
\par\end{centering}
\caption{\label{fig:shallow_water} Illustration of quantities for the shallow
water model (\ref{eq:shallow_water}).}
\end{figure}

See also \href{https://en.wikipedia.org/wiki/Shallow_water_equations}{Shallow\_water\_equations on wikipedia.}.
Let $x=\left(x_{1},x_{2}\right)\in\mathbb{R}^{2}$ be local coordinates
on the horizontal plane near the equator. $x_{1}$ is the longitude
and $x_{2}$ the latitude. The function $\left(h\left(x,t\right)+H\right)\in\mathbb{R}$
with $H>0$ represents the depth of water (or of a layer of hot water)
at position $x$ and time $t\in\mathbb{R}$. The vector $u\left(x,t\right)=\left(u_{1}\left(x,t\right),u_{2}\left(x,t\right)\right)\in\mathbb{R}^{2}$
represents the (horizontal) velocity of this water. Water is submitted
to gravity ($g=9.81\,m/s^{2}$ is the g-force) and since the earth
is rotating with frequency $\Omega$, there is also an effective Coriolis
force. The Navier-Stokes equations with shallow water assumptions
give
\begin{align}
\partial_{t}h+\mathrm{div}\left(\left(h+H\right)u\right) & =0\label{eq:shallow_water}\\
\partial_{t}u+u\cdot\mathrm{grad}\left(u\right)= & -g\mathrm{grad}\left(h\right)-fn\wedge u\nonumber 
\end{align}
with $f\left(x\right)=2\Omega\cdot n\left(x\right)\in\mathbb{R}$
and $n\left(x\right)$ being the unit normal vector at position $x$.
See Figure \ref{fig:shallow_water}.

\paragraph{Linearization:}

The idea of Matsuno is to linearize the equations (\ref{eq:shallow_water})
in the vicinity of $x_{2}=0$ (the equator), $u=0$ (small velocities),
$h=0$ (small fluctuations). We assume
\[
f\left(x\right)=\beta x_{2},\quad\beta>0.
\]
Then (\ref{eq:shallow_water}) at first order give the following linear
equations
\begin{align}
\partial_{t}h & =-H\mathrm{div}\left(u\right)\label{eq:shallow_water-1}\\
\partial_{t}u= & -g\mathrm{grad}\left(h\right)-\beta x_{2}\left(\begin{array}{c}
-u_{2}\\
u_{1}
\end{array}\right)\nonumber 
\end{align}

With $c=\sqrt{gH}$ and the change of variables
\[
t'=\sqrt{c\beta}t,\quad x'=\sqrt{\frac{\beta}{c}}x,\quad h'=\sqrt{\frac{\beta}{c}}h,\quad u'=\frac{1}{c}u,
\]
we obtain the dimensionless equations, written without $'$ (equivalently
we put $H=1$, $g=1$, $\beta=1$):
\begin{align*}
\partial_{t}h & =-\partial_{x_{1}}u_{1}-\partial_{x_{2}}u_{2}\\
\partial_{t}u_{1} & =-\partial_{x_{1}}h+x_{2}u_{2}\\
\partial_{t}u_{2} & =-\partial_{x_{2}}h-x_{2}u_{1}
\end{align*}
We will write
\[
\Psi=\left(\begin{array}{c}
h\\
u_{1}\\
u_{2}
\end{array}\right)\in L^{2}\left(\mathbb{R}_{x_{1},x_{2},t}^{3}\right)\otimes\mathbb{C}^{3}.
\]
Then
\[
i\partial_{t}\Psi=\left(\begin{array}{ccc}
0 & -i\partial_{x_{1}} & -i\partial_{x_{2}}\\
-i\partial_{x_{1}} & 0 & ix_{2}\\
-i\partial_{x_{2}} & -ix_{2} & 0
\end{array}\right)\Psi.
\]
Since the coefficients do not depend on $x_{1}$ one can assume the
Fourier mode in $x_{1}$:
\[
\Psi\left(x_{1},x_{2},t\right)=e^{i\mu x_{1}}\psi\left(x_{2},t\right)
\]
with Fourier variable $\mu\in\mathbb{R}$ and $\psi\in L^{2}\left(\mathbb{R}_{x_{2},t}^{2}\right)\otimes\mathbb{C}^{3}$.
In other words, $\mu$ is the spatial frequency in $x_{1}$ (and $\lambda_{1}=\frac{2\pi}{\mu}$
is the wave length).

For simplicity we replace $\left(x_{2},\xi_{2}\right)$ by $\left(x,\xi\right)$.
This gives the Matsuno model:

\begin{cBoxA}{}
\begin{defn}
The \textbf{<<Matsuno model>>} is the system of equations for $\psi:\left(t,x\right)\in\mathbb{R}^{2}\rightarrow\psi\left(t,x\right)\in\mathbb{C}^{3}$
given by
\[
i\partial_{t}\psi=\hat{H}_{\mu}\psi
\]
with the operator
\begin{equation}
\hat{H}_{\mu}=\left(\begin{array}{ccc}
0 & \mu & \hat{\xi}\\
\mu & 0 & i\hat{x}\\
\hat{\xi} & -i\hat{x} & 0
\end{array}\right)=\mathrm{Op}\left(H_{\mu}\right),\quad\in\mathrm{Herm}\left(L^{2}\left(\mathbb{R}_{x}\right)\otimes\mathbb{C}^{3}\right)\label{eq:mastuno}
\end{equation}
and its symbol
\begin{equation}
H_{\mu}\left(x,\xi{}_{2}\right)=\left(\begin{array}{ccc}
0 & \mu & \xi\\
\mu & 0 & ix\\
\xi & -ix & 0
\end{array}\right)\in\mathrm{Herm}\left(\mathbb{C}^{3}\right)\label{eq:symbol_matsuno}
\end{equation}
and $\hat{\xi}=\mathrm{Op}_{1}\left(\xi\right):=-i\partial_{x}$,
$\hat{x}=\mathrm{Op}_{1}\left(x\right):=x$.
\end{defn}

\end{cBoxA}

\subsection{Spectral index $\mathcal{N}$}

The following proposition describes the spectrum of the operator $\hat{H}_{\mu}$
with respect to the $\mu$ parameter.

\begin{figure}
\selectlanguage{french}%
\begin{centering}
\includegraphics[scale=0.15]{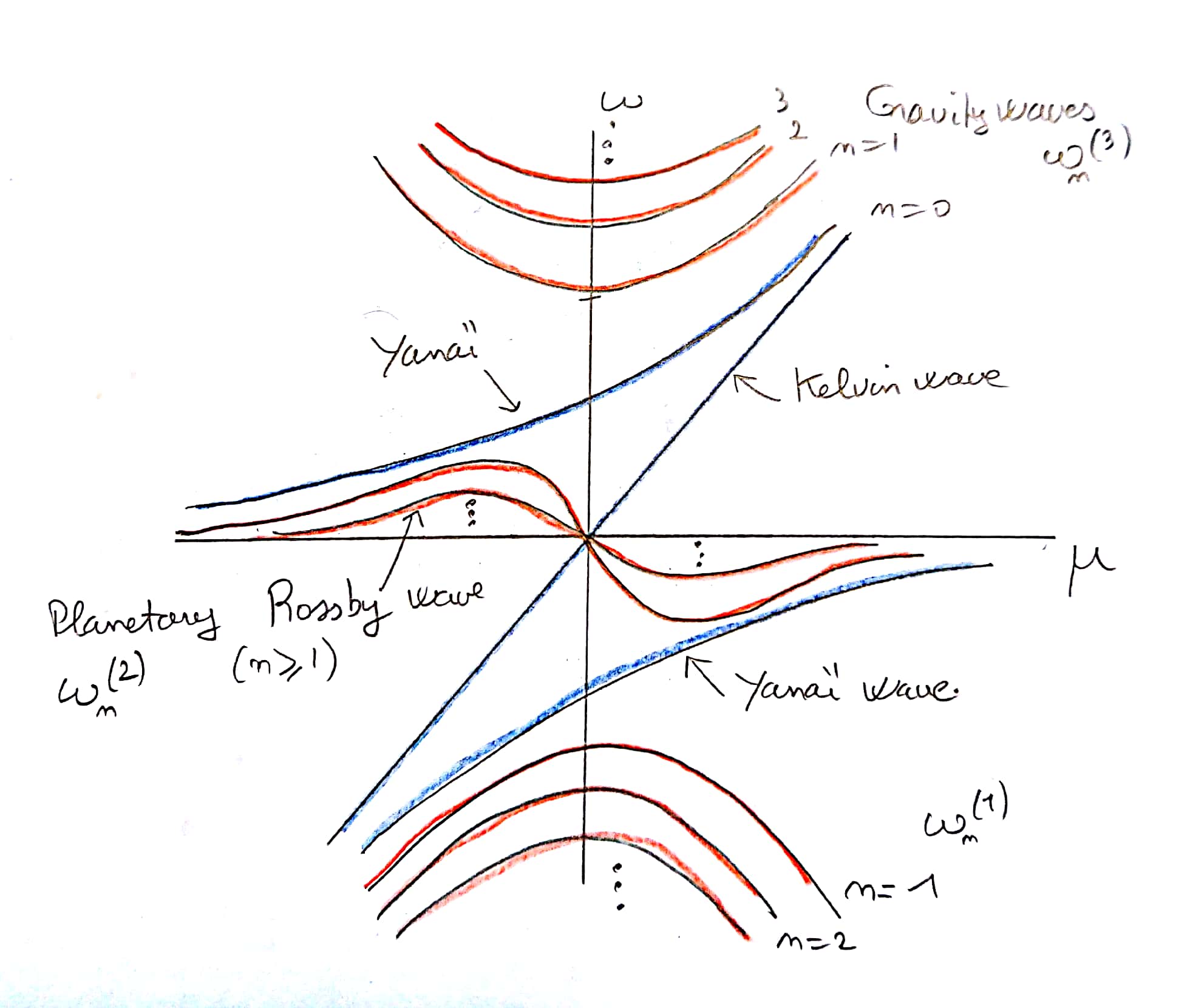}\foreignlanguage{english}{}
\par\end{centering}
\selectlanguage{english}%
\caption{\foreignlanguage{french}{\label{fig:matsuno}\foreignlanguage{english}{Representations of eigenvalues
$\omega_{n}^{\left(j\right)}\left(\mu\right)$, Eq.(\ref{eq:matsuno}).
We observe a spectral index of $\mathcal{N}=+2$ levels.}}}
\end{figure}

\begin{cBoxB}{}
\begin{prop}
\textbf{\label{prop:Spectre-de-.-1}<<Spectrum of $\hat{H}_{\mu}$>>}.
\cite{matsuno1966quasi} For each $\mu\in\mathbb{R}$, the operator
$\hat{H}_{\mu}$, (\ref{eq:mastuno}), has a discrete spectrum in
$L^{2}\left(\mathbb{R}_{x}\right)\otimes\mathbb{C}^{3}$ given by
\begin{equation}
\hat{H}_{\mu}\phi_{n}^{\left(j\right)}=\omega_{n}^{\left(j\right)}\phi_{n}^{\left(j\right)},\quad j=1,2,3,\quad n\geq1,\label{eq:matsuno}
\end{equation}
with $\omega_{n}^{\left(j\right)},j=1,2,3$ solutions of the equation
of degree 3 in $\omega$:
\begin{equation}
\omega^{3}-\left(\mu^{2}+2n+1\right)\omega-\mu=0,\label{eq:3eme}
\end{equation}
called \textbf{gravity waves} for $j=1,3$ and \textbf{Rossby planetary
waves} for $j=2$.

In addition there are the solutions
\[
\hat{H}_{\mu}\phi_{K}=\mu\phi_{K}\quad:\text{Kelvin mode}
\]
\[
\hat{H}_{\mu}\phi_{Y}^{\pm}=\omega_{\pm}\phi_{Y}^{\pm}\quad:\text{Yanaï mode}
\]
with $\omega_{\pm}=\frac{1}{2}\left(\mu\pm\sqrt{\mu^{2}+4}\right)$
solutions of $\left(\omega^{2}-\mu\omega-1\right)=0$. We observe
in figure \ref{fig:matsuno} that when $\mu$ increases, there is
\[
\mathcal{N}=+2
\]
eigenvalues that are going upward.
\end{prop}

\end{cBoxB}

\paragraph{Remarks on the physics of equatorial waves:}

(from oral explanations by \href{http://perso.ens-lyon.fr/antoine.venaille/}{Antoine Venaille}).

\begin{figure}
\begin{centering}
\includegraphics[scale=0.1]{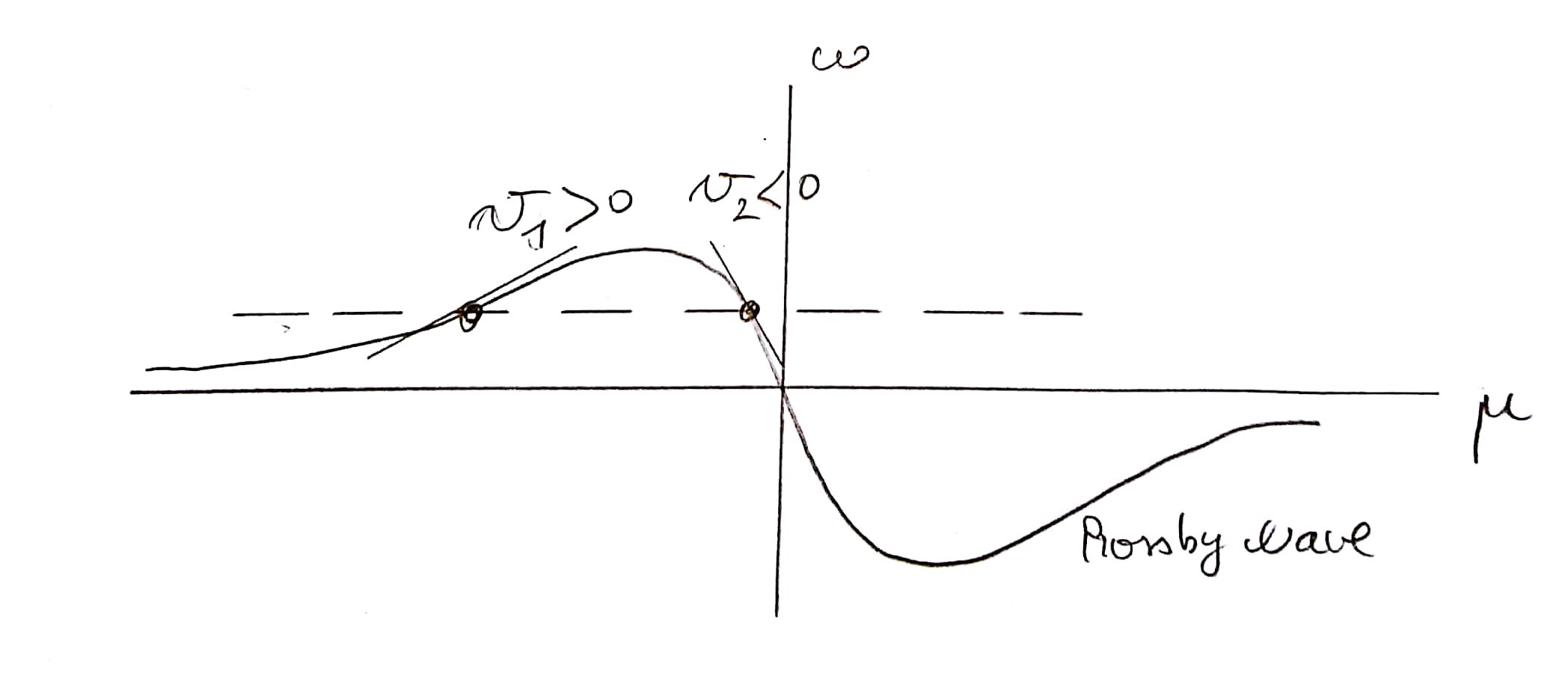}
\par\end{centering}
\caption{\foreignlanguage{french}{\label{fig:rossby}\foreignlanguage{english}{At a given frequency
$\omega$ , there are two components with group velocities$v_{1}>0$,
$v_{2}<0$ but $\left|v_{2}\right|\gg\left|v_{1}\right|$. This explains
why energy accumulates on the east coasts, see figure \ref{fig:rossby-1}.}}}
\end{figure}

\textbullet{} The Matsuno model applies either to the ocean or the
atmosphere. It can for instance describe the dynamics of the upper
oceanic layer called the \href{https://en.wikipedia.org/wiki/Thermocline}{thermocline}
($1$ km depth), above the \href{https://en.wikipedia.org/wiki/Abyssal_plain}{abyss}
($4$ km). It can also describe the dynamics of the \href{https://en.wikipedia.org/wiki/Troposphere}{troposphere}
($10$ km) below the \href{https://en.wikipedia.org/wiki/Stratosphere}{stratosphere}
($50$ km).
\begin{itemize}
\item The Matsuno model applies to the ocean (warm water layer of thickness
$\sim1$ km) or to the atmosphere (boundary layer between troposphere
and stratosphere around $10$ km).
\item The \href{https://en.wikipedia.org/wiki/El_Ni\%C3\%B1o\%E2\%80\%93Southern_Oscillation}{El Nino}
phenomenon in the atmosphere-ocean climate system is triggered by
a trapped oceanic Kelvin wave propagating across the Pacific ocean.
It is symmetric in $x_{2}$, of wavelength $\lambda_{1}=2\pi/\mu$
and propagates towards Peru. More precisely El Nino is a phenomena
that couple ocean and atmosphere. The Kelvin oceanic mode is an essential
ingredient for the apparition of high temperature anomalies on the
Peru coast and has global consequences.
\item From satellites, Yanai modes can sometimes be observed in the form
of regular cloud pattern asymmetric with respect to the equator. These
clouds reflect the patterns of vertical velocity fields, related to
horizontal temperature anomalies.
\item The group velocity of the wave $\Psi$ according to $x_{1}$ corresponds
to the derivative of the curves $\omega\left(\mu\right)$: $v_{g}=\frac{\partial\omega}{\partial\mu}$.
For the Rossby waves, at fixed $\omega$, we observe a component with
strong group velocity $v_{1}=\frac{\partial\omega}{\partial\mu}<0$
to the east and another component more weak $v_{2}=\frac{\partial\omega}{\partial\mu}>0$
to the west. See figure \ref{fig:rossby}. There is thus an accumulation
of energy (then dissipation) on the east coasts of the continents,
very visible, ex: Gulf Stream on the figure \ref{fig:rossby-1}.
\end{itemize}

\begin{proof}
We will see from (\ref{eq:val_p-1}) that the operator $\hat{H}_{\mu}$
is \href{https://en.wikipedia.org/wiki/Elliptic_operator}{elliptic}.
So \footnote{We can do without this argument by noticing at the end of the computation
that the found eigenvectors form a basis of the Hilbert space.} $\hat{H}_{\mu}$ has discrete spectrum. Any vector $\psi\in L^{2}\left(\mathbb{R}\right)\otimes\mathbb{C}^{3}$
can be written
\[
\psi=\sum_{n\geq0}\left(\begin{array}{c}
a_{n}\varphi_{n}\\
b_{n}\varphi_{n}\\
c_{n}\varphi_{n}
\end{array}\right),\qquad a_{n},b_{n},c_{n}\in\mathbb{C}.
\]

We introduce
\[
a=\frac{1}{\sqrt{2}}\left(\hat{x}_{2}+i\hat{\xi}_{2}\right),\quad a^{\dagger}=\frac{1}{\sqrt{2}}\left(\hat{x}_{2}-i\hat{\xi}_{2}\right),
\]
\[
\Leftrightarrow\hat{x}_{2}=\frac{1}{\sqrt{2}}\left(a+a^{\dagger}\right),\quad\hat{\xi}_{2}=-i\partial_{x}=\frac{i}{\sqrt{2}}\left(a^{\dagger}-a\right).
\]

We have
\[
\hat{H}_{\mu}\psi=\omega\psi\Leftrightarrow\sum_{n\geq0}\left(\begin{array}{ccc}
-\omega & \mu & \frac{i}{\sqrt{2}}\left(a^{\dagger}-a\right)\\
\mu & -\omega & \frac{i}{\sqrt{2}}\left(a+a^{\dagger}\right)\\
\frac{i}{\sqrt{2}}\left(a^{\dagger}-a\right) & \frac{-i}{\sqrt{2}}\left(a+a^{\dagger}\right) & -\omega
\end{array}\right)\left(\begin{array}{c}
a_{n}\varphi_{n}\\
b_{n}\varphi_{n}\\
c_{n}\varphi_{n}
\end{array}\right)=0
\]
\[
\Leftrightarrow\begin{cases}
\sum_{n\geq0}-\omega a_{n}\varphi_{n}+\mu b_{n}\varphi_{n}+\frac{i}{\sqrt{2}}\left(a^{\dagger}-a\right)c_{n}\varphi_{n}=0\\
\sum_{n\geq0}\mu a_{n}\varphi_{n}-\omega b_{n}\varphi_{n}+\frac{i}{\sqrt{2}}\left(a^{\dagger}+a\right)c_{n}\varphi_{n}=0\\
\sum_{n\geq0}\frac{i}{\sqrt{2}}\left(a^{\dagger}-a\right)a_{n}\varphi_{n}-\frac{i}{\sqrt{2}}\left(a^{\dagger}+a\right)b_{n}\varphi_{n}-\omega c_{n}\varphi_{n}=0
\end{cases}
\]
\[
\Leftrightarrow\begin{cases}
-\omega a_{n}+\mu b_{n}+\frac{i}{\sqrt{2}}\left(c_{n-1}\sqrt{n}-c_{n+1}\sqrt{n+1}\right)=0 & ,\forall n\\
\mu a_{n}-\omega b_{n}+\frac{i}{\sqrt{2}}\left(c_{n-1}\sqrt{n}+c_{n+1}\sqrt{n+1}\right)=0\\
\frac{i}{\sqrt{2}}\left(a_{n-1}\sqrt{n}-a_{n+1}\sqrt{n+1}\right)-\frac{i}{\sqrt{2}}\left(b_{n-1}\sqrt{n}+b_{n+1}\sqrt{n+1}\right)-\omega c_{n}=0
\end{cases}
\]
\[
\Leftrightarrow\begin{cases}
\left(\mu-\omega\right)\left(a_{n}+b_{n}\right)+i\sqrt{2}c_{n-1}\sqrt{n}=0\\
\left(\mu+\omega\right)\left(a_{n}-b_{n}\right)+i\sqrt{2}c_{n+1}\sqrt{n+1}=0\\
\left(a_{n-1}-b_{n-1}\right)\sqrt{n}-\left(a_{n+1}+b_{n+1}\right)\sqrt{n+1}+i\sqrt{2}\omega c_{n}=0
\end{cases}
\]
We introduce
\[
s_{n}:=a_{n}+b_{n},\quad d_{n}:=b_{n}-a_{n}
\]
\[
\Leftrightarrow a_{n}=\frac{1}{2}\left(s_{n}-d_{n}\right),\quad b_{n}=\frac{1}{2}\left(s_{n}+d_{n}\right)
\]
Then

\[
\Leftrightarrow\begin{cases}
\left(\mu-\omega\right)s_{n'}+i\sqrt{2n'}c_{n'-1}=0 & \forall n'\geq0.\\
-\left(\mu+\omega\right)d_{n'}+i\sqrt{2\left(n'+1\right)}c_{n'+1}=0\\
-d_{n'-1}\sqrt{n'}-s_{n'+1}\sqrt{n'+1}+i\sqrt{2}\omega c_{n'}=0
\end{cases}
\]
We consider different cases.
\begin{enumerate}
\item If $\mu=\omega$, this gives $c_{n'}=0$ for $n'\geq0$ and so $d_{n'}=0$
for $n'\geq0$ and $s_{n'}=0$ for $n'\geq1$. There is a solution
with $s_{0}=1$, called \textquotedbl\textbf{Kelvin Wave}\textquotedbl .
\item If $\mu=-\omega$, this gives $c_{n'}=0,s_{n'}=0,d_{n'}=0$ for all
$n'\geq0$, so there is no solution.
\item If $\mu-\omega\neq0$ and $\mu+\omega\neq0$ then
\begin{align*}
s_{n'+1} & =\frac{-i\sqrt{2\left(n'+1\right)}}{\left(\mu-\omega\right)}c_{n'},\quad d_{n'-1}=\frac{i\sqrt{2n'}}{\left(\mu+\omega\right)}c_{n'}
\end{align*}
hence for every $n'\geq0$
\begin{align*}
-\frac{i\sqrt{2}n'}{\left(\mu+\omega\right)}c_{n'}-\frac{-i\sqrt{2}\left(n'+1\right)}{\left(\mu-\omega\right)}c_{n'}+i\sqrt{2}\omega c_{n'} & =0\\
\Leftrightarrow\left(-n'\left(\mu-\omega\right)+\left(n'+1\right)\left(\mu+\omega\right)+\omega\left(\mu^{2}-\omega^{2}\right)\right)c_{n'} & =0\\
\Leftrightarrow\left(\omega^{3}-\omega\left(2n'+1+\mu^{2}\right)-\mu\right)c_{n'} & =0
\end{align*}
If moreover $c_{n}\neq0$ then $\omega^{3}-\omega\left(2n+1+\mu^{2}\right)-\mu=0$.
\begin{enumerate}
\item If $c_{0}=1$ then $\omega^{3}-\omega\left(1+\mu^{2}\right)-\mu=\left(\omega+\mu\right)\left(\omega^{2}-\mu\omega-1\right)=0$
giving $\left(\omega^{2}-\mu\omega-1\right)=0$ (since $\omega=-\mu$
is excluded) and the \textbf{<<Yanai} \textbf{waves}>>:
\[
\omega_{\pm}=\frac{1}{2}\left(\mu\pm\sqrt{\mu^{2}+4}\right)
\]
with components $c_{0}=1$, $c_{n'}=0$ for $n'\geq1$. $s_{0}=0$,
$s_{1}=\frac{-i\sqrt{2}}{\left(\mu-\omega\right)}$ and $s_{n'}=0$
for $n'\geq2$. $d_{n'}=0$ for $n'\geq0$. This determines $a_{n'},b_{n'}$.
\item If $c_{n}=1$ and $c_{n'}=0$ for $n'\neq n$, then $\omega_{n}^{\left(j\right)}$,
$j=1,2,3$, are solutions of (\ref{eq:3eme}). This determines the
components $d_{n'},s_{n'},a_{n'},b_{n'}$, called \textbf{gravity
waves} for $j=1,3$ and \textbf{Rossby planetary waves }for $j=2$.
\end{enumerate}
\end{enumerate}
\end{proof}

\subsection{Topological Chern index $\mathcal{C}$}

We can first consult the section \ref{sec:Espaces-fibr=0000E9s-vectoriels}
which introduces the notion of topology of a complex vector bundle
of rank 1 on the sphere $S^{2}$.

\begin{cBoxB}{}
\begin{prop}
\textbf{\label{prop:Aspects-topologiques-du}``Topological aspects
of the $H_{\mu}\left(x,\xi\right)$ (\ref{eq:symbole_H_mu})}''.
\cite{Delplace_Venaille_2018} The eigenvalues of the matrix $H_{\mu}\left(x,\xi\right)\in\mathrm{Herm}\left(\mathbb{C}^{3}\right)$
are
\begin{align}
\omega^{\left(1\right)}\left(\mu,x,\xi\right) & =-\sqrt{\mu^{2}+x^{2}+\xi^{2}}\label{eq:val_p-1}\\
\omega^{\left(2\right)}\left(\mu,x,\xi\right) & =0\\
\omega^{\left(3\right)}\left(\mu,x,\xi\right) & =+\sqrt{\mu^{2}+x^{2}+\xi^{2}}
\end{align}
There is therefore a degeneracy at $\left(\mu,x,\xi\right)=\left(0,0,0\right)$.
For $\left(\mu,x,\xi\right)\in S^{2}\subset\mathbb{R}^{3}$, and $j=1,2,3$,
the eigenspace $F^{\left(j\right)}\left(\mu,x,\xi\right)\subset\mathbb{C}^{2}$
associated with the eigenvalue $\omega^{\left(j\right)}\left(\mu,x,\xi\right)$
defines a complex vector bundle of rank 1 above $S^{2}$, whose topological
indices of Chern $\mathcal{C}_{j}$ are respectively 
\[
\mathcal{C}_{1}=+2,\quad\mathcal{C}_{2}=0,\quad\mathcal{C}_{3}=-2.
\]
\end{prop}

\end{cBoxB}

\begin{figure}
\begin{centering}
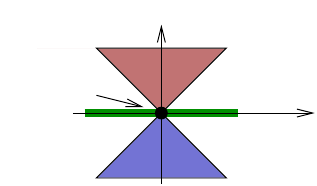
\par\end{centering}
\caption{Domains that represent the eigenvalues $\omega^{\left(j\right)}\left(\mu,x,\xi\right)$
for $\mu$ fixed and all possible values of $\left(x,\xi\right)\in\mathbb{R}^{2}$,
$j=1,2,3$. We have $\omega^{\left(1\right)}\protect\leq-\left|\mu\right|$,
$\omega^{\left(2\right)}=0$, $\omega^{\left(3\right)}\protect\geq\left|\mu\right|$.}
\end{figure}

\begin{proof}
The proof is similar to that of the proposition\ref{prop:Aspects-topologiques-du-1}.
We will use two different methods to compute $\mathcal{C}_{j}$. First
using zeros of a global section (\ref{eq:C_section}) and secondly
using curvature integral (\ref{eq:curvature_integral_S2}). The eigenvalues
and eigenvectors of the matrix $H_{\mu}\left(x,\xi\right)$ are\footnote{Obtained with xcas, by writing:\texttt{ H:={[}{[}0,mu,xi{]},{[}mu,0,i{*}x{]},{[}xi,-i{*}x,0{]}{]};
eigenvals(H); eigenvects(H);}}

\[
\omega_{1}\left(\mu,x,\xi\right)=-r,\quad\omega_{2}\left(\mu,x,\xi\right)=0,\quad\omega_{3}\left(\mu,x,\xi\right)=r,
\]
with $r=\sqrt{\mu^{2}+x^{2}+\xi^{2}}$. Eigenvectors are
\begin{equation}
U_{1}=\left(\begin{array}{c}
\mu^{2}+\xi^{2}\\
i\xi x-\mu r\\
-i\mu x-\xi r
\end{array}\right),\quad U_{2}=\left(\begin{array}{c}
-x\\
i\xi\\
-i\mu
\end{array}\right),\quad U_{3}=\left(\begin{array}{c}
\mu^{2}+\xi^{2}\\
i\xi x+\mu r\\
-i\mu x+\xi r
\end{array}\right).\label{eq:U1U2U3}
\end{equation}
We have
\[
H_{\mu}\left(x,\xi\right)=\sum_{j=1}^{3}\omega_{j}\Pi_{j},
\]
with orthogonal spectral projectors
\[
\Pi_{j}=\frac{U_{j}\langle U_{j}|.\rangle}{\left\Vert U_{j}\right\Vert ^{2}}.
\]
\[
F_{j}\left(\mu,x,\xi\right)=\mathrm{Im}\Pi_{j}\left(\mu,x,\xi\right)\subset\mathbb{C}^{3}
\]
is the eigenspace and defines a complex bundle of rank 1 on $\mathbb{R}^{3}\backslash\left\{ 0\right\} $
noted $F_{j}$.

\paragraph{Computation of $\mathcal{C}_{1}$}

Let $u_{0}=\left(\begin{array}{c}
0\\
0\\
1
\end{array}\right)\in\mathbb{C}^{3}$ a fixed vector and 
\[
s_{1}\left(\mu,x,\xi\right)=\Pi_{1}u_{0}\in\mathbb{C}^{3}
\]
 that defines a global section of the bundle $F_{1}$. We consider
the sphere $S^{2}=\left\{ \left(\mu,x,\xi\right),r=1\right\} $. We
have
\begin{equation}
s_{1}\left(\mu,x,\xi\right)=\Pi_{1}u_{0}\eq{\ref{eq:U1U2U3}}\frac{U_{1}}{\left\Vert U_{1}\right\Vert ^{2}}\left(i\mu x-\xi\right)\label{eq:s1-2}
\end{equation}
with $\left\Vert U_{1}\right\Vert ^{2}\eq{\ref{eq:U1U2U3}}2\left(\mu^{2}+\xi^{2}\right)$
\[
\left\Vert s_{1}\right\Vert ^{2}=\frac{\left(\left(\mu x\right)^{2}+\xi^{2}\right)}{2\left(\mu^{2}+\xi^{2}\right)}.
\]
We observe that $\left\Vert s_{1}\right\Vert ^{2}$ vanishes at two
points $\left(\mu,x,\xi\right)=\left(\pm1,0,0\right)\in S^{2}$. Near
the point $\left(\mu,x,\xi\right)=\left(1,0,0\right)$, writing $x+i\xi=\epsilon e^{i\theta}$,
we have
\begin{align*}
s_{1} & \eq{\ref{eq:s1-2}}\frac{1}{2}\left(\begin{array}{c}
1\\
-1\\
0
\end{array}\right)\left(i\mu x-\xi\right)+o\left(\epsilon\right)\\
 & =\frac{i}{2}\left(\begin{array}{c}
1\\
-1\\
0
\end{array}\right)\epsilon e^{i\theta}+o\left(\epsilon\right)
\end{align*}
There appears the function $e^{i\theta}\in S^{1}\rightarrow e^{i\theta}\in S^{1}$
with degree $+1$. Near the point $\left(\mu,x,\xi\right)=\left(-1,0,0\right)$,
writing $\xi+ix=\epsilon e^{i\theta}$ (that respects the orientation
on $S^{2}$), we have
\begin{align*}
s_{1} & \eq{\ref{eq:s1-2}}\frac{1}{2}\left(\begin{array}{c}
1\\
-1\\
0
\end{array}\right)\left(i\mu x-\xi\right)+o\left(\epsilon\right)\\
 & =-\frac{1}{2}\left(\begin{array}{c}
1\\
-1\\
0
\end{array}\right)\epsilon e^{i\theta}+o\left(\epsilon\right)
\end{align*}
There appears the function $e^{i\theta}\in S^{1}\rightarrow e^{i\theta}\in S^{1}$
of degree $+1$. In total we deduce that
\[
\mathcal{C}_{1}=+1+1=+2.
\]

\paragraph{Computation of $\mathcal{C}_{2}$}

We have for $\left(\mu,x,\xi\right)\in S^{2}$ on $\left\Vert U_{2}\right\Vert =1$
so $U_{2}$ defines a non-zero global section of fiber $F_{2}$ therefore
\[
\mathcal{C}_{2}=0.
\]

\paragraph{Computation of $\mathcal{C}_{3}$}

Let we use $\sum_{j=1}^{3}\mathcal{C}_{j}=0$ (because the bundle
$\mathbb{C}^{3}\rightarrow S^{2}$ is trivial) giving directly $\mathcal{C}_{3}=-2$,
we calculate $\mathcal{C}_{3}$ as we did for $\mathcal{C}_{1}$:

Let $u_{0}=\left(\begin{array}{c}
0\\
0\\
1
\end{array}\right)\in\mathbb{C}^{3}$ a fixed vector and
\[
s_{3}\left(\mu,x,\xi\right)=\Pi_{3}u_{0}\in\mathbb{C}^{3}
\]
which defines a global section of the bundle $F_{3}$. We consider
the sphere $S^{2}=\left\{ \left(\mu,x,\xi\right),r=1\right\} $. We
have
\begin{equation}
s_{3}\left(\mu,x,\xi\right)=\Pi_{3}u_{0}\eq{\ref{eq:U1U2U3}}\frac{U_{3}}{\left\Vert U_{3}\right\Vert ^{2}}\left(i\mu x+\xi\right)\label{eq:s1-2-1}
\end{equation}
with $\left\Vert U_{3}\right\Vert ^{2}\eq{\ref{eq:U1U2U3}}2\left(\mu^{2}+\xi^{2}\right),$
\[
\left\Vert s_{3}\right\Vert ^{2}=\frac{\left(\left(\mu x\right)^{2}+\xi^{2}\right)}{2\left(\mu^{2}+\xi^{2}\right)}.
\]
We observe that $\left\Vert s_{3}\right\Vert ^{2}$ vanishes in two
points $\left(\mu,x,\xi\right)=\left(\pm1,0,0\right)\in S^{2}$. Near
the point $\left(\mu,x,\xi\right)=\left(1,0,0\right)$, setting $x+i\xi=\epsilon e^{i\theta}$,
we have
\begin{align*}
s_{3} & \eq{\ref{eq:s1-2-1}}\frac{1}{2}\left(\begin{array}{c}
1\\
1\\
0
\end{array}\right)\left(i\mu x+\xi\right)+o\left(\epsilon\right)\\
 & =\frac{i}{2}\left(\begin{array}{c}
1\\
1\\
0
\end{array}\right)\epsilon e^{-i\theta}+o\left(\epsilon\right)
\end{align*}
There is the function $e^{i\theta}\rightarrow e^{-i\theta}$ of degree
$-1$. Near the point $\left(\mu,x,\xi\right)=\left(-1,0,0\right)$,
writing $\xi+ix=\epsilon e^{i\theta}$ (that respects the orientation
on $S^{2}$), we have
\begin{align*}
s_{1} & \eq{\ref{eq:s1-2-1}}\frac{1}{2}\left(\begin{array}{c}
1\\
1\\
0
\end{array}\right)\left(i\mu x+\xi\right)+o\left(\epsilon\right)\\
 & =\frac{1}{2}\left(\begin{array}{c}
1\\
1\\
0
\end{array}\right)\epsilon e^{-i\theta}+o\left(\epsilon\right)
\end{align*}
There appears function $e^{i\theta}\rightarrow e^{-i\theta}$ of degree
$-1$. In total we deduce that
\[
\mathcal{C}_{3}=-1-1=-2.
\]

\paragraph{2nd method of computation using curvature integral:}

We will use spherical coordinates $\left(\theta,\varphi\right)\in]0,\pi[\times]0,2\pi[$
on the sphere $S^{2}$ defined by
\begin{align*}
x & =\cos\theta\\
\xi & =\sin\theta\cos\varphi\\
\mu & =\sin\theta\sin\varphi.
\end{align*}
A unit vector $v\left(\theta,\varphi\right)$ in the fiber $F_{1}$
over the unit sphere $S^{2}$ is given by
\[
v=\frac{U_{1}}{\left\Vert U_{1}\right\Vert }\eq{\ref{eq:U1U2U3}}\frac{1}{\sqrt{2\left(\mu^{2}+\xi^{2}\right)}}\left(\begin{array}{c}
\mu^{2}+\xi^{2}\\
i\xi x-\mu r\\
-i\mu x-\xi r
\end{array}\right)=\frac{1}{\sqrt{2}}\left(\begin{array}{c}
\sin\theta\\
i\cos\theta\cos\varphi-\sin\varphi\\
-i\cos\theta\sin\varphi-\cos\varphi
\end{array}\right)
\]
Then
\begin{equation}
\frac{\partial v}{\partial\theta}=\frac{1}{\sqrt{2}}\left(\begin{array}{c}
\cos\theta\\
-i\sin\theta\cos\varphi\\
i\sin\theta\sin\varphi
\end{array}\right),\qquad\frac{\partial v}{\partial\varphi}=\frac{1}{\sqrt{2}}\left(\begin{array}{c}
0\\
-i\cos\theta\sin\varphi-\cos\varphi\\
-i\cos\theta\cos\varphi+\sin\varphi
\end{array}\right).\label{eq:dvdtheta}
\end{equation}
We use the curvature integral formula (\ref{eq:curvature_integral_S2})
that gives
\[
\mathcal{C}_{1}=\frac{1}{2\pi}\int_{S^{2}}i\Omega
\]
with the curvature two form
\begin{align*}
\Omega\eq{\ref{eq:iOmega}}\langle dv\wedge|dv\rangle & =\left(\langle\frac{\partial v}{\partial\theta}|\frac{\partial v}{\partial\varphi}\rangle-\langle\frac{\partial v}{\partial\varphi}|\frac{\partial v}{\partial\theta}\rangle\right)d\theta\wedge d\varphi\\
 & =2i\mathrm{Im}\left(\langle\frac{\partial v}{\partial\theta}|\frac{\partial v}{\partial\varphi}\rangle\right)d\theta\wedge d\varphi\\
 & \eq{\ref{eq:dvdtheta}}-i\sin\theta d\theta\wedge d\varphi
\end{align*}
We get
\begin{align*}
\mathcal{C}_{1} & =\frac{1}{2\pi}\int_{S^{2}}i\Omega=\frac{1}{2\pi}\int_{\theta=0}^{\pi}\int_{\varphi=0}^{2\pi}\left(\sin\theta\right)d\theta d\varphi=2
\end{align*}
\end{proof}

\subsection{Conclusion on the model (\ref{eq:H_mu})}

Formulation given in \cite{fred-boris,fred-boris-02bis,fred-boris01}
in a more general context: for $\left|\mu\right|\gg1$, there are
three groups of levels $j=1,2,3$ in the spectrum of $\hat{H}_{\mu}$.
When changing $\mu=-\infty\rightarrow+\infty$ each group has a variation
$\Delta\mathcal{N}_{j}\in\mathbb{Z}$ of the number of levels. We
have the formula
\[
\Delta\mathcal{N}_{j}=-\mathcal{C}_{j}
\]
where $\mathcal{C}_{j}$ is the Chern index of the bundle $F_{j}\rightarrow S^{2}$.

Another possible formulation: In the model defined by (\ref{eq:H_mu}),
one observes from the symbol, a vector bundle $F_{1}$ (or $F_{1}\oplus F_{2}$)
whose index of Chern is $\mathcal{C}=+2$ and we observe that there
is $\mathcal{N}=+2$ levels that transits (upwards) in the spectrum
of the operator. We see in Section \ref{sec:The-model}, Theorem \ref{thm:Formule-de-l'indice.-1},
that this equality
\[
\mathcal{N}=\mathcal{C}
\]
 is a special case of a more general result, called the\textbf{ index
formula}, valid for a continuous family of symbols and for spaces
and bundles of larger dimensions.

\appendix

\section{\label{sec:Quantification,-op=0000E9rateurs-pseud}Quantization,
pseudo-differential-operators, semi-classical analysis on $\mathbb{R}^{2d}$}

\subsection{Quantization and pseudo-differential-operators (PDO)}

References for this Section are \cite{nicola_rodino_livre_11}\cite{zworski_book_2012}\cite{martinez-01}.

We denote $x\in\mathbb{R}^{n}$ the ``\textbf{position}'' and $\xi\in\mathbb{R}^{n}$
its dual variable, called ``\textbf{momentum}''. Let $\epsilon>0$
be a small parameter called \textbf{semi-classical parameter}.

\begin{cBoxA}{}
\begin{defn}
If $a\left(x,\xi\right)\in\mathcal{S}\left(\mathbb{R}^{n}\times\mathbb{R}^{n};\mathbb{C}\right)$
is a function on \textbf{phase space} $T^{*}\mathbb{R}^{n}=\mathbb{R}^{2n}$
called \textbf{symbol}, we associate a \textbf{pseudo-differential
operator} \textbf{(PDO) }denoted \textbf{$\hat{a}=\mathrm{Op}_{\epsilon}\left(a\right)$}
defined on a function $\psi\in\mathcal{S}\left(\mathbb{R}^{n}\right)$
by
\begin{equation}
\left(\hat{a}\psi\right)\left(x\right)=\left(\mathrm{Op}_{\epsilon}\left(a\right)\psi\right)\left(x\right)=\frac{1}{\left(2\pi\epsilon\right)^{n}}\int a\left(\frac{x+y}{2},\xi\right)e^{i\xi\cdot\left(x-y\right)/\epsilon}\psi\left(y\right)dyd\xi\label{eq:def_Op}
\end{equation}
The operation
\[
\mathrm{Op}_{\epsilon}:\quad a\rightarrow\hat{a}=\mathrm{Op}_{\epsilon}\left(a\right)
\]
that gives an operator $\hat{a}$ from a symbol $a$ is called \textbf{Weyl
quantization.}
\end{defn}

\end{cBoxA}

\begin{rem}
For example,
\begin{itemize}
\item For a function $V\left(x\right)$ (function of $x$ only) we get that
$\mathrm{Op}_{\epsilon}\left(V\left(x\right)\right)=V\left(x\right)$,
is the multiplication operator by $V$. For example $\hat{x}_{j}=\mathrm{Op}_{\epsilon}\left(x_{j}\right)$
is called the \href{https://en.wikipedia.org/wiki/Position_operator}{position operator}.
\item We have $\hat{\xi}_{j}=\mathrm{Op}_{\epsilon}\left(\xi_{j}\right)=-i\epsilon\frac{\partial.}{\partial x^{j}}$
called the \textbf{\href{https://en.wikipedia.org/wiki/Momentum_operator}{momentum operator}}
and for a function $W:\mathbb{R}^{n}\rightarrow\mathbb{R}$ we have
$\mathrm{Op}_{\epsilon}\left(W\left(\xi\right)\right)=W\left(\left(\mathrm{Op}_{\epsilon}\left(\xi_{j}\right)\right)_{j}\right)$,
hence $\mathrm{Op}_{\epsilon}\left(\left|\xi\right|^{2}\right)=\sum_{j}\left(\mathrm{Op}_{\epsilon}\left(\xi_{j}\right)\right)^{2}=-\epsilon^{2}\Delta$.
The \textbf{Schrödinger or Hamiltonian operator }$\hat{H}$ in quantum
mechanics is obtained from the Hamilton function $H\left(x,\xi\right)$
by Weyl quantization:
\[
H\left(x,\xi\right)=\frac{\left|\xi^{2}\right|}{2m}+V\left(x\right)\quad\rightarrow\quad\hat{H}=\mathrm{Op}\left(H\right)=-\frac{\epsilon^{2}}{2m}\Delta+V\left(x\right)
\]
\item We have $\mathrm{Op}_{\epsilon}\left(\overline{a}\right)=\left(\mathrm{Op}_{\epsilon}\left(a\right)\right)^{\dagger}$
(the $L^{2}$-\href{https://en.wikipedia.org/wiki/Hermitian_adjoint}{adjoint}).
\end{itemize}
\end{rem}

\subsection{Algebra of operators PDO}

The following proposition shows that the product of two PDO is a PDO

\begin{cBoxB}{}
\begin{prop}
\label{prop:}

\begin{itemize}
\item \cite[sec.4.3]{zworski_book_2012}\textbf{<<Composition of PDO and
star product of symbols}>>: For any $a,b\in\mathcal{S}\left(\mathbb{R}^{2n}\right)$
we have for $\epsilon\ll1$
\begin{equation}
\mathrm{Op}_{\epsilon}\left(a\right)\circ\mathrm{Op}_{\epsilon}\left(b\right)=\mathrm{Op}_{\epsilon}\left(a\star b\right)\label{eq:composition_OPD}
\end{equation}
with $a\star b\in\mathcal{S}\left(\mathbb{R}^{2n}\right)$ given by
\begin{align*}
a\star b & =\left(e^{i\epsilon\hat{A}}\left(a\left(x,\xi\right)b\left(y,\eta\right)\right)\right)_{y=x,\eta=\xi}\\
 & =ab+\epsilon\frac{1}{2i}\left\{ a,b\right\} +\epsilon^{2}\ldots
\end{align*}
and $\hat{A}=\frac{1}{2}\left(\partial_{x}\partial_{\eta}-\partial_{\xi}\partial_{y}\right)$.
\item <<\textbf{Commutator of PDO and \href{https://en.wikipedia.org/wiki/Poisson_bracket}{Poisson brackets}
of symbols}>>:
\begin{equation}
\left[\left(-\frac{i}{\epsilon}\right)\mathrm{Op}_{\epsilon}\left(a\right),\left(-\frac{i}{\epsilon}\right)\mathrm{Op}_{\epsilon}\left(b\right)\right]=\left(-\frac{i}{\epsilon}\right)\mathrm{Op}_{\epsilon}\left(\left\{ a,b\right\} \right)\:\left(1+O\left(\epsilon\right)\right)\label{eq:commut_OPD}
\end{equation}
i.e.:
\[
\left[a,b\right]_{\star}:=a\star b-b\star a=i\epsilon\left\{ a,b\right\} +O\left(\epsilon^{3}\right)
\]
\item \textbf{Trace of PDO}:
\begin{equation}
\mathrm{Tr}\left(\mathrm{Op}_{\epsilon}\left(a\right)\right)=\frac{1}{\left(2\pi\epsilon\right)^{n}}\int_{\mathbb{R}^{2n}}a\left(x,\xi\right)dxd\xi\label{eq:Trace_OPD}
\end{equation}
\item \textbf{Theorem of boundedness}. see \cite[Section 1.4]{nicola_rodino_livre_11}.
\end{itemize}
\end{prop}

\end{cBoxB}

\begin{example}
In dimension $n=1$, we compute directly that $x\left(-i\epsilon\frac{d}{dx}\right)\psi-\left(-i\epsilon\frac{d}{dx}\right)\left(x\psi\right)=i\epsilon\psi$
and $\left\{ x,\xi\right\} =1$. This gives $\left[\hat{x},\hat{\xi}\right]=i\epsilon\mathrm{Id}$
or 
\[
\left[\left(-\frac{i}{\epsilon}\right)\mathrm{Op}_{\epsilon}\left(x\right),\left(-\frac{i}{\epsilon}\right)\mathrm{Op}_{\epsilon}\left(\xi\right)\right]=\left(-\frac{i}{\epsilon}\right)\mathrm{Op}_{\epsilon}\left(\left\{ x,\xi\right\} \right)
\]
in accordance with (\ref{eq:commut_OPD}).
\end{example}

\subsection{Classes of symbols}

The relations of proposition \ref{prop:} are a little bit formal.
In order to make them useful, one has to control the remainders in
terms of operator norm. For this we need to make some assumption on
the symbols that express their ``slow variation at the Plank scale
$dxd\xi\sim\epsilon$>> (i.e. uncertainty principle). We call class
of symbol the set of symbols that forms an algebra for the operator
of composition $\star$. For example, the following classes of symbols
have been introduced by Hörmander \cite{hormander_3}. Let $M$ be
a smooth compact manifold. For $x\in\mathbb{R}^{n}$, we denote $\left\langle x\right\rangle :=\left(1+\left|x\right|^{2}\right)^{1/2}\in\mathbb{R}^{+}$,
called the \href{https://hsm.stackexchange.com/questions/12887/what-is-the-origin-of-the-japanese-bracket}{Japanese bracket}.

\begin{cBoxA}{}
\begin{defn}
\label{def:Let--called}Let $m\in\mathbb{R}$ called the \textbf{order}\index{order}.
Let $0\leq\delta<\frac{1}{2}<\rho\leq1$. The \textbf{class of symbols}\index{class of symbols}
$S_{\rho,\delta}^{m}$ contains smooth functions $a\in C^{\infty}\left(T^{*}M\right)$
such that on any charts of $U\subset M$ with coordinates $x=\left(x_{1},\ldots x_{n}\right)$
and associated dual coordinates $\xi=\left(\xi_{1},\ldots\xi_{n}\right)$
on $T_{x}^{*}U$, any multi-index $\alpha,\beta\in\mathbb{N}^{n}$,
there is a constant $C_{\alpha,\beta}$ such that 
\begin{equation}
\left|\partial_{\xi}^{\alpha}\partial_{x}^{\beta}a\left(x,\xi\right)\right|\leq C_{\alpha,\beta}\left\langle \xi\right\rangle ^{m-\rho\left|\alpha\right|+\delta\left|\beta\right|}\label{eq_symbol-1}
\end{equation}

The case $\rho=1$,$\delta=0$ is very common. We denote $S^{m}:=S_{1,0}^{m}$.
\end{defn}

\end{cBoxA}

For example on a chart, $p\left(x,\xi\right)=\left\langle \xi\right\rangle ^{m}$
is a symbol $p\in S^{m}$.

If $m\leq m'$ then $S^{m}\subset S^{m'}$ . We have $S^{-\infty}:=\bigcap_{m\in\mathbb{R}}S^{m}=\mathcal{S}\left(T^{*}M\right)$.
\begin{rem}
The geometric meaning of Definition \ref{def:Let--called} may be
not very clear a priori. Hörmander improved the geometrical meaning
in \cite{hormander1979weyl,hormander_3} by introducing an associated
metric on phase space $T^{*}M$. See also \cite{nicola_rodino_livre_11},
\cite{faure_tsujii_Ruelle_resonances_density_2016}.
\end{rem}

\section{\label{sec:Espaces-fibr=0000E9s-vectoriels}Vector bundles and topology}

Some references for this appendix are Fedosov \cite{fedosov96}, page
11, Hatcher \cite{hatcher_ktheory} p.14.

We will give precise definitions in Section \ref{subsec:Vector-bundles-in}.
We begin in Section \ref{subsec:Topologie-d'un-fibr=0000E9} and \ref{subsec:Topologie-d'un-fibr=0000E9-1}
by a description of vector bundles based on examples and sufficient
to understand the case of dimension $n=1$ used in this paper.

A complex (or real) vector bundle $F\rightarrow B$ of rank $r$ is
a collection of complex (or real) vector spaces $F_{x}$ of dimension
$r$, called fiber, and continuously parametrized by points $x$ on
a manifold $B$, called ``base space''. Locally over $U\subset B$,
$F$ is isomorphic to a direct product $U\times\mathbb{C}^{r}$.

\subsection{\label{subsec:Topologie-d'un-fibr=0000E9}Topology of a real vector
bundle of rank 1 on $S^{1}$}

\subsubsection{Construction of a real vector bundle of rank $1$ on $S^{1}$}

The simplest example is the case where the base space is the circle
$B=S^{1}$ and the rank is $r=1$, i.e. each fiber is isomorphic (as
a vector space) to the real line $\mathbb{R}$.

One can easily imagine two examples of real fiber space of rank $1$
on $S^{1}$:
\begin{itemize}
\item The \textbf{trivial bundle} $S^{1}\times\mathbb{R}$ that we obtain
from the trivial bundle $\left[0,1\right]\times\mathbb{R}$ on the
segment $x\in\left[0,1\right]$ (i.e. direct product) and identifying
the points $\left(0,t\right)\sim\left(1,t\right)$, for all $t\in\mathbb{R}$.
\item The \textbf{Moebius bundle}, which is obtained from the bundle $\left[0,1\right]\times\mathbb{R}$
on the segment $x\in\left[0,1\right]$, identifying $\left(0,t\right)\sim\left(1,-t\right)$,
$t\in\mathbb{R}$.
\end{itemize}
\begin{center}
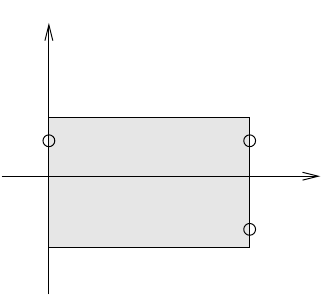
\par\end{center}

The Moebius bundle is not isomorphic to the trivial bundle. One way
to justify this is that in the case of the trivial bundle, the complement
of the null section $\left(s\left(x\right)=0,\forall x\right)$ has
two connected components, whereas for the bundle of Moebius, the complement
has only one component. (Make a paper construction that is cut with
scissors according to $s\left(x\right)=0$ to observe this).
\begin{center}
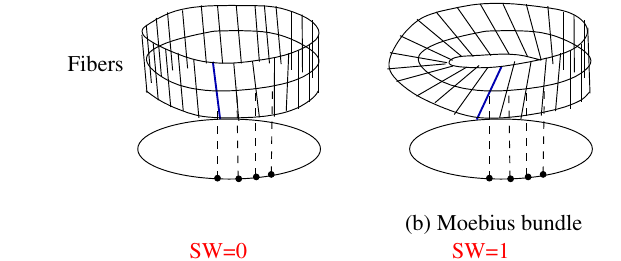
\par\end{center}

\begin{cBoxB}{}
\begin{thm}
Any real vector bundle $F\rightarrow S^{1}$ of rank 1 is isomorphic
to the trivial bundle or to the Möebius's bundle. In other words,
there are only two classes of equivalences:
\[
\mbox{Vect}_{\mathbb{R}}^{1}\left(S^{1}\right)=\left\{ 0,1\right\} 
\]
associated with the Stiefel-Whitney index $SW=0$: trivial bundle,
$SW=1$: bundle of Moebius.
\end{thm}

\end{cBoxB}

\begin{proof}
Starting from any bundle $F\rightarrow S^{1}$ of rank $1$, we cut
the base space $S^{1}$ at a point, and we are left with the bundle
$\left[0,1\right]\times\mathbb{R}$ over $x\in\left[0,1\right]$.
To reconstruct the initial bundle $F$, there are two possibilities:
for all $t\in\mathbb{R},$ identify $\left(0,t\right)\sim\left(1,t\right)$,
or $\left(0,t\right)\sim\left(1,-t\right)$, which gives the trivial
or Moebius bundle respectively.
\end{proof}
\begin{rem}
~

\textbullet{} the Stiefel-Whitney index $SW=0,1$ gives the number
of half turns that the fibers make above the base space $S^{1}$.
The case $SW=2$ (one full turn) is isomorphic to the trivial bundle.
We therefore agree that the index $SW\in\mathbb{Z}/\left(2\mathbb{Z}\right)$,
i.e. $SW$ is an integer modulo $2$. It is interesting to have the
additive structure on the SW indices ($1+1=0$ for example).

\textbullet Note that in the space $\mathbb{R}^{3}$, a ribbon making
a turn, i.e. $SW=2$, can not be deformed continuously towards the
trivial bundle. \footnote{Because if we cut this ribbon on the section $s=0$, we obtain two
ribbons interlaced, whereas the same cut for a trivial ribbon gives
two separate ribbons}. This restriction is due to the embedding in the space $\mathbb{R}^{3}$
(in $\mathbb{R}^{4}$, this would be possible), and is not an intrinsic
property of the bundle that is nevertheless trivial.
\end{rem}

\subsubsection{Topology of a real vector bundle of rank $1$ over $S^{1}$ from
the zeros of a section}

\begin{cBoxA}{}
\begin{defn}
\label{def:section} If $F\rightarrow B$ is a vector space, a \textbf{global
section} of the bundle is an application (continuous or $C^{\infty}$)
$s:B\rightarrow F$ such that each base point $x\in B$ is maped to
a point in the fiber $s\left(x\right)\in F_{x}$. We note
\begin{equation}
C^{\infty}\left(B,F\right)\label{eq:sections_F}
\end{equation}
the space of the smooth sections of the bundle $F$.
\end{defn}

\end{cBoxA}

\begin{center}
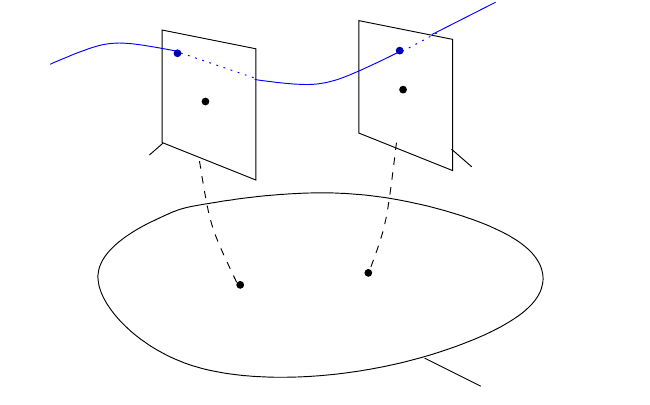
\par\end{center}

We call zeros of the section $s$ the points $x\in B$ such that $s\left(x\right)=0$.
Let us first consider the very simple and instructive case of a real
bundle of rank $1$ on $S^{1}$. A section is locally like a real
value numerical function, so generically, it vanishes transversely
at isolated points. Note that \textquotedbl generic\textquotedbl{}
means \textquotedbl except for exceptional case\textquotedbl . The
following figure shows that we have the following result:

\begin{cBoxB}{}
\begin{thm}
\label{th:zero_rang_1} If $F\rightarrow S^{1}$ is a real bundle
of rank $1$ on $S^{1}$, and $s$ is a ``generic'' section, then
the topological index $SW\left(F\right)$ is given by
\[
SW\left(F\right)=\sum_{x\mbox{ t.q. }s\left(x\right)=0}\sigma_{s}\left(x\right)
\]
where $\sigma_{s}\left(x\right)=1$ for a generic zero of section
$s$. The sum is obtained modulo $2$, and so $SW\left(F\right)\in\mathbb{Z}_{2}=\left\{ 0,1\right\} $.
The result is independent of the chosen section $s$.
\end{thm}

\end{cBoxB}

\begin{center}
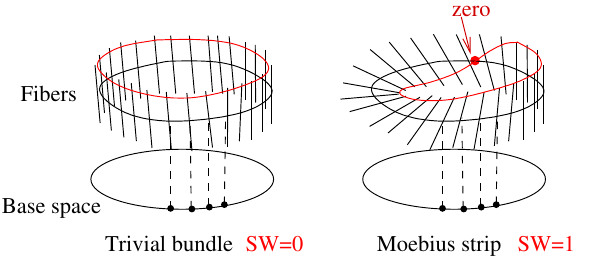
\par\end{center}

\subsection{\label{subsec:Topologie-d'un-fibr=0000E9-1}Topology of a complex
rank $1$ vector bundle over $S^{2}$}

We proceed similarly to the previous Section \ref{subsec:Topologie-d'un-fibr=0000E9}.

\subsubsection{\label{subsec:Construction-of-a}Construction of a complex vector
bundle of rank $1$ on $S^{2}$}

Let's first see how to build a complex fiber space of rank $1$ over
$S^{2}$. We cut the sphere $S^{2}$ along the equator $S^{1}$, obtaining
two hemispheres $H_{1}$ and $H_{2}$. We get two trivial bundles
$F_{1}=H_{1}\times\mathbb{C}$ and $F_{2}=H_{2}\times\mathbb{C}$
on each hemisphere. To construct a bundle on $S^{2}$, it is enough
to decide how to \textquotedbl connect\textquotedbl{} or \textquotedbl identify\textquotedbl{}
the fibers of $F_{1}$ above the equator with those of $F_{2}$. Note
$\theta\in S^{1}$ the angle \footnote{Here, we note $S^{1}$ the circle. $\theta\in S^{1}$ is therefore
marked with an angle $\theta\in\left[0,2\pi\right]$.} (longitude) that characterizes a point on the equator. Note $\varphi\left(\theta\right)\in S^{1}$
the angle which means that the fiber $F_{2}\left(\theta\right)$ is
identified to the fiber $F_{1}\left(\theta\right)$ after a rotation
of angle $\varphi\left(\theta\right)$: a $v\in F_{1}\left(\theta\right)\equiv\mathbb{C}$
is identified with the $e^{i\varphi\left(\theta\right)}v\in F_{2}\left(\theta\right)$.
After gluing that way the two hemispheres and the fibers above the
equator, we obtain a complex vector bundle $F\rightarrow S^{2}$ of
rank 1. Thus the bundle $F$ that we have just built is defined by
its clutching function on the equator
\[
\varphi:\theta\in S^{1}\rightarrow\varphi\left(\theta\right)\in S^{1}
\]
It is a continuous and periodic function so: $\varphi\left(2\pi\right)\equiv\varphi\left(0\right)\left[2\pi\right]$,
or
\begin{equation}
\varphi\left(2\pi\right)=\varphi\left(0\right)+2\pi\mathcal{C},\qquad\mathcal{C}\in\mathbb{Z},\label{eq:def_degre-1}
\end{equation}
with the integer \emph{$\mathcal{C}\in\mathbb{Z}$} that represents
the number of revolutions that $\varphi$ makes when $\theta$ goes
around. We call $\mathcal{C}$ the degree of the application\textbf{
$\varphi:S^{1}\rightarrow S^{1}$}. It is clear that two functions
$\varphi,\varphi'$ are homotopic if and only if they have the same
degree $\mathcal{C}=\mathcal{C}'$, and therefore the bundles $F$
and $F'$ are isomorphic if and only if $\mathcal{C}=\mathcal{C}'$.

\begin{center}
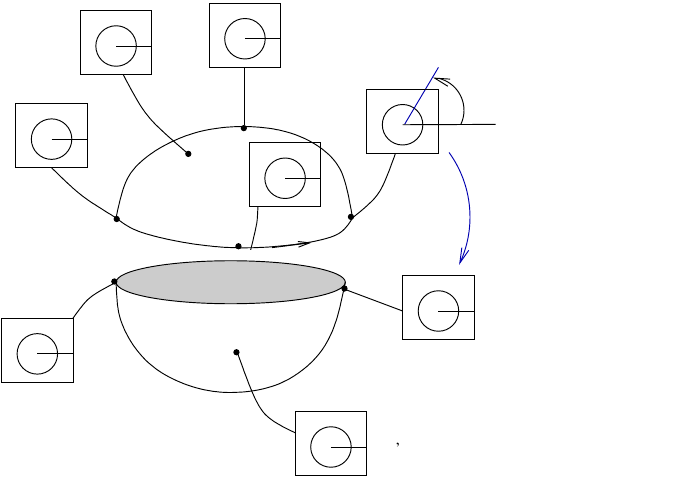
\par\end{center}

\begin{center}
\includegraphics[scale=0.2]{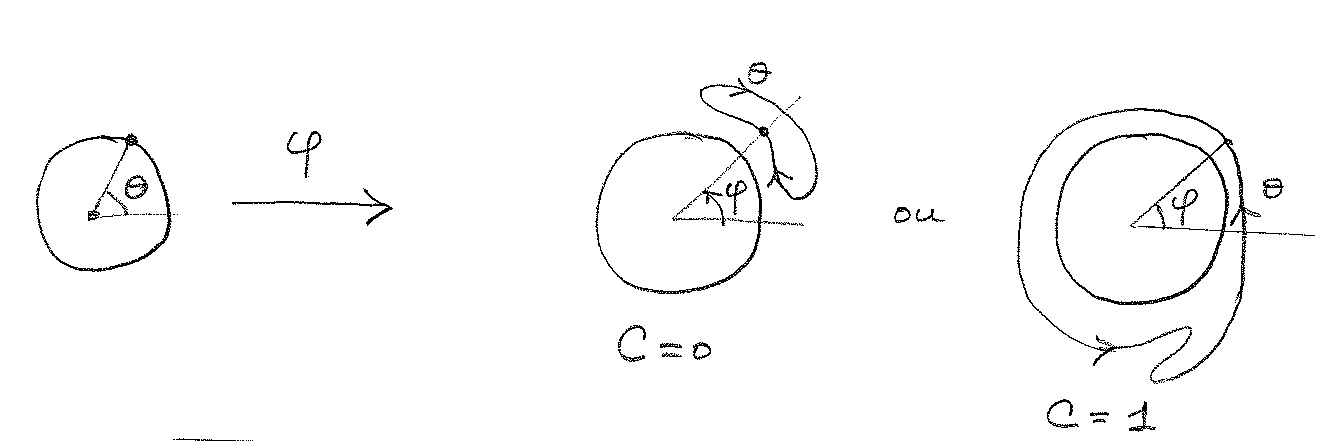}
\par\end{center}

\begin{cBoxB}{}
\begin{thm}
\label{th:fibre_S2}Any complex fiber bundle $F\rightarrow S^{2}$
of rank $1$ is isomorphic to a bundle constructed as above with a
clutching function $\varphi$ on the equator. Its topology is characterized
by an integer $\mathcal{C}\in\mathbb{Z}$ called (1st) Chern index
given by $\mathcal{C}=\mathrm{deg}\left(\varphi\right)$. In other
words the equivalence class of rank 1 complex vector bundle on $S^{2}$
is
\[
\mbox{Vect}_{\mathbb{C}}^{1}\left(S^{2}\right)=\mathbb{Z}
\]
\end{thm}

\end{cBoxB}

\begin{proof}
We must show that every bundle $F$ is isomorphic to a bundle constructed
as above. Starting from a given bundle $F$, we cut the base space
$S^{2}$ along the equator denoted $S^{1}$ to obtain two bundles
$F_{1}\rightarrow H_{1}$ and $F_{2}\rightarrow H_{2}$. Each of these
bundles is trivial because\cite[corrollaire 1.8 p.21]{hatcher_ktheory}
the base spaces are disks (contractile spaces). The bundle $F$ is
thus defined by its clutching function above the equator $S^{1}$,
$\varphi:S^{1}\rightarrow S^{1}$.
\end{proof}
Consider the example of the tangent bundle $TS^{2}$ of the sphere.
$TS^{2}$ can be identified with a complex bundle of rank $1$ because
$S^{2}$ is orientable.

\begin{cBoxB}{}
\begin{thm}
\label{th:fibre_TS2_C}The tangent bundle $TS^{2}$ has Chern index
\begin{equation}
\mathcal{C}\left(TS^{2}\right)=+2\label{eq:C_S2}
\end{equation}
and is therefore non trivial.
\end{thm}

\end{cBoxB}

\begin{proof}
We will calculate the degree $C$ of its recollection function defined
by Eq. (\ref{eq:def_degre}). We proceed as in the proof above. We
trivialize the bundle above $H_{1}$, and $H_{2}$, and we deduce
the degree $C$ of the gluing function. See figure that represents
the two hemispheres seen from above and below with a vector field
on each. We find $\mathcal{C}=+2$.

\begin{center}
\scalebox{0.8}[0.8]{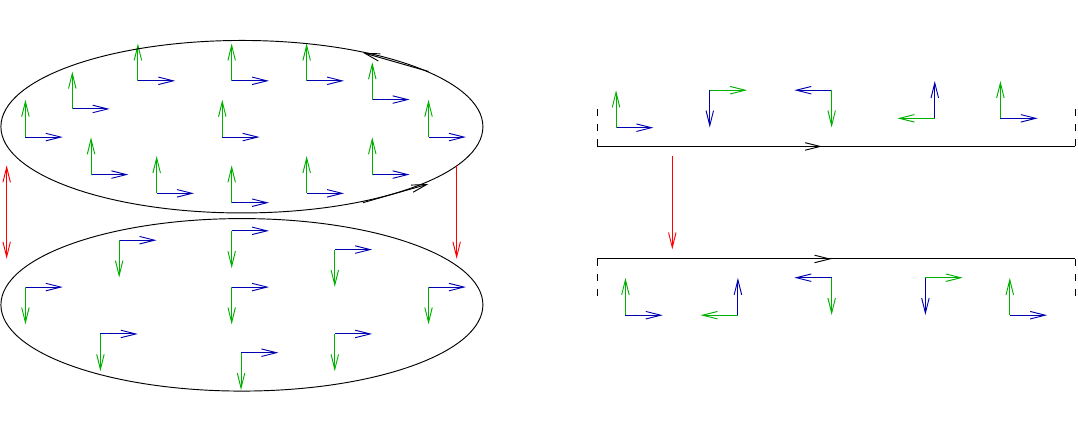}
\par\end{center}

\end{proof}
\begin{rem}
The trivial bundle $S^{2}\times\mathbb{C}$ has the Chern index $\mathcal{C}=0$.
\end{rem}

\subsubsection{Topology of the rank 1 vector bundle on $S^{2}$ from the zeros of
a section}

There is a result analogous to Thm. \ref{th:zero_rang_1} for a complex
bundle $F\rightarrow S^{2}$ of rank $1$ on $S^{2}$. Before establishing
it, let us notice that a section $s$ of such a bundle is locally
like a function with two variables and with values \LyXZeroWidthSpace \LyXZeroWidthSpace in
$\mathbb{C}$, so generically, it vanishes transversely at isolated
points. If $\theta\in S^{1}$ parameterizes a small circle of points
$x_{\theta}$ around a zero $x\in S^{2}$ of $s$,then by hypothesis,
the value of the section $s\left(x_{\theta}\right)\in F_{x_{\theta}}\equiv\mathbb{C}$
is non-zero for all $x_{\theta}$, and we write $\varphi\in S^{1}$
his argument. For each zero $x$ of the section $s$ is therefore
associated an application $\varphi:\theta\rightarrow\varphi\left(\theta\right)$
whose degree, also called index of the zero (defined by Eq. (\ref{eq:def_degre})),
will be noted $\sigma_{s}\left(x\right)\in\mathbb{Z}$. Generically,
$\sigma_{s}\left(x\right)=\pm1$. (Note that the sign of $\sigma_{s}\left(x\right)$
depends on the chosen orientation of the base space and the fiber.
In the case of the tangent bundle on $S^{2}$, these two orientations
are not independent, and the result $\sigma_{s}\left(x\right)$ becomes
independent of the choice of orientation).

\begin{center}
\includegraphics[scale=0.2]{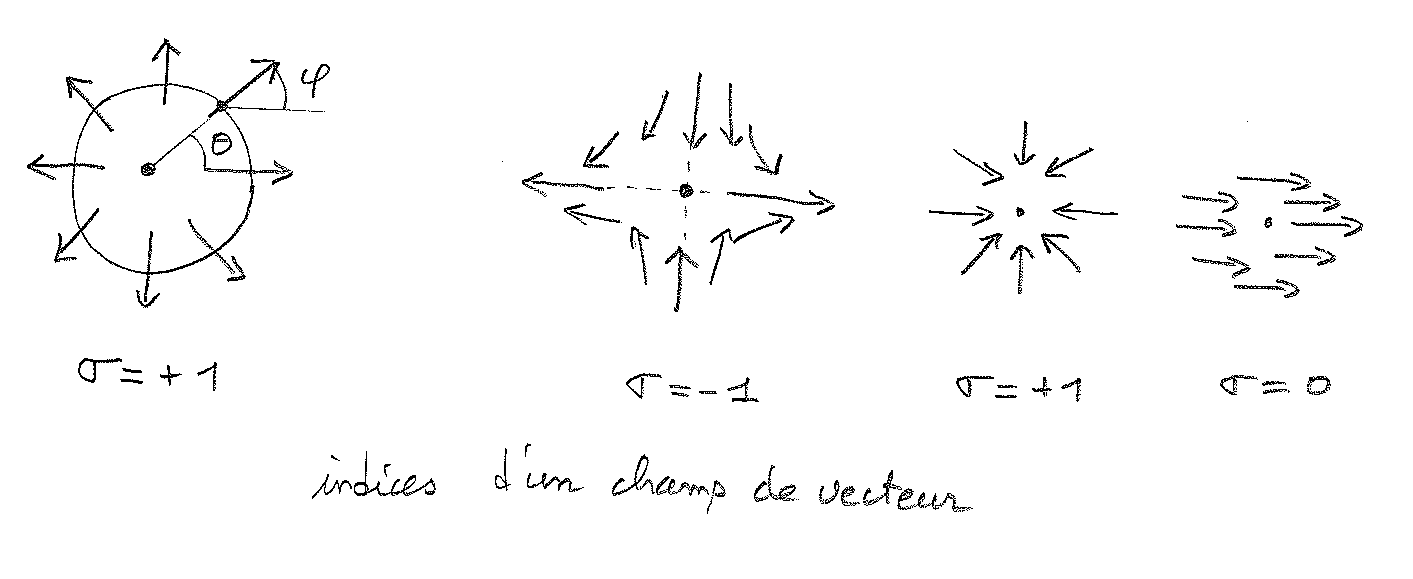}
\par\end{center}

\begin{cBoxB}{}
\begin{thm}
If $F\rightarrow S^{2}$ is a complex bundle of rank $1$ on $S^{2}$,
and $s$ is a \textquotedbl generic\textquotedbl{} section, then
the topological index of Chern $\mathcal{C}\left(F\right)$ is given
by
\begin{equation}
\mathcal{C}\left(F\right)=\sum_{x\mbox{ t.q. }s\left(x\right)=0}\sigma_{s}\left(x\right)\qquad\in\mathbb{Z}\label{eq:C_section}
\end{equation}

where $\sigma_{s}\left(x\right)=\pm1$ characterizes the degree of
zero. The result is independent of the chosen section $s$ .
\end{thm}

\end{cBoxB}

\begin{proof}
In the proof of the theorem \ref{th:fibre_S2}, we have constructed
sections $v_{1},v_{2}$ for the respectively bundles $F\rightarrow H_{1}$,
$F\rightarrow H_{2}$, that never vanish. If we modify these sections
$v_{1},v_{2}$ to make them coincide on the equator for the purpose
of constructing a global section $s$ of the bundle $F\rightarrow S^{2}$,
we can get do this except in points isolated, which will be the zeros
of $s$, and one realizes that the sum of the indices will be equal
to the degree of the clutching function $\varphi$ therefore equal
to $C\left(F\right)$.
\end{proof}

\paragraph{Example of the bundle $TS^{2}$}

The following figure shows a vector field on the $S^{2}$ sphere.
It is a global section of the tangent bundle. This vector field has
two zeros with indices $+1$ each. Thus we find $\mathcal{C}\left(TS^{2}\right)=+2$,
i.e. Eq. (\ref{eq:C_S2}).

\begin{center}
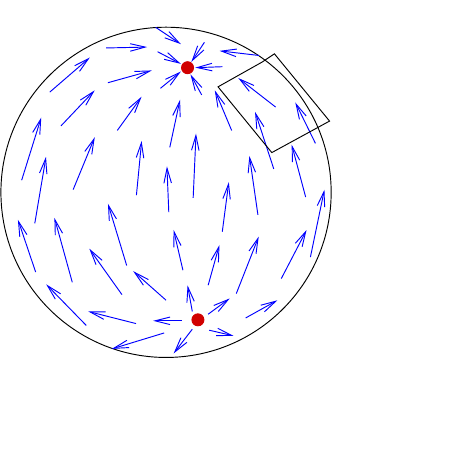
\par\end{center}
\begin{rem}
If we want to give an explicit computation we need an explicit global
section (or vector field on $TS^{2}$). We can take the fixed vector
in $\mathbb{R}^{3}$: $V=\left(0,0,1\right)$ oriented along the $z$
axis. Then for a given point $x\in S^{2}$ we choose:

\begin{equation}
s\left(x\right)=P_{x}V\in T_{x}S^{2}\label{eq:Choix_V}
\end{equation}
where $P_{x}:\mathbb{R}^{3}\rightarrow T_{x}S^{2}$ is the orthogonal
projector given by $P_{x}=\mathrm{Id}-|x\rangle\langle x|.\rangle$.
We get 
\begin{equation}
s\left(x\right)=V-x\langle x|V\rangle=\left(-x_{3}x_{1},-x_{3}x_{2},1-x_{3}^{2}\right).\label{eq:s_on_TS2}
\end{equation}
The vector field $s\left(x\right)$ vanishes at the north and south
pole. At distance $\epsilon$ of north pole $\left(0,0,1\right)$,
we use local oriented coordinates $\left(x_{1},x_{2}\right)\equiv\epsilon e^{i\theta}$
and get $s\left(x\right)=\left(-x_{1},-x_{2},0\right)+O\left(\epsilon^{2}\right)=-e^{i\theta}+O\left(\epsilon^{2}\right)$.
The map $e^{i\theta}\in S^{1}\rightarrow-e^{i\theta}\in S^{1}$ has
degree $1$ hence the zero has index $\sigma=+1$. At distance $\epsilon$
of south pole $\left(0,0,-1\right)$, we use local oriented coordinates
$\left(x_{2},x_{1}\right)\equiv\epsilon e^{i\theta}$ and get $s\left(x\right)=\left(x_{1},x_{2},0\right)+O\left(\epsilon^{2}\right)=e^{i\theta}+O\left(\epsilon^{2}\right)$.
The map $e^{i\theta}\in S^{1}\rightarrow e^{i\theta}\in S^{1}$ has
degree $1$ hence the zero has again index $\sigma=+1$. Formula (\ref{eq:C_section})
gives 
\[
\mathcal{C}\left(TS^{2}\right)=+1+1=+2.
\]
\end{rem}

\subsubsection{Topology of the rank 1 vector bundle on $S^{2}$ from a curvature
integral in differential geometry}

Let $F\rightarrow S^{2}$ be a complex vector bundle of rank 1 over
$S^{2}$. Let us assume\footnote{This is the case in the model of Section \ref{sec:The-model} and
every vector bundle can be realized like this, see \cite{fedosov96}.} that there exists a fixed vector space $\mathbb{C}^{d}$ such that
for every $x\in S^{2}$, the fiber $F_{x}\subset\mathbb{C}^{d}$ is
a linear subspace of $\mathbb{C}^{d}$ for some $d\geq1$. For every
point $x\in S^{2}$, let us denote $P_{x}:\mathbb{C}^{d}\rightarrow\mathbb{C}^{d}$
the orthogonal projector onto $F_{x}$. Then if $s\in C^{\infty}\left(S^{2};F\right)$
is a smooth section we can consider $s\in C^{\infty}\left(S^{2};\mathbb{C}^{d}\right)$
as a $d$ multi-components function on $S^{2}$. If $V\in T_{x}S^{2}$
is a tangent vector at point $x\in S^{2}$, the derivative $V\left(s\right)\in\mathbb{C}^{d}$
can be projected onto $F_{x}$. We get
\[
\left(D_{V}s\right)\left(x\right):=P_{x}V\left(s\right)\in F_{x}
\]
called the \textbf{covariant derivative} of $s$ along $V$ at point
$x$. It measures the variations of $s$ within the fibers $F$. Since
$V\left(s\right)=ds\left(V\right)$ where $ds$ means the differential\footnote{In local coordinates $x=\left(x_{1},x_{2}\right)\in\mathbb{R}^{2}$
on $S^{2}$, if $f\left(x_{1},x_{2}\right)$ is a function, then its
\textbf{differential} is written
\[
df=\sum_{k}\left(\frac{\partial f}{\partial x_{k}}\right)dx_{k}
\]
and a \textbf{tangent vector} is written $V=\sum_{k}V_{k}\frac{\partial}{\partial x_{k}}$
. Then since $df\left(V\right)=V\left(f\right)$ gives in particular
for the function $x_{k}$ that $dx_{k}\left(\frac{\partial}{\partial x_{l}}\right)=\frac{\partial x_{k}}{\partial x_{l}}=\delta_{k=l}$,
we get that $df\left(V\right)=\sum_{k}\left(\frac{\partial f}{\partial x_{k}}\right)V_{k}$.} , we usually write
\[
Ds:=Pds
\]
for the \textbf{covariant derivative} or \textbf{Levi-Civita connection}
(in differential geometry, $Pds\in C^{\infty}\left(S^{2};\Lambda^{1}\otimes F\right)$
is a one form valued in $F$).

Suppose that $U\subset S^{2}$ and for every point $x\in U$ one has
$v\left(x\right)\in F_{x}$ a unitary vector that depends smoothly
on $x\in U$. This is called a local \textbf{unitary trivialization}
of $F\rightarrow U$ (as in the proof of Theorem \ref{th:fibre_TS2_C}).
Since the fiber $F_{x}$ is dimension $1$, the vector $v\left(x\right)$
is a unitary basis of $F_{x}$ and if $V\in T_{x}S^{2}$, the covariant
derivative $\left(D_{V}v\right)\left(x\right)=P_{x}V\left(v\right)\in F_{x}$
can expressed in this basis with one complex component:
\[
\left(D_{V}v\right)\left(x\right)=\underbrace{\left(A\left(x\right)\right)\left(V\right)}_{\in\mathbb{R}}v\left(x\right)
\]
where $A\left(x\right)=i\mathcal{A}\left(x\right)$ is $i\mathbb{R}$
valued\footnote{$A$ is imaginary valued from the fact that $\langle v|v\rangle=1$
hence 
\[
0=d\langle v|v\rangle=\langle Dv|v\rangle+\langle v|Dv\rangle=2\mathrm{Re}\left(\langle v|Av\rangle\right)=2\mathrm{Re}\left(A\right).
\]
} linear form on $T_{x}S^{2}$ (a cotangent vector) called\footnote{If $s\in C\left(S^{2};F\right)$ is an arbitrary section, then locally
one can write $s\left(x\right)=\phi\left(x\right)v\left(x\right)$
with some complex component $\phi\left(x\right)\in\mathbb{C}$. Then
\begin{align*}
Ds & =D\left(\phi v\right)=\left(d\phi\right)v+\phi Dv=\left(d\phi\right)v+\phi Av\\
 & =\left(d\phi+\phi A\right)v=\sum_{k}\left(\frac{\partial\phi}{\partial x_{k}}+A_{k}\phi\right)\left(dx_{k}\right)v,
\end{align*}
with $A=\sum_{k}A_{k}dx_{k}$. Writing $A=i\mathcal{A}$, it shows
that the components of the covariant derivatice $Ds$ with respect
to the unitary trivialization $v\left(x\right)$ and local coordinates
$\left(x_{k}\right)_{k}$ on $U$ are $\left(\frac{\partial\phi}{\partial x_{k}}+i\mathcal{A}_{k}\phi\right)_{k}$.
In quantum physics books it is common to see the expression $\left(\frac{\partial\phi}{\partial x_{k}}+i\mathcal{A}_{k}\phi\right)_{k}$
for a definition of the ``covariant derivative'' or ``minimal coupling'',
e.g. \cite[p.31]{zuber}.} \textbf{connection one form}. In short, 
\begin{equation}
Dv=Av.\label{eq:Dv}
\end{equation}
Let
\begin{equation}
\Omega:=dA\label{eq:def_Omega}
\end{equation}
be the two form\footnote{In local coordinates if $A=\sum_{k}A_{k}dx_{k}$ is a one form with
components $A_{k}\left(x\right)$ then $dA=\sum_{k,l}\frac{\partial A_{k}}{\partial x_{l}}dx_{l}\wedge dx_{k}$.} called the \textbf{curvature }of the connection.

\begin{cBoxB}{}
\begin{lem}
Let $F\rightarrow S^{2}$ be a rank 1 complex vector bundle with $F_{x}\subset\mathbb{C}^{d}$.
Let $v\left(x\right)\in F_{x}$ a given local unitary trivialization
and $Dv=Av$ with $A$ the connection one form and $\Omega$ the curvature
two form. Then
\[
A=\langle v|dv\rangle_{\mathbb{C}^{d}}=\sum_{k}\langle v|\frac{\partial v}{\partial x_{k}}\rangle dx_{k}
\]
\begin{equation}
\Omega=\langle dv|\wedge dv\rangle=\sum_{k,l}\langle\frac{\partial v}{\partial x_{k}}|\frac{\partial v}{\partial x_{l}}\rangle dx_{k}\wedge dx_{l}.\label{eq:iOmega}
\end{equation}
and $\Omega$ does not depend on the trivialization, hence is globally
defined on $S^{2}$. Finally the topological Chern index $\mathcal{C}$
defined in (\ref{eq:def_degre-1}) is given by the \textbf{curvature
integral}
\begin{equation}
\mathcal{C}=\frac{1}{2\pi}\iint_{S^{2}}i\Omega.\label{eq:curvature_integral_S2}
\end{equation}
\end{lem}

\end{cBoxB}

\begin{proof}
The orthogonal projector is given by
\[
P_{x}=|v_{x}\rangle\langle v_{x}|.\rangle
\]
hence the covariant derivative is given by $Ds=Pds=P_{x}=|v_{x}\rangle\langle v_{x}|ds\rangle$
and since by definition $Dv=Av$ we get $A=\langle v|dv\rangle$ and
\[
\Omega=dA=d\left(\sum_{k}\langle v|\frac{\partial v}{\partial x_{k}}\rangle dx_{k}\right)=\sum_{k,l}\langle\frac{\partial v}{\partial x_{l}}|\frac{\partial v}{\partial x_{k}}\rangle dx_{l}\wedge dx_{k}+\underbrace{\sum_{k,l}\langle v|\frac{\partial^{2}v}{\partial x_{l}\partial x_{k}}\rangle dx_{l}\wedge dx_{k}}_{=0}.
\]
The second term vanishes since $\left(\frac{\partial^{2}v}{\partial x_{l}\partial x_{k}}\right)_{k,l}$
is a symmetric array and $\left(dx_{l}\wedge dx_{k}\right)_{k,l}$
is antisymmetric. If we replace $v$ by another trivialization $v'\left(x\right)=e^{i\alpha\left(x\right)}v\left(x\right)$
(this is called a \textbf{Gauge transformation}) then
\begin{align*}
A' & =\langle v'|dv'\rangle=e^{-i\alpha}\sum_{k}\langle v|\frac{\partial e^{i\alpha}v}{\partial x_{k}}\rangle dx_{k}=\sum_{k}\left(i\frac{\partial\alpha}{\partial x_{k}}\right)dx_{k}+\langle v|\frac{\partial v}{\partial x_{k}}\rangle dx_{k}\\
 & =id\alpha+A
\end{align*}
is changed but
\[
\Omega'=dA'=idd\alpha+dA=\Omega
\]
is unchanged because $dd\alpha=\sum_{k,l}\left(\frac{\partial^{2}\alpha}{\partial x_{l}\partial x_{k}}\right)dx_{l}\wedge dx_{k}=0$.

As in Section \ref{subsec:Construction-of-a}, let $H_{1},H_{2}$
be the north and south hemispheres of $S^{2}$ and suppose that for
every point $x\in H_{1}$, $v_{1}\left(x\right)\in F_{x}$ is a unitary
vector that depends smoothly on $x$, i.e. $v_{1}$ is a trivialization
of $F\rightarrow H_{1}$. Suppose that $v_{2}$ is a trivialization
of $F\rightarrow H_{2}$ (as in the proof of Theorem \ref{th:fibre_TS2_C}).
Let $x\equiv\left(\theta,\varphi\right)$ denotes the spherical coordinates
on $S^{2}$. For a given $0\leq\theta\leq\frac{\pi}{2}$ on Hemisphere
$H_{1}$, let $\gamma_{\theta}:\varphi\in[0,2\pi]\rightarrow\gamma_{\theta}\left(\varphi\right)\in S^{2}$
be the closed path. Let $\psi_{\theta}^{\left(1\right)}\left(0\right)\in F_{\theta,0}$
and 
\[
\psi_{\theta}^{\left(1\right)}\left(\varphi\right)=e^{i\alpha_{\theta}^{\left(1\right)}\left(\varphi\right)}v_{1}\left(\theta,\varphi\right)\in F_{\theta,\varphi}
\]
 obtained for $0\leq\varphi\leq2\pi$ by \textbf{parallel transport,
}i.e. under the condition of zero covariant derivative
\[
\frac{D\psi_{\theta}^{\left(1\right)}}{d\varphi}=0\Leftrightarrow\frac{D\left(e^{i\alpha_{\theta}^{\left(1\right)}}v_{1}\right)}{d\varphi}=0\underset{(\ref{eq:Dv})}{\Leftrightarrow}i\frac{d\alpha_{\theta}^{\left(1\right)}}{d\varphi}v_{1}+Av_{1}=0
\]
giving that
\begin{equation}
\alpha_{\theta}^{\left(1\right)}\left(2\pi\right)-\alpha_{\theta}^{\left(1\right)}\left(0\right)=\int_{\gamma_{\theta}}iA\eq{\ref{eq:def_Omega},\mathrm{Stokes}}\iint_{H_{\theta}}i\Omega\label{eq:alpha1}
\end{equation}
where $H_{\theta}=\left\{ \left(\theta',\varphi'\right),\theta'\geq\theta,\varphi'\in\left[0,2\pi\right]\right\} \subset S^{2}$
is a surface with boundary $\gamma_{\theta}$. The angle $\alpha_{\theta}^{\left(1\right)}\left(2\pi\right)$
is called the \textbf{holonomy} of the connection on the closed path
$\gamma_{\theta}$ and also called \textbf{Berry's phase} after the
paper of M. Berry \cite{berry1} that shows its natural manifestation
in quantum mechanics, see also \cite{faure_cours_thm_adiabatique_2018}.
We can do the same on the south hemisphere $H_{2}$ with $v_{2}$
and angles $\alpha_{\theta}^{\left(2\right)}$, giving at $\theta=0$,
\begin{equation}
\alpha_{0}^{\left(2\right)}\left(2\pi\right)-\alpha_{0}^{\left(2\right)}\left(0\right)=-\iint_{H_{2}}i\Omega\label{eq:alpha_2}
\end{equation}
with opposite signe because the orientation of $\gamma_{0}$ is reversed.
In particular, on the equator $\theta=0$ that belongs to both Hemisphere,
we have for every $\varphi$ that
\begin{align*}
v_{2}\left(0,\varphi\right) & =e^{i\beta\left(\varphi\right)}v_{1}\left(0,\varphi\right)
\end{align*}
and by definition of Chern index $\mathcal{C}$,
\[
\beta\left(2\pi\right)\eq{\ref{eq:def_degre-1}}\beta\left(0\right)+2\pi\mathcal{C}
\]
Also
\[
\psi_{0}^{\left(1\right)}\left(\varphi\right)=e^{i\alpha_{0}^{\left(1\right)}\left(\varphi\right)}v_{1}\left(0,\varphi\right),\quad\psi_{0}^{\left(2\right)}\left(\varphi\right)=e^{i\alpha_{0}^{\left(2\right)}\left(\varphi\right)}v_{2}\left(0,\varphi\right),
\]
and since the parallel transport preserves the angles, $\psi_{0}^{\left(2\right)}\left(\varphi\right)=e^{ic}\psi_{0}^{\left(1\right)}\left(\varphi\right)$
with a constant $c$ (independent on $\varphi$). Finally we get
\begin{align*}
v_{2}\left(0,\varphi\right) & =e^{i\beta\left(\varphi\right)}v_{1}\left(0,\varphi\right)=e^{i\left(\beta\left(\varphi\right)-\alpha_{0}^{\left(1\right)}\left(\varphi\right)\right)}\psi_{0}^{\left(1\right)}\left(\varphi\right)=e^{i\left(\beta\left(\varphi\right)-\alpha_{0}^{\left(1\right)}\left(\varphi\right)-c\right)}\psi_{0}^{\left(2\right)}\left(\varphi\right)\\
 & =e^{i\left(\beta\left(\varphi\right)-\alpha_{0}^{\left(1\right)}\left(\varphi\right)-c+\alpha_{0}^{\left(2\right)}\left(\varphi\right)\right)}v_{2}\left(0,\varphi\right)
\end{align*}
hence
\[
\beta\left(\varphi\right)=\alpha_{0}^{\left(1\right)}\left(\varphi\right)+c-\alpha_{0}^{\left(2\right)}\left(\varphi\right)
\]
and
\begin{align*}
\mathcal{C} & =\frac{1}{2\pi}\left(\beta\left(2\pi\right)-\beta\left(0\right)\right)=\frac{1}{2\pi}\left(\left(\alpha_{0}^{\left(1\right)}\left(2\pi\right)-\alpha_{0}^{\left(1\right)}\left(0\right)\right)-\left(\alpha_{0}^{\left(2\right)}\left(2\pi\right)-\alpha_{0}^{\left(2\right)}\left(0\right)\right)\right)\\
 & \eq{\ref{eq:alpha1},\ref{eq:alpha_2}}\frac{1}{2\pi}\left(\iint_{H_{1}}i\Omega+\iint_{H_{2}}i\Omega\right)=\frac{1}{2\pi}\iint_{S^{2}}i\Omega
\end{align*}
\end{proof}
\begin{rem}
Formula (\ref{eq:curvature_integral_S2}) is a special case of a more
general Chern-Weil formula formula (\ref{eq:Chern_index}) given below
for a general vector bundle $F\rightarrow S^{2n}$ of rank $r$.
\end{rem}

\begin{example}
For the special case of the tangent bundle $TS^{2}$, with fiber $T_{x}S^{2}\subset\mathbb{R}^{3}$,
if $i\Omega$ is the (2 form) Gauss curvature of the sphere (that
is, the curvature of the tangent bundle $TS^{2}$, which is the solid
angle), the \href{https://en.wikipedia.org/wiki/Gauss\%E2\%80\%93Bonnet_theorem}{Gauss-Bonnet formula}
gives:
\begin{equation}
\mathcal{C}=\frac{1}{2\pi}\int_{S^{2}}i\Omega=\frac{4\pi}{2\pi}=2\label{eq:Gauss-Bonnet}
\end{equation}
as in (\ref{eq:C_S2}).
\end{example}

\subsection{\label{subsec:Vector-bundles-in}General vector bundles over sphere
$S^{k}$}

\subsubsection{Definitions}

\begin{cBoxA}{}
\begin{defn}
We say that $\left(F,\pi,B\right)$ is a \textbf{complex vector bundle
of rank $r$} if $F,B$ are manifolds, $\pi:F\rightarrow B$ a map
such that there exists a covering $\left(U_{i}\right)_{i}$ of $B$
and diffeomorphisms $\varphi_{i}:\pi^{-1}\left(U_{i}\right)\rightarrow U_{i}\times\mathbb{C}^{r}$
such that
\end{defn}

\begin{enumerate}
\item $\pi:\pi^{-1}\left(U_{i}\right)\rightarrow U_{i}$ is the composition
of $\varphi_{i}$ with projection onto $U_{i}$
\item if $U_{i}\cap U_{j}\neq\emptyset$ then $\varphi_{i}\varphi_{j}^{-1}:\left(U_{i}\cap U_{j}\right)\times\mathbb{C}^{r}\rightarrow\left(U_{i}\cap U_{j}\right)\times\mathbb{C}^{r}$
is given by $\left(x,u\right)\rightarrow\left(x,f_{ij}\left(x\right)u\right)$
with $f_{ij}\left(x\right)\in GL\left(r,\mathbb{C}\right)$.
\end{enumerate}
\end{cBoxA}

We say that $\varphi_{i}$ are \textbf{trivialization functions,}
and $f_{ij}$ are \textbf{transition functions.}

\begin{cBoxB}{}
\begin{prop}
The transition functions satisfy the \textbf{cocycle conditions:}

\[
f_{ji}=f_{ij}^{-1},\forall x\in U_{i}\cap U_{j}\qquad f_{ij}f_{jk}f_{ki}=1,\,\forall x\in U_{i}\cap U_{j}\cap U_{k}
\]
Conversely functions $f_{ij}$ with cocycle conditions, define a unique
vector bundle.
\end{prop}

\end{cBoxB}

\begin{proof}
$f_{ji}^{-1}=\left(\varphi_{j}\varphi_{i}^{-1}\right)^{-1}=\varphi_{i}\varphi_{j}^{-1}=f_{ij}$.
And $f_{ij}f_{jk}f_{ki}=\left(\varphi_{i}\varphi_{j}^{-1}\right)\left(\varphi_{j}\varphi_{k}^{-1}\right)\left(\varphi_{k}\varphi_{i}^{-1}\right)=1$.

\noindent\fcolorbox{red}{white}{\begin{minipage}[t]{1\columnwidth - 2\fboxsep - 2\fboxrule}%
\begin{defn}
Two vector bundles $\left(F,\pi,B\right)$ and $\left(F',\pi',B\right)$
(with same base $B$) are \textbf{isomorphic} if there exists $h:F\rightarrow F'$
which preserves the fibers and such that $h:F_{x}\rightarrow F'_{x}$
is an isomorphism of linear spaces.
\end{defn}

\end{minipage}}
\end{proof}
We write $\mathrm{Vect}_{\mathbb{C}}^{r}\left(B\right)$ for the isomorphism
class of complex vector bundles of rank $r$ over $B$.

\begin{cBoxB}{}
\begin{prop}
Two vector bundles $F$ and $F'$ are isomorphic if and only if there
exists functions $h_{i}:U_{i}\rightarrow GL\left(n,\mathbb{C}\right)$
such that
\[
f'_{ij}=h_{i}f_{ij}h_{j}^{-1}
\]
where $f_{ij}$, $f'_{ij}$ are the transition functions.
\end{prop}

\end{cBoxB}

\begin{proof}
If $h$ is an isomorphism, define $h_{i}=\varphi'_{i}h\varphi_{j}^{-1}$.
Conversely, define $h=\left(\varphi'_{i}\right)^{-1}h_{i}\varphi_{i}$
on $U_{i}$ which does not depend on $i$.
\end{proof}

\subsubsection{\label{subsec:Complex-Vector-bundles}Complex Vector bundles over
spheres $S^{k}$}

Reference: Hatcher \cite{hatcher_ktheory} p.22. We treat the case
where the base space is a sphere
\[
B=S^{k}:=\left\{ \left(x_{1},\ldots x_{k+1}\right)\in\mathbb{R}^{k+1},\quad\sum_{j}x_{j}^{2}=1\right\} .
\]
The sphere $S^{k}=D_{1}^{k}\bigcup D_{2}^{k}$ can be decomposed in
two disks (or hemispheres), the north hemisphere $D_{1}^{k}$ where
$x_{k+1}\geq0$ and the south hemisphere $D_{2}^{k}$ where $x_{k+1}\leq0$.
The common set is the equator $S^{k-1}=D_{1}^{k}\cap D_{2}^{k}=\left\{ x\in\mathbb{R}^{k+1},x_{k+1}=0\right\} $
which is also a sphere $S^{k-1}$. So a vector bundle is described
by the transition function at the equator: $f_{21}:S^{k-1}\rightarrow GL\left(r,\mathbb{C}\right)$,
which is called the \textbf{clutching function}. Let us denote $\left[f_{21}\right]$
the homotopy class of the map $f_{21}$. The set of homotopy classes
is $\left[S^{k-1},GL\left(r,\mathbb{C}\right)\right]\equiv\left[S^{k-1},U\left(r\right)\right]=:\pi_{k-1}\left(U\left(r\right)\right)$
is called \href{https://en.wikipedia.org/wiki/Homotopy_group}{homotopy group}
of $U\left(r\right)$.

\begin{cBoxB}{}
\begin{prop}
Two vector bundles $F\rightarrow S^{k}$, $F'\rightarrow S^{k}$ are
isomorphic if and only if their clutching functions are homotopic
$\left[f_{21}\right]=\left[f'_{21}\right]$. In other words the group
of equivalence classes of vector bundles coincide with the homotopy
groups:
\[
\mathrm{Vect}_{\mathbb{C}}^{r}\left(S^{k}\right)\equiv\pi_{k-1}\left(U\left(r\right)\right).
\]
\end{prop}

\end{cBoxB}

\paragraph{Homotopy groups of spheres}

The groups $\mathrm{Vect}^{r}\left(S^{k}\right)=\pi_{k-1}\left(U\left(r\right)\right)$
can be obtained from \href{https://en.wikipedia.org/wiki/Homotopy_groups_of_spheres}{homotopy groups of the spheres}
$\pi_{m}\left(S^{n}\right)$ from the fact that 
\begin{equation}
U\left(r\right)/U\left(r-1\right)\equiv S^{2r-1}.\label{eq:quotient}
\end{equation}
This is obtained by observing that the unit sphere in $\mathbb{C}^{r}$
is $S^{2r-1}$ and thus, for $f\in U\left(r\right)$ and $e_{r}=\left(0,\ldots,0,1\right)\in\mathbb{C}^{r}$
we have $f\left(e_{r}\right)\in S^{2r-1}\subset\mathbb{C}^{r}$ that
characterizes $f$ up to $U\left(r-1\right)$, i.e. its action on
$\mathbb{C}^{r-1}$. See table \ref{tab:Groupes-homo_sphere}. See
\href{https://pi.math.cornell.edu/~hatcher/VBKT/VBpage.html}{Hatcher's book}.
\begin{center}
\begin{table}
\begin{centering}
\begin{tabular}{|c|c|c|c|c|c|c|c|}
\hline 
$\pi_{n}\left(S^{m}\right)$ & $\pi_{1}$ & $\pi_{2}$ & $\pi_{3}$ & $\pi_{4}$ & $\pi_{5}$ & $\pi_{6}$ & $\pi_{7}$\tabularnewline
\hline 
\hline 
$S^{1}$ & $\boxed{\mathbb{Z}}$ & $0$ & $0$ & $0$ & $0$ & $0$ & $0$\tabularnewline
\hline 
$S^{2}$ & $0$ & $\boxed{\mathbb{Z}}$ & $\mathbb{Z}$ & $\mathbb{Z}_{2}$ & $\mathbb{Z}_{2}$ & $\mathbb{Z}_{12}$ & $\mathbb{Z}_{2}$\tabularnewline
\hline 
$S^{3}$ & $0$ & $0$ & $\boxed{\mathbb{Z}}$ & $\mathbb{Z}_{2}$ & $\mathbb{Z}_{2}$ & $\mathbb{Z}_{12}$ & $\mathbb{Z}_{6}$\tabularnewline
\hline 
$S^{4}$ & $0$ & $0$ & $0$ & $\boxed{\mathbb{Z}}$ & $\mathbb{Z}_{2}$ & $\mathbb{Z}_{2}$ & $\mathbb{Z}\times\mathbb{Z}_{12}$\tabularnewline
\hline 
$S^{5}$ & $0$ & $0$ & $0$ & $0$ & $\boxed{\mathbb{Z}}$ & $\mathbb{Z}_{2}$ & $\mathbb{Z}_{2}$\tabularnewline
\hline 
$S^{6}$ & $0$ & $0$ & $0$ & $0$ & $0$ & $\boxed{\mathbb{Z}}$ & $\mathbb{Z}_{2}$\tabularnewline
\hline 
$S^{7}$ & $0$ & $0$ & $0$ & $0$ & $0$ & $0$ & $\boxed{\mathbb{Z}}$\tabularnewline
\hline 
\end{tabular}
\par\end{centering}
\caption{\foreignlanguage{french}{\label{tab:Groupes-homo_sphere}\foreignlanguage{english}{Homotopy
groups of the spheres $\pi_{m}\left(S^{n}\right)$}}}
\end{table}
\par\end{center}

We have
\[
\pi_{n}\left(S^{n}\right)=\mathbb{Z}
\]
which is the \href{https://en.wikipedia.org/wiki/Degree_of_a_continuous_mapping\#From_Sn_to_Sn}{degree}
and is computed as follows.

\begin{cBoxA}{}
\begin{defn}
\label{def:le--d'une}The \textbf{\href{https://fr.wikipedia.org/wiki/Degr\%C3\%A9_d\%27une_application}{degree}}
of a map $f:S^{m}\rightarrow S^{m}$ is
\begin{equation}
\mathrm{deg}\left(f\right):=\sum_{x\in f^{-1}\left(y\right)}\mathrm{sign}\left(\mathrm{det}\left(D_{x}f\right)\right)\in\mathbb{Z},\label{eq:def_degre}
\end{equation}
which is independent of the choice of the generic point $y\in S^{m}$.
In the case $f:S^{1}\rightarrow S^{1}$, the degree $\mathrm{deg}\left(f\right)$
is also called <<\href{https://en.wikipedia.org/wiki/Winding_number}{winding number}
of $f$>>.
\end{defn}

\end{cBoxA}

For $m<n$ we have
\[
\pi_{m}\left(S^{n}\right)=0,
\]
because the image of $f:S^{m}\rightarrow S^{n}$ is not onto and therefore
gives $f:\mathbb{R}^{m}\rightarrow S^{n}$ which can be retracted
to a point because $\mathbb{R}^{m}$ is contractible. For $m>n$,
the \href{https://en.wikipedia.org/wiki/Homotopy_groups_of_spheres}{homotopy groups of the spheres}
$\pi_{n}\left(S^{m}\right)$ are quite complicated and are not all
known.

\paragraph{Homotopy groups of $U\left(r\right)$}

From the fibration (\ref{eq:quotient}) and table \ref{tab:Groupes-homo_sphere}
we deduce table \ref{tab:Groupes-d'=0000E9quivalences-de}. See \cite{hatcher_ktheory},
\cite{hatcher_topology}.
\begin{center}
\begin{table}
\begin{centering}
\begin{tabular}{|c|c|c|c|c|c|c|}
\hline 
$\pi_{k}\left(U\left(r\right)\right)$ & $\pi_{1}$ & $\pi_{2}$ & $\pi_{3}$ & $\pi_{4}$ & $\pi_{5}$ & $\pi_{6}$\tabularnewline
\hline 
\hline 
$U\left(1\right)$ & $\boxed{\mathbb{Z}}$ & $0$ & $0$ & $0$ & $0$ & $0$\tabularnewline
\hline 
$U\left(2\right)$ & $\mathbb{Z}$ & $\boxed{0}$ & $\boxed{\mathbb{Z}}$ & $\mathbb{Z}_{2}$ & $\mathbb{Z}_{2}$ & $\mathbb{Z}_{12}$\tabularnewline
\hline 
$U\left(3\right)$ & $\mathbb{Z}$ & $0$ & $\mathbb{Z}$ & $\boxed{0}$ & $\boxed{\mathbb{Z}}$ & $\mathbb{Z}_{6}$\tabularnewline
\hline 
$U\left(4\right)$ & $\mathbb{Z}$ & $0$ & $\mathbb{Z}$ & $0$ & $\mathbb{Z}$ & $\boxed{0}$\tabularnewline
\hline 
$U\left(5\right)$ & $\mathbb{Z}$ & $0$ & $\mathbb{Z}$ & $0$ & $\mathbb{Z}$ & $0$\tabularnewline
\hline 
 & $\vdots$ & $\vdots$ & $\vdots$ & $\vdots$ & $\vdots$ & $\vdots$\tabularnewline
\hline 
$\tilde{K}\left(S^{k}\right)$ & $\mathbb{Z}$ & $0$ & $\mathbb{Z}$ & $0$ & $\mathbb{Z}$ & $0$\tabularnewline
\hline 
\end{tabular}$\qquad$%
\begin{tabular}{|c|c|c|c|c|c|c|}
\hline 
$\mathrm{Vect}^{r}\left(S^{k}\right)$ & $S^{2}$ & $S^{3}$ & $S^{4}$ & $S^{5}$ & $S^{6}$ & $S^{7}$\tabularnewline
\hline 
\hline 
$\mathrm{Vect^{1}}$ & $\boxed{\mathbb{Z}}$ & $0$ & $0$ & $0$ & $0$ & $0$\tabularnewline
\hline 
$\mathrm{Vect^{2}}$ & $\mathbb{Z}$ & $\boxed{0}$ & $\boxed{\mathbb{Z}}$ & $\mathbb{Z}_{2}$ & $\mathbb{Z}_{2}$ & $\mathbb{Z}_{12}$\tabularnewline
\hline 
$\mathrm{Vect^{3}}$ & $\mathbb{Z}$ & $0$ & $\mathbb{Z}$ & $\boxed{0}$ & $\boxed{\mathbb{Z}}$ & $\mathbb{Z}_{6}$\tabularnewline
\hline 
$\mathrm{Vect^{4}}$ & $\mathbb{Z}$ & $0$ & $\mathbb{Z}$ & $0$ & $\mathbb{Z}$ & $\boxed{0}$\tabularnewline
\hline 
$\mathrm{Vect^{5}}$ & $\mathbb{Z}$ & $0$ & $\mathbb{Z}$ & $0$ & $\mathbb{Z}$ & $0$\tabularnewline
\hline 
 & $\vdots$ & $\vdots$ & $\vdots$ & $\vdots$ & $\vdots$ & $\vdots$\tabularnewline
\hline 
$\tilde{K}\left(S^{k}\right)$ & $\mathbb{Z}$ & $0$ & $\mathbb{Z}$ & $0$ & $\mathbb{Z}$ & $0$\tabularnewline
\hline 
\end{tabular}
\par\end{centering}
\caption{\label{tab:Groupes-d'=0000E9quivalences-de}Equivalence groups of
complex vector bundles of rank $r$ over sphere $S^{k}$.\foreignlanguage{french}{
$\mathrm{Vect}^{r}\left(S^{k}\right)=\pi_{k-1}\left(U\left(r\right)\right)$.}}
\end{table}
\par\end{center}

\paragraph{Observations on table \ref{tab:Groupes-d'=0000E9quivalences-de}}
\begin{itemize}
\item $\mathrm{Vect^{2}}\left(S^{5}\right)\equiv\mathbb{Z}_{2}=\left\{ 0,1\right\} $:
means that there is only one class of non trivial bundles of rank
$2$ over $S^{5}$.
\item $\mathrm{Vect^{r}}\left(S^{3}\right)\equiv0,\forall r\geq1,$ means
that complex vector bundles over $S^{3}$ are all trivial.
\item $\mathrm{Vect^{1}}\left(S^{k\geq2}\right)\equiv0$ means that all
vector bundles of rank $1$ over $S^{k\geq2}$ are all trivial.
\end{itemize}
A remarkable observation is the following theorem: (K-theory\footnote{The symbol $K\left(X\right)$ comes from <<Klassen>> in german,
by A. Grothendieck 1957, see \href{https://inis.iaea.org/collection/NCLCollectionStore/_Public/38/098/38098190.pdf}{Lectures of Karoubi}.
The symbol $C\left(X\right)$ was already used.})

\begin{cBoxB}{}
\begin{thm}
\textbf{\label{thm:Bott-Periodicity-Theorem}<<Bott periodicity Theorem
1959>>}. If $2r\geq k$ then $\mathrm{Vect^{r}}\left(S^{k}\right)$
is independent on $r$. We denote $\tilde{K}\left(S^{k}\right):=\mathrm{Vect}^{r}\left(S^{k}\right)$
called group of K-theory. Moreover there is the periodicity property:
\begin{align*}
\tilde{K}\left(S^{k+2}\right) & =\tilde{K}\left(S^{k}\right)=\mathbb{Z}\text{ if }k\text{ is even}\\
 & =0\text{ if }k\text{ is odd}
\end{align*}
\end{thm}

\end{cBoxB}

For the proof, see \cite{hatcher_ktheory}.

\subsubsection{Topological Chern index $\mathcal{C}$ of a complex vector bundle
$F\rightarrow S^{2n}$ of rank $r\protect\geq n$}

From the table \ref{tab:Groupes-d'=0000E9quivalences-de}, if $F\rightarrow S^{2n}$
is a complex vector bundle of rank $r$, with $r\geq n$, then its
isomorphism class is characterized by an integer $\mathcal{C}\in\mathbb{Z}$
called \textbf{topological Chern index}.

Here is an explicit expression for $\mathcal{C}$. The equivalence
class of the bundle $F$ is characterized by the homotopy class of
the clutching function at the equator $g=f_{21}$,
\begin{equation}
g:S^{2n-1}\rightarrow U\left(r\right).\label{eq:clutching_g}
\end{equation}
which is the transition function from north hemisphere to south hemisphere.
\begin{center}
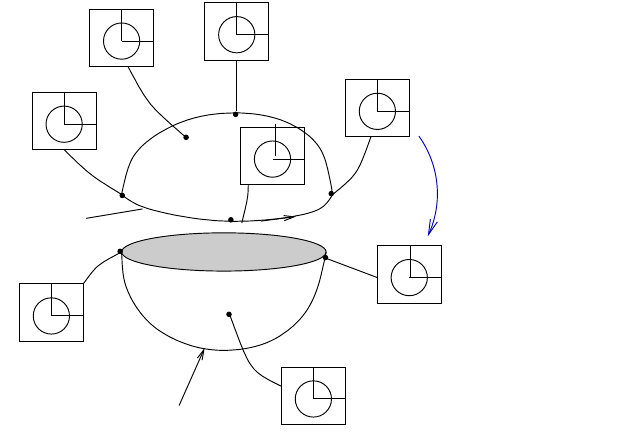
\par\end{center}

If $r>n$, we can continuously deform $g$ so that $\forall x\in S^{2n-1},g_{x}\left(e_{r}\right)=e_{r}$,
where $\left(e_{1},\ldots e_{r}\right)$ is the canonical basis of
\noun{$\mathbb{C}^{r}$}. Cf \cite[Section III.1.B, p.271]{booss_85}.
Then $g$ restricted to $\mathbb{C}^{r-1}\subset\mathbb{C}^{r}$ gives
a function $g:S^{2n-1}\rightarrow U\left(r-1\right)$. By iteration
we get the case $r=n$ with a clutching function $g:S^{2n-1}\rightarrow U\left(n\right)$.
Then using $g$ we define the function
\begin{equation}
f:\begin{cases}
S^{2n-1} & \rightarrow S^{2n-1}\subset\mathbb{C}^{n}\\
x & \rightarrow g_{x}\left(e_{1}\right)
\end{cases}\label{eq:def_f-1}
\end{equation}

The degree $\mathrm{deg}\left(f\right)$ has been defined in Definition
\ref{def:le--d'une}.

\begin{cBoxB}{}
\begin{thm}
\label{thm:def_indice_topologique}(Bott 1958)\cite[Section III.1.B, p.271]{booss_85}.
Let $F\rightarrow S^{2n}$ be a complex vector bundle of rank $r\geq n$.
The topological index
\begin{equation}
\mathcal{C}:=\frac{\mathrm{deg}\left(f\right)}{\left(n-1\right)!}.\label{eq:def_D-1}
\end{equation}
is an integer $\mathcal{C}\in\mathbb{Z}$ (not only a rational number!)
and characterizes the topology of $F$. Namely, if $F\rightarrow S^{2n}$
and $F'\rightarrow S^{2n}$ are fiber bundles of same rank $r\geq n$
with the same index $\mathcal{C}$ then $F$ and $F'$ are isomorphic.
\end{thm}

\end{cBoxB}

\begin{rem}
If the vector bundle $F\rightarrow S^{2n}$ has a (arbitrary) connection,
the \href{https://en.wikipedia.org/wiki/Chern\%E2\%80\%93Weil_homomorphism}{Chern-Weil theory}
permits to express the topological index $\mathcal{C}$ from the curvature
$\Omega$ of the connection, considered as a imaginary valued 2-form
on $S^{2n}$ as follows. We first define $\mathrm{Ch}\left(F\right)$
called the \textbf{Chern Character} which is a differential form on
$S^{2n}$:
\begin{align*}
\mathrm{Ch}\left(F\right) & :=\mathrm{Tr}\left(\exp\left(\frac{i\Omega}{2\pi}\right)\right)\\
 & =\mathrm{Tr}\left(1+\frac{i\Omega}{2\pi}+\frac{1}{2!}\left(\frac{i\Omega}{2\pi}\right)\wedge\left(\frac{i\Omega}{2\pi}\right)+\ldots\right)\\
 & =\mathrm{Ch}_{0}\left(F\right)+\mathrm{Ch}_{2}\left(F\right)+\ldots
\end{align*}
We denote $\mathrm{Ch}_{2n}\left(F\right)$ its component of exterior
degree $2n$ which is a volume form on $S^{2n}$:
\[
\mathrm{Ch}_{2n}\left(F\right)=\frac{1}{n!}\left(\frac{i\Omega}{2\pi}\right)^{\wedge n}
\]
Then 
\begin{equation}
\mathcal{C}=\int_{S^{2n}}\mathrm{Ch}_{2n}\left(F\right).\label{eq:Chern_index}
\end{equation}
Formula (\ref{eq:Chern_index}) is a generalization of Gauss-Bonnet
formula (\ref{eq:Gauss-Bonnet}). For example, for a rank 1 complex
vector bundle $F\rightarrow S^{2}$, i.e. $n=1$, we have $\mathrm{Ch}_{2}\left(F\right)=\frac{i\Omega}{2\pi}$
and (\ref{eq:Chern_index}) gives (\ref{eq:curvature_integral_S2}).
\end{rem}

\subsubsection{A normal form bundle $F_{n}\rightarrow S^{2n-1}$ in each K-isomorphism
class}

We have seen in Theorem \ref{thm:Bott-Periodicity-Theorem} that for
$r\geq n$ then the isomorphism class of complex vector bundles of
rank $r$ over $S^{2n-1}$ is $\mathrm{Vect}^{r}\left(S^{k}\right)\equiv\mathbb{Z}$.
In this section we provide and \textbf{explicit model for the generator}
in this class, i.e. giving the topological index $\mathcal{C}=+1\in\mathrm{Vect}^{r}\left(S^{k}\right)\equiv\mathbb{Z}$.

These models can be considered as \href{https://en.wikipedia.org/wiki/Canonical_form}{canonical forms}
(or normal forms). We will consider $S^{2n-1}:=\left\{ \left(z_{1},z_{2},\ldots z_{n}\right)\in\mathbb{C}^{n}\text{ s.t. }\sum_{j=1}^{n}\left|z_{j}\right|=1\right\} $
as the unit sphere. 

\begin{cBoxA}{}
\begin{defn}
\label{def:Fn}\cite[section 1.2]{Puttmann_2003}''\textbf{Normal
form bundles}''. For $n\in\mathbb{N}^{*}$, we define a normal (canonical)
vector bundle $F_{n}\rightarrow S^{2n}$ of rank $r=2^{n-1}$ from
the normal (canonical) form clutching function
\[
g_{n}:S^{2n-1}\rightarrow U\left(2^{n-1}\right)
\]
by
\begin{align}
g_{1} & :\begin{cases}
S^{1}\subset\mathbb{C} & \rightarrow U\left(1\right)\subset\mathbb{C}\\
z & \rightarrow z
\end{cases}\label{eq:def_g1}
\end{align}
and iteration
\begin{equation}
g_{n+1}\left(z_{1},\underbrace{z_{2},\ldots z_{n+1}}_{z}\right)=\left(\begin{array}{cc}
z_{1}\mathrm{Id}_{2^{n-1}} & -\left(g_{n}\left(z\right)\right)^{\dagger}\\
g_{n}\left(z\right) & \overline{z_{1}}\mathrm{Id}_{2^{n-1}}
\end{array}\right)\label{eq:Bott_map}
\end{equation}
where $\mathrm{Id}_{2^{n-1}}$ denotes the $2^{n-1}\times2^{n-1}$
identity matrix.
\end{defn}

\end{cBoxA}

\begin{rem}
The map $\mathcal{B}:g_{n}\rightarrow g_{n+1}$ in (\ref{eq:Bott_map})
is called the \textbf{Bott map}, see \cite[section 1.1]{Puttmann_2003}
and references therein. Here are the first few expressions of $g_{n}$:
\begin{equation}
g_{1}\left(z_{1}\right)=z_{1},\quad g_{2}\left(z_{1},z_{2}\right)=\left(\begin{array}{cc}
z_{1} & -\overline{z_{2}}\\
z_{2} & \overline{z_{1}}
\end{array}\right),\quad g_{3}\left(z_{1},z_{2},z_{3}\right)=\left(\begin{array}{cccc}
z_{1} & 0 & -\overline{z_{2}} & -\overline{z_{3}}\\
0 & z_{1} & z_{3} & -z_{2}\\
z_{2} & -\overline{z_{3}} & \overline{z_{1}} & 0\\
z_{3} & \overline{z_{2}} & 0 & \overline{z_{1}}
\end{array}\right),\ldots\label{eq:g_n_examples}
\end{equation}
\end{rem}

~
\begin{rem}
These normal forms $g_{n}$ correspond to \href{https://en.wikipedia.org/wiki/Hurwitz\%27s_theorem_(composition_algebras)}{Hurwitz Radon matrices}
\cite{eckmann6hurwitz} and are related to \href{https://en.wikipedia.org/wiki/Gamma_matrices}{gamma matrices},
of generalized \href{https://en.wikipedia.org/wiki/Higher-dimensional_gamma_matrices}{ gamma matrices}
.
\end{rem}

\begin{cBoxB}{}
\begin{prop}
The normal form bundle $F_{n}\rightarrow S^{2n}$ of rank $r=2^{n-1}$
in Definition \ref{def:Fn} is a \textbf{generator of the K-theory
group} $\tilde{K}\left(S^{2n}\right)\equiv\mathrm{Vect}^{r}\left(S^{2n}\right)\equiv\mathbb{Z}$,
hence has topological index
\begin{equation}
\mathcal{C}=+1.\label{eq:c_1}
\end{equation}
\end{prop}

\end{cBoxB}

For the proof, see \cite[section 1.1]{Puttmann_2003} and references
therein. Here let us observe that taking the first column in (\ref{eq:g_n_examples})
and removing zero elements we get the vector $\left(\begin{array}{c}
z_{1}\\
\vdots\\
z_{n}
\end{array}\right)\in S^{2n-1}\subset\mathbb{C}^{n}$. Equivalently, with $\delta_{1}=\left(1,0,\ldots0\right)\in\mathbb{C}^{n}$,
the map $g_{n}\delta_{1}:S^{2n-1}\rightarrow\mathbb{C}^{2^{2n-1}}$
retracts to the identity map $\mathrm{Id}:\mathbb{C}^{n}\rightarrow\mathbb{C}^{n}$. 

\subsubsection{Quantization of the normal form bundle}

In the next Proposition, we consider the normal clutching function
given in (\ref{eq:Bott_map}) as a function
\[
g_{n}:\left(z_{1},z_{2},\ldots z_{n}\right)\in\mathbb{C}^{n}\rightarrow U\left(2^{n-1}\right)
\]
see examples (\ref{eq:g_n_examples}). By writing $z_{j}=x_{j}+i\xi_{j}$,
with $j=1\ldots n$, we get a function $g_{n}\left(x_{1},\xi_{1},x_{2},\xi_{2},\ldots x_{n},\xi_{n}\right)\in U\left(2^{n-1}\right)$
considered as a symbol on $\mathbb{R}^{2n}$ valued in unitary matrices.
Following definition (\ref{eq:def_Op}) we quantize this symbol, giving
an operator
\[
\hat{g}_{n}:=\mathrm{Op}_{1}\left(g_{n}\right):\mathcal{S}\left(\mathbb{R}^{2n};\mathbb{C}^{2^{n-1}}\right)\rightarrow\mathcal{S}\left(\mathbb{R}^{2n};\mathbb{C}^{2^{n-1}}\right).
\]
In fact this operation is quite simple since the symbol is linear:
to get the operator $\hat{g}_{n}$ from the symbol $g_{n}$, we only
have to replace each complex variable $z_{j}=x_{j}+i\xi_{j}$ by $\mathrm{Op}_{1}\left(z_{j}\right)=\mathrm{Op}_{1}\left(x_{j}\right)+i\mathrm{Op}_{1}\left(\xi_{j}\right)=:\sqrt{2}a_{j}$
where $a_{j}$ is called the \href{https://en.wikipedia.org/wiki/Creation_and_annihilation_operators}{annihilation operator}.
For example from (\ref{eq:g_n_examples}), we get
\[
\hat{g}_{1}=\sqrt{2}a_{1},\quad\hat{g}_{2}=2\left(\begin{array}{cc}
a_{1} & -a_{2}^{\dagger}\\
a_{2} & a_{1}^{\dagger}
\end{array}\right),\quad\hat{g}_{3}=2^{3/2}\left(\begin{array}{cccc}
a_{1} & 0 & -a_{2}^{\dagger} & -a_{3}^{\dagger}\\
0 & a_{1} & a_{3} & -a_{2}\\
a_{2} & -a_{3}^{\dagger} & a_{1}^{\dagger} & 0\\
a_{3} & a_{2}^{\dagger} & 0 & a_{1}^{\dagger}
\end{array}\right),\ldots\ldots
\]

\begin{cBoxA}{}
\begin{prop}
\textbf{``Normal form quantum operator''. }The operator $\hat{g}_{n}:=\mathrm{Op}_{1}\left(g_{n}\right)$
is \href{https://en.wikipedia.org/wiki/Fredholm_operator}{Fredholm}
with index
\begin{equation}
\mathrm{Ind}\left(\hat{g}_{n}\right)=+1.\label{eq:ind_gn}
\end{equation}
\end{prop}

\end{cBoxA}

\begin{proof}
For the symbols we compute $\left(g_{1}^{\dagger}g_{1}\right)\left(z\right)\eq{\ref{eq:def_g1}}\left|z\right|^{2}$
and
\[
g_{n+1}^{\dagger}g_{n+1}\eq{\ref{eq:Bott_map}}\left(\begin{array}{cc}
\left|z_{1}\right|^{2}+g_{n}^{\dagger}g_{n} & 0\\
0 & \left|z_{1}\right|^{2}+g_{n}g_{n}^{\dagger}
\end{array}\right).
\]
Recursively we deduce that for any $n\in\mathbb{N}^{*}$, $g_{n+1}^{\dagger}g_{n+1}$
vanishes only at $z=0$, the operators $\hat{g}_{n}^{\dagger}\hat{g}_{n}$
and $\hat{g}_{n}$ are elliptic hence Fredholm \cite[thm3 p.185]{booss_85}.
From (\ref{eq:Bott_map}) we compute recursively that
\[
\mathrm{Ker}\left(\hat{g}_{n}\right)=\mathrm{Span}\left\{ \left(\begin{array}{c}
\varphi_{0}\\
0\\
\vdots
\end{array}\right)\right\} ,\quad\mathrm{Ker}\left(\hat{g}_{n}^{\dagger}\right)=\left\{ 0\right\} ,
\]
where $\varphi_{0}$ is the Gaussian function (\ref{eq:phi_0}) spanning
the kernel of $a_{1}:=\frac{1}{\sqrt{2}}\left(\mathrm{Op}_{1}\left(x_{j}\right)+i\mathrm{Op}_{1}\left(\xi_{j}\right)\right)$.
We deduce that the index is \cite[thm2 p.16]{booss_85}
\begin{equation}
\mathrm{Ind}\left(\hat{g}_{n}\right)=\mathrm{dim}\mathrm{Ker}\left(\hat{g}_{n}\right)-\mathrm{dim}\mathrm{Ker}\left(\hat{g}_{n}^{\dagger}\right)=1-0=1.\label{eq:indA-1}
\end{equation}
\end{proof}

\subsubsection{The index formula on Euclidean space of Fedosov-Hörmander}

For the previous canonical vector bundle $F_{n}\rightarrow S^{2n}$
with topological index $\mathcal{C}$ and clutching function $g_{n}$
we have observed that
\begin{equation}
\mathrm{Ind}\left(\mathrm{Op}_{1}\left(g_{n}\right)\right)\eq{\ref{eq:ind_gn},\ref{eq:c_1}}\mathcal{C},\label{eq:index_f1}
\end{equation}
and that $\mathcal{C}=+1$, meaning that this vector bundle $F_{n}$
is the generator of its equivalence class in $K$-theory. Since both
indices $\mathrm{Ind}\left(\mathrm{Op}_{1}\left(g_{n}\right)\right)$
and $\mathcal{C}$ are additive under direct sum of vector bundles
in $K$-theory, we deduce the next Theorem showing that (\ref{eq:index_f1})
is generally true. 

We consider $F\rightarrow S^{2n}$, a general complex vector bundle
of rank $r$ with topological index $\mathcal{C}\in\mathbb{Z}$ as
defined in (\ref{eq:def_D-1}) and clutching function $g:S^{2n-1}\rightarrow U\left(r\right)$
on the equator $S^{2n-1}$ as defined in (\ref{eq:clutching_g}).
We extend $g$ from $S^{2n-1}$ to 1-homogeneous function on $\mathbb{R}^{2n}\backslash\left\{ 0\right\} $
by $g\left(z\right):=\left|z\right|g\left(\frac{z}{\left|z\right|}\right)$
and consider this extension as a symbol $g:\mathbb{R}^{2n}\backslash\left\{ 0\right\} \rightarrow\mathrm{GL}\left(r\right)$.
Quantization (\ref{eq:def_Op}) gives an operator $\mathrm{Op}_{1}\left(g\right)$.

\begin{cBoxB}{}
\begin{thm}
\cite[thm 7.3 p. 422]{hormander1979weyl}\cite[Thm 1, page 252]{booss_85}\textbf{``The
index formula on Euclidean space of Fedosov-Hörmander''}. Let $F\rightarrow S^{2n}$
be a complex vector bundle of rank $r$ with topological index $\mathcal{C}\in\mathbb{Z}$
and clutching function $g:S^{2n-1}\rightarrow U\left(r\right)$ on
the equator $S^{2n-1}$. We have
\begin{equation}
\mathrm{Ind}\left(\mathrm{Op}_{1}\left(g\right)\right)=\mathcal{C}.\label{eq:index_formula_Euclidean}
\end{equation}
\end{thm}

\end{cBoxB}

\bibliographystyle{plain}
\bibliography{/home/faure/articles/articles}

\end{document}

%% file: images/hyp.pdftex_t
\begin{picture}(0,0)%
\includegraphics{images/hyp}%
\end{picture}%
\setlength{\unitlength}{4144sp}%
\begingroup\makeatletter\ifx\SetFigFont\undefined%
\gdef\SetFigFont#1#2#3#4#5{%
  \reset@font\fontsize{#1}{#2pt}%
  \fontfamily{#3}\fontseries{#4}\fontshape{#5}%
  \selectfont}%
\fi\endgroup%
\begin{picture}(3555,3259)(1201,-4184)
\put(3556,-2896){\makebox(0,0)[lb]{\smash{{\SetFigFont{12}{14.4}{\familydefault}{\mddefault}{\updefault}{\color[rgb]{0,0,0}$\mu$}%
}}}}
\put(2431,-1096){\makebox(0,0)[lb]{\smash{{\SetFigFont{12}{14.4}{\familydefault}{\mddefault}{\updefault}{\color[rgb]{0,0,0}$x,\xi$}%
}}}}
\put(3376,-2806){\makebox(0,0)[lb]{\smash{{\SetFigFont{12}{14.4}{\familydefault}{\mddefault}{\updefault}{\color[rgb]{0,0,0}$2$}%
}}}}
\put(1216,-2806){\makebox(0,0)[lb]{\smash{{\SetFigFont{12}{14.4}{\familydefault}{\mddefault}{\updefault}{\color[rgb]{0,0,0}$-2$}%
}}}}
\put(2386,-2806){\makebox(0,0)[lb]{\smash{{\SetFigFont{12}{14.4}{\familydefault}{\mddefault}{\updefault}{\color[rgb]{0,0,0}$0$}%
}}}}
\put(2881,-2806){\makebox(0,0)[lb]{\smash{{\SetFigFont{12}{14.4}{\familydefault}{\mddefault}{\updefault}{\color[rgb]{0,0,0}$1$}%
}}}}
\put(3511,-1906){\makebox(0,0)[lb]{\smash{{\SetFigFont{12}{14.4}{\familydefault}{\mddefault}{\updefault}{\color[rgb]{0,0,0}a spectral gap}%
}}}}
\put(3511,-2131){\makebox(0,0)[lb]{\smash{{\SetFigFont{12}{14.4}{\familydefault}{\mddefault}{\updefault}{\color[rgb]{0,0,0}of $H_\mu (x,\xi)$}%
}}}}
\put(3511,-1681){\makebox(0,0)[lb]{\smash{{\SetFigFont{12}{14.4}{\familydefault}{\mddefault}{\updefault}{\color[rgb]{0,0,0}$[-C,C]$ is}%
}}}}
\put(1801,-2806){\makebox(0,0)[lb]{\smash{{\SetFigFont{12}{14.4}{\familydefault}{\mddefault}{\updefault}{\color[rgb]{0,0,0}$-1$}%
}}}}
\end{picture}%

%% file: images/hyp_spect.pdftex_t
\begin{picture}(0,0)%
\includegraphics{images/hyp_spect}%
\end{picture}%
\setlength{\unitlength}{4144sp}%
\begingroup\makeatletter\ifx\SetFigFont\undefined%
\gdef\SetFigFont#1#2#3#4#5{%
  \reset@font\fontsize{#1}{#2pt}%
  \fontfamily{#3}\fontseries{#4}\fontshape{#5}%
  \selectfont}%
\fi\endgroup%
\begin{picture}(4347,3451)(-239,-3644)
\put(3781,-3211){\makebox(0,0)[lb]{\smash{{\SetFigFont{12}{14.4}{\familydefault}{\mddefault}{\updefault}{\color[rgb]{0,0,0}$2$}%
}}}}
\put(-224,-2716){\makebox(0,0)[lb]{\smash{{\SetFigFont{12}{14.4}{\familydefault}{\mddefault}{\updefault}{\color[rgb]{0,0,0}$-C$}%
}}}}
\put(-179,-916){\makebox(0,0)[lb]{\smash{{\SetFigFont{12}{14.4}{\familydefault}{\mddefault}{\updefault}{\color[rgb]{0,0,0}$C$}%
}}}}
\put(136,-3211){\makebox(0,0)[lb]{\smash{{\SetFigFont{12}{14.4}{\familydefault}{\mddefault}{\updefault}{\color[rgb]{0,0,0}$-2$}%
}}}}
\put(2836,-3211){\makebox(0,0)[lb]{\smash{{\SetFigFont{12}{14.4}{\familydefault}{\mddefault}{\updefault}{\color[rgb]{0,0,0}$1$}%
}}}}
\put(991,-3211){\makebox(0,0)[lb]{\smash{{\SetFigFont{12}{14.4}{\familydefault}{\mddefault}{\updefault}{\color[rgb]{0,0,0}$-1$}%
}}}}
\put(2071,-376){\makebox(0,0)[lb]{\smash{{\SetFigFont{12}{14.4}{\familydefault}{\mddefault}{\updefault}{\color[rgb]{0,0,0}$\omega$}%
}}}}
\put(4006,-1951){\makebox(0,0)[lb]{\smash{{\SetFigFont{12}{14.4}{\familydefault}{\mddefault}{\updefault}{\color[rgb]{0,0,0}$\mu$}%
}}}}
\end{picture}%

%% file: images/def_N.pdftex_t
\begin{picture}(0,0)%
\includegraphics{images/def_N}%
\end{picture}%
\setlength{\unitlength}{4144sp}%
\begingroup\makeatletter\ifx\SetFigFont\undefined%
\gdef\SetFigFont#1#2#3#4#5{%
  \reset@font\fontsize{#1}{#2pt}%
  \fontfamily{#3}\fontseries{#4}\fontshape{#5}%
  \selectfont}%
\fi\endgroup%
\begin{picture}(5350,2994)(-509,-3289)
\put(4051,-1726){\makebox(0,0)[lb]{\smash{{\SetFigFont{12}{14.4}{\familydefault}{\mddefault}{\updefault}{\color[rgb]{0,0,0}$\mu$}%
}}}}
\put(3781,-3211){\makebox(0,0)[lb]{\smash{{\SetFigFont{12}{14.4}{\familydefault}{\mddefault}{\updefault}{\color[rgb]{0,0,0}$2$}%
}}}}
\put(2971,-2266){\makebox(0,0)[lb]{\smash{{\SetFigFont{12}{14.4}{\familydefault}{\mddefault}{\updefault}{\color[rgb]{0,0,0}$n_{\mathrm{out}}=-2$}%
}}}}
\put(136,-3211){\makebox(0,0)[lb]{\smash{{\SetFigFont{12}{14.4}{\familydefault}{\mddefault}{\updefault}{\color[rgb]{0,0,0}$-2$}%
}}}}
\put(2836,-3211){\makebox(0,0)[lb]{\smash{{\SetFigFont{12}{14.4}{\familydefault}{\mddefault}{\updefault}{\color[rgb]{0,0,0}$1+\alpha$}%
}}}}
\put(3961,-1321){\makebox(0,0)[lb]{\smash{{\SetFigFont{12}{14.4}{\rmdefault}{\mddefault}{\updefault}{\color[rgb]{0,0,0}$\mathcal{N}=+2$}%
}}}}
\put(-449,-916){\makebox(0,0)[lb]{\smash{{\SetFigFont{12}{14.4}{\familydefault}{\mddefault}{\updefault}{\color[rgb]{0,0,0}$C-\alpha$}%
}}}}
\put(-494,-2716){\makebox(0,0)[lb]{\smash{{\SetFigFont{12}{14.4}{\familydefault}{\mddefault}{\updefault}{\color[rgb]{0,0,0}$-C+\alpha$}%
}}}}
\put(766,-3211){\makebox(0,0)[lb]{\smash{{\SetFigFont{12}{14.4}{\familydefault}{\mddefault}{\updefault}{\color[rgb]{0,0,0}$-1-\alpha$}%
}}}}
\put(2026,-511){\makebox(0,0)[lb]{\smash{{\SetFigFont{12}{14.4}{\familydefault}{\mddefault}{\updefault}{\color[rgb]{0,0,0}$\omega$}%
}}}}
\put(1711,-1096){\makebox(0,0)[lb]{\smash{{\SetFigFont{12}{14.4}{\familydefault}{\mddefault}{\updefault}{\color[rgb]{0,0,.69}$n=2$}%
}}}}
\put(361,-2086){\makebox(0,0)[lb]{\smash{{\SetFigFont{12}{14.4}{\familydefault}{\mddefault}{\updefault}{\color[rgb]{0,0,0}$n_{\mathrm{in}}=0$}%
}}}}
\put(1126,-2221){\makebox(0,0)[lb]{\smash{{\SetFigFont{12}{14.4}{\rmdefault}{\mddefault}{\updefault}{\color[rgb]{0,0,.69}$n=0$}%
}}}}
\put(2296,-1951){\makebox(0,0)[lb]{\smash{{\SetFigFont{12}{14.4}{\familydefault}{\mddefault}{\updefault}{\color[rgb]{0,0,.69}$n=-1$}%
}}}}
\put(2251,-2356){\makebox(0,0)[lb]{\smash{{\SetFigFont{12}{14.4}{\familydefault}{\mddefault}{\updefault}{\color[rgb]{0,0,.69}$n=-2$}%
}}}}
\put(1621,-1546){\makebox(0,0)[lb]{\smash{{\SetFigFont{12}{14.4}{\familydefault}{\mddefault}{\updefault}{\color[rgb]{0,0,.69}$n=1$}%
}}}}
\end{picture}%

%% file: images/contact.pdftex_t
\begin{picture}(0,0)%
\includegraphics{images/contact}%
\end{picture}%
\setlength{\unitlength}{3947sp}%
\begingroup\makeatletter\ifx\SetFigFont\undefined%
\gdef\SetFigFont#1#2#3#4#5{%
  \reset@font\fontsize{#1}{#2pt}%
  \fontfamily{#3}\fontseries{#4}\fontshape{#5}%
  \selectfont}%
\fi\endgroup%
\begin{picture}(2649,2469)(-11186,-2197)
\put(-9899, 89){\makebox(0,0)[lb]{\smash{{\SetFigFont{12}{14.4}{\rmdefault}{\mddefault}{\updefault}{\color[rgb]{0,0,0}$\omega$}%
}}}}
\put(-8699,-1186){\makebox(0,0)[lb]{\smash{{\SetFigFont{12}{14.4}{\rmdefault}{\mddefault}{\updefault}{\color[rgb]{0,0,0}$\mu$}%
}}}}
\end{picture}%

%% file: images/2_contacts.pdftex_t
\begin{picture}(0,0)%
\includegraphics{images/2_contacts}%
\end{picture}%
\setlength{\unitlength}{3947sp}%
\begingroup\makeatletter\ifx\SetFigFont\undefined%
\gdef\SetFigFont#1#2#3#4#5{%
  \reset@font\fontsize{#1}{#2pt}%
  \fontfamily{#3}\fontseries{#4}\fontshape{#5}%
  \selectfont}%
\fi\endgroup%
\begin{picture}(4261,1206)(589,-2305)
\put(4726,-1711){\makebox(0,0)[lb]{\smash{{\SetFigFont{12}{14.4}{\rmdefault}{\mddefault}{\updefault}{\color[rgb]{0,0,0}$\mu$}%
}}}}
\put(2626,-1786){\makebox(0,0)[lb]{\smash{{\SetFigFont{12}{14.4}{\rmdefault}{\mddefault}{\updefault}{\color[rgb]{0,0,0}$\mu$}%
}}}}
\put(1051,-2236){\makebox(0,0)[lb]{\smash{{\SetFigFont{12}{14.4}{\rmdefault}{\mddefault}{\updefault}{\color[rgb]{0,0,0}$1$}%
}}}}
\put(1426,-2236){\makebox(0,0)[lb]{\smash{{\SetFigFont{12}{14.4}{\rmdefault}{\mddefault}{\updefault}{\color[rgb]{0,0,0}$+$}%
}}}}
\put(1801,-2236){\makebox(0,0)[lb]{\smash{{\SetFigFont{12}{14.4}{\rmdefault}{\mddefault}{\updefault}{\color[rgb]{0,0,0}$1$}%
}}}}
\put(2701,-2236){\makebox(0,0)[lb]{\smash{{\SetFigFont{12}{14.4}{\rmdefault}{\mddefault}{\updefault}{\color[rgb]{0,0,0}$=$}%
}}}}
\put(3826,-2236){\makebox(0,0)[lb]{\smash{{\SetFigFont{12}{14.4}{\rmdefault}{\mddefault}{\updefault}{\color[rgb]{0,0,0}$0$}%
}}}}
\put(676,-1486){\makebox(0,0)[lb]{\smash{{\SetFigFont{12}{14.4}{\rmdefault}{\mddefault}{\updefault}{\color[rgb]{0,0,0}$F$}%
}}}}
\put(2251,-1486){\makebox(0,0)[lb]{\smash{{\SetFigFont{12}{14.4}{\rmdefault}{\mddefault}{\updefault}{\color[rgb]{0,0,0}$F$}%
}}}}
\end{picture}%

%% file: images/mol_cd4_bis.pdftex_t
\begin{picture}(0,0)%
\includegraphics{images/mol_cd4_bis}%
\end{picture}%
\setlength{\unitlength}{3947sp}%
\begingroup\makeatletter\ifx\SetFigFont\undefined%
\gdef\SetFigFont#1#2#3#4#5{%
  \reset@font\fontsize{#1}{#2pt}%
  \fontfamily{#3}\fontseries{#4}\fontshape{#5}%
  \selectfont}%
\fi\endgroup%
\begin{picture}(3053,2290)(-1746,-2413)
\end{picture}%

%% file: images/bandes_energie_exact_bis.pdftex_t
\begin{picture}(0,0)%
\includegraphics{images/bandes_energie_exact_bis}%
\end{picture}%
\setlength{\unitlength}{3947sp}%
\begingroup\makeatletter\ifx\SetFigFont\undefined%
\gdef\SetFigFont#1#2#3#4#5{%
  \reset@font\fontsize{#1}{#2pt}%
  \fontfamily{#3}\fontseries{#4}\fontshape{#5}%
  \selectfont}%
\fi\endgroup%
\begin{picture}(3320,2715)(3886,-2530)
\put(6001,-2161){\makebox(0,0)[lb]{\smash{{\SetFigFont{12}{14.4}{\familydefault}{\mddefault}{\updefault}{\color[rgb]{0,0,0}$\mathcal{N}_1$}%
}}}}
\put(6001,-1486){\makebox(0,0)[lb]{\smash{{\SetFigFont{12}{14.4}{\familydefault}{\mddefault}{\updefault}{\color[rgb]{0,0,0}$\mathcal{N}_2$}%
}}}}
\put(6001,-586){\makebox(0,0)[lb]{\smash{{\SetFigFont{12}{14.4}{\familydefault}{\mddefault}{\updefault}{\color[rgb]{0,0,0}$\mathcal{N}_3$}%
}}}}
\put(5701,-2461){\makebox(0,0)[lb]{\smash{{\SetFigFont{12}{14.4}{\familydefault}{\mddefault}{\updefault}{\color[rgb]{0,0,0}$J$}%
}}}}
\put(3901,-436){\makebox(0,0)[lb]{\smash{{\SetFigFont{12}{14.4}{\rmdefault}{\mddefault}{\updefault}{\color[rgb]{0,0,0}$\omega$}%
}}}}
\put(3976, 14){\makebox(0,0)[lb]{\smash{{\SetFigFont{12}{14.4}{\familydefault}{\mddefault}{\updefault}{\color[rgb]{0,0,0}energy levels}%
}}}}
\put(3976,-136){\makebox(0,0)[lb]{\smash{{\SetFigFont{12}{14.4}{\familydefault}{\mddefault}{\updefault}{\color[rgb]{0,0,0}that moves from one band to another}%
}}}}
\end{picture}%

%% file: images/spectre_model.pdftex_t
\begin{picture}(0,0)%
\includegraphics{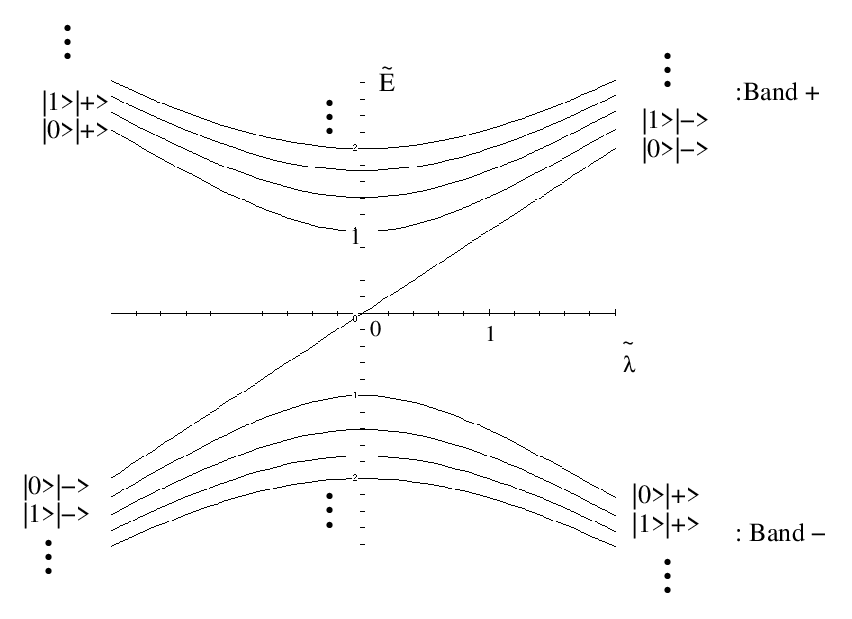}%
\end{picture}%
\setlength{\unitlength}{3947sp}%
\begingroup\makeatletter\ifx\SetFigFont\undefined%
\gdef\SetFigFont#1#2#3#4#5{%
  \reset@font\fontsize{#1}{#2pt}%
  \fontfamily{#3}\fontseries{#4}\fontshape{#5}%
  \selectfont}%
\fi\endgroup%
\begin{picture}(4050,2712)(376,-4936)
\put(3226,-3811){\makebox(0,0)[lb]{\smash{{\SetFigFont{12}{14.4}{\rmdefault}{\mddefault}{\updefault}{\color[rgb]{0,0,0}$\mu$}%
}}}}
\put(3301,-2836){\makebox(0,0)[lb]{\smash{{\SetFigFont{12}{14.4}{\rmdefault}{\mddefault}{\updefault}{\color[rgb]{0,0,0}$\omega^+_1$}%
}}}}
\put(3301,-2611){\makebox(0,0)[lb]{\smash{{\SetFigFont{12}{14.4}{\rmdefault}{\mddefault}{\updefault}{\color[rgb]{0,0,0}$\omega^+_2$}%
}}}}
\put(3226,-3136){\makebox(0,0)[lb]{\smash{{\SetFigFont{12}{14.4}{\rmdefault}{\mddefault}{\updefault}{\color[rgb]{0,0,0}$\omega_0$}%
}}}}
\put(3376,-4636){\makebox(0,0)[lb]{\smash{{\SetFigFont{12}{14.4}{\rmdefault}{\mddefault}{\updefault}{\color[rgb]{0,0,0}$\omega^-_2$}%
}}}}
\put(3376,-4411){\makebox(0,0)[lb]{\smash{{\SetFigFont{12}{14.4}{\rmdefault}{\mddefault}{\updefault}{\color[rgb]{0,0,0}$\omega^-_1$}%
}}}}
\put(2122,-2453){\makebox(0,0)[lb]{\smash{{\SetFigFont{12}{14.4}{\rmdefault}{\mddefault}{\updefault}{\color[rgb]{0,0,0}$\omega$}%
}}}}
\end{picture}%

%% file: images/cone.pdftex_t
\begin{picture}(0,0)%
\includegraphics{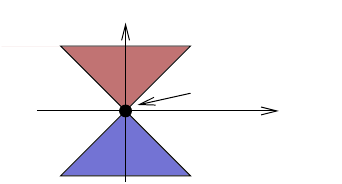}%
\end{picture}%
\setlength{\unitlength}{4144sp}%
\begingroup\makeatletter\ifx\SetFigFont\undefined%
\gdef\SetFigFont#1#2#3#4#5{%
  \reset@font\fontsize{#1}{#2pt}%
  \fontfamily{#3}\fontseries{#4}\fontshape{#5}%
  \selectfont}%
\fi\endgroup%
\begin{picture}(2603,1386)(2509,-2008)
\put(3466,-781){\makebox(0,0)[lb]{\smash{{\SetFigFont{12}{14.4}{\familydefault}{\mddefault}{\updefault}{\color[rgb]{0,0,0}$E$}%
}}}}
\put(4546,-1636){\makebox(0,0)[lb]{\smash{{\SetFigFont{12}{14.4}{\familydefault}{\mddefault}{\updefault}{\color[rgb]{0,0,0}$\mu$}%
}}}}
\put(4051,-1366){\makebox(0,0)[lb]{\smash{{\SetFigFont{12}{14.4}{\familydefault}{\mddefault}{\updefault}{\color[rgb]{0,0,0}degeneracy}%
}}}}
\put(4051,-1096){\makebox(0,0)[lb]{\smash{{\SetFigFont{12}{14.4}{\familydefault}{\mddefault}{\updefault}{\color[rgb]{0,0,0}$E_+(\mu,x,\xi)$}%
}}}}
\put(3961,-1906){\makebox(0,0)[lb]{\smash{{\SetFigFont{12}{14.4}{\familydefault}{\mddefault}{\updefault}{\color[rgb]{0,0,0}$E_-(\mu,x,\xi)$}%
}}}}
\end{picture}%

%% file: images/shallow_water.pdftex_t
\begin{picture}(0,0)%
\includegraphics{images/shallow_water}%
\end{picture}%
\setlength{\unitlength}{4144sp}%
\begingroup\makeatletter\ifx\SetFigFont\undefined%
\gdef\SetFigFont#1#2#3#4#5{%
  \reset@font\fontsize{#1}{#2pt}%
  \fontfamily{#3}\fontseries{#4}\fontshape{#5}%
  \selectfont}%
\fi\endgroup%
\begin{picture}(7044,2011)(-3161,-3824)
\put(3421,-3751){\makebox(0,0)[lb]{\smash{{\SetFigFont{12}{14.4}{\rmdefault}{\mddefault}{\updefault}{\color[rgb]{0,0,0}$x$}%
}}}}
\put(-764,-3301){\makebox(0,0)[lb]{\smash{{\SetFigFont{12}{14.4}{\rmdefault}{\mddefault}{\updefault}{\color[rgb]{0,0,0}$x_1$}%
}}}}
\put(-2024,-1996){\makebox(0,0)[lb]{\smash{{\SetFigFont{12}{14.4}{\rmdefault}{\mddefault}{\updefault}{\color[rgb]{0,0,0}$x_2$}%
}}}}
\put(-2609,-3751){\makebox(0,0)[lb]{\smash{{\SetFigFont{12}{14.4}{\rmdefault}{\mddefault}{\updefault}{\color[rgb]{0,0,0}Horizontal plane}%
}}}}
\put(946,-3796){\makebox(0,0)[lb]{\smash{{\SetFigFont{12}{14.4}{\rmdefault}{\mddefault}{\updefault}{\color[rgb]{0,0,0}Vertical slice}%
}}}}
\put(2206,-2086){\makebox(0,0)[lb]{\smash{{\SetFigFont{12}{14.4}{\rmdefault}{\mddefault}{\updefault}{\color[rgb]{0,0,0}$h(x,t)$}%
}}}}
\put(766,-1996){\makebox(0,0)[lb]{\smash{{\SetFigFont{12}{14.4}{\rmdefault}{\mddefault}{\updefault}{\color[rgb]{0,0,0}$z$}%
}}}}
\put(1216,-2086){\makebox(0,0)[lb]{\smash{{\SetFigFont{12}{14.4}{\rmdefault}{\mddefault}{\updefault}{\color[rgb]{0,0,0}fluctuations}%
}}}}
\put(1049,-2881){\makebox(0,0)[lb]{\smash{{\SetFigFont{12}{14.4}{\rmdefault}{\mddefault}{\updefault}{\color[rgb]{0,0,0}Ocean}%
}}}}
\put(2123,-2854){\makebox(0,0)[lb]{\smash{{\SetFigFont{12}{14.4}{\rmdefault}{\mddefault}{\updefault}{\color[rgb]{0,0,0}$H$}%
}}}}
\put(-1304,-2536){\makebox(0,0)[lb]{\smash{{\SetFigFont{12}{14.4}{\rmdefault}{\mddefault}{\updefault}{\color[rgb]{0,0,0}$u=(u_1,u_2)$}%
}}}}
\put(-1529,-2266){\makebox(0,0)[lb]{\smash{{\SetFigFont{12}{14.4}{\rmdefault}{\mddefault}{\updefault}{\color[rgb]{0,0,0}velocity}%
}}}}
\end{picture}%

%% file: images/cone2.pdftex_t
\begin{picture}(0,0)%
\includegraphics{images/cone2}%
\end{picture}%
\setlength{\unitlength}{4144sp}%
\begingroup\makeatletter\ifx\SetFigFont\undefined%
\gdef\SetFigFont#1#2#3#4#5{%
  \reset@font\fontsize{#1}{#2pt}%
  \fontfamily{#3}\fontseries{#4}\fontshape{#5}%
  \selectfont}%
\fi\endgroup%
\begin{picture}(2412,1410)(2236,-2008)
\put(4546,-1636){\makebox(0,0)[lb]{\smash{{\SetFigFont{12}{14.4}{\familydefault}{\mddefault}{\updefault}{\color[rgb]{0,0,0}$\mu$}%
}}}}
\put(3961,-1906){\makebox(0,0)[lb]{\smash{{\SetFigFont{12}{14.4}{\familydefault}{\mddefault}{\updefault}{\color[rgb]{0,0,.69}$\omega^{(1)}(\mu,x,\xi)$}%
}}}}
\put(2251,-1276){\makebox(0,0)[lb]{\smash{{\SetFigFont{12}{14.4}{\familydefault}{\mddefault}{\updefault}{\color[rgb]{0,0,0}degeneracy}%
}}}}
\put(3781,-1411){\makebox(0,0)[lb]{\smash{{\SetFigFont{12}{14.4}{\familydefault}{\mddefault}{\updefault}{\color[rgb]{0,.56,0}$\omega^{(2)}(\mu,x,\xi)$}%
}}}}
\put(3466,-781){\makebox(0,0)[lb]{\smash{{\SetFigFont{12}{14.4}{\familydefault}{\mddefault}{\updefault}{\color[rgb]{0,0,0}$\omega$}%
}}}}
\put(3961,-1051){\makebox(0,0)[lb]{\smash{{\SetFigFont{12}{14.4}{\familydefault}{\mddefault}{\updefault}{\color[rgb]{.56,0,0}$\omega^{(3)}(\mu,x,\xi)$}%
}}}}
\end{picture}%

%% file: images/ruban.pdftex_t
\begin{picture}(0,0)%
\includegraphics{images/ruban}%
\end{picture}%
\setlength{\unitlength}{4144sp}%
\begingroup\makeatletter\ifx\SetFigFont\undefined%
\gdef\SetFigFont#1#2#3#4#5{%
  \reset@font\fontsize{#1}{#2pt}%
  \fontfamily{#3}\fontseries{#4}\fontshape{#5}%
  \selectfont}%
\fi\endgroup%
\begin{picture}(2454,2241)(889,-3043)
\put(1081,-961){\makebox(0,0)[lb]{\smash{{\SetFigFont{12}{14.4}{\familydefault}{\mddefault}{\updefault}{\color[rgb]{0,0,0}$t$}%
}}}}
\put(3196,-2356){\makebox(0,0)[lb]{\smash{{\SetFigFont{12}{14.4}{\familydefault}{\mddefault}{\updefault}{\color[rgb]{0,0,0}$x$}%
}}}}
\put(2836,-2311){\makebox(0,0)[lb]{\smash{{\SetFigFont{12}{14.4}{\familydefault}{\mddefault}{\updefault}{\color[rgb]{0,0,0}$1$}%
}}}}
\put(2926,-2581){\makebox(0,0)[lb]{\smash{{\SetFigFont{12}{14.4}{\familydefault}{\mddefault}{\updefault}{\color[rgb]{0,0,0}$-t$}%
}}}}
\put(991,-2311){\makebox(0,0)[lb]{\smash{{\SetFigFont{12}{14.4}{\familydefault}{\mddefault}{\updefault}{\color[rgb]{0,0,0}$0$}%
}}}}
\put(946,-1906){\makebox(0,0)[lb]{\smash{{\SetFigFont{12}{14.4}{\familydefault}{\mddefault}{\updefault}{\color[rgb]{0,0,0}$t$}%
}}}}
\put(2881,-1906){\makebox(0,0)[lb]{\smash{{\SetFigFont{12}{14.4}{\familydefault}{\mddefault}{\updefault}{\color[rgb]{0,0,0}$t$}%
}}}}
\end{picture}%

%% file: images/moebius.pdftex_t
\begin{picture}(0,0)%
\includegraphics{images/moebius}%
\end{picture}%
\setlength{\unitlength}{3947sp}%
\begingroup\makeatletter\ifx\SetFigFont\undefined%
\gdef\SetFigFont#1#2#3#4#5{%
  \reset@font\fontsize{#1}{#2pt}%
  \fontfamily{#3}\fontseries{#4}\fontshape{#5}%
  \selectfont}%
\fi\endgroup%
\begin{picture}(4962,2068)(961,-2926)
\put(976,-2311){\makebox(0,0)[lb]{\smash{{\SetFigFont{12}{14.4}{\rmdefault}{\mddefault}{\updefault}{\color[rgb]{0,0,0}Base space $S^1$}%
}}}}
\put(1501,-2686){\makebox(0,0)[lb]{\smash{{\SetFigFont{12}{14.4}{\rmdefault}{\mddefault}{\updefault}{\color[rgb]{0,0,0}(a) trivial bundle $S^1 \times \mathrm{R}$}%
}}}}
\put(2626,-2461){\makebox(0,0)[lb]{\smash{{\SetFigFont{12}{14.4}{\rmdefault}{\mddefault}{\updefault}{\color[rgb]{0,0,0}$x$}%
}}}}
\put(2476,-2011){\makebox(0,0)[lb]{\smash{{\SetFigFont{12}{14.4}{\rmdefault}{\mddefault}{\updefault}{\color[rgb]{0,0,.69}$F_x$}%
}}}}
\put(4736,-2477){\makebox(0,0)[lb]{\smash{{\SetFigFont{12}{14.4}{\rmdefault}{\mddefault}{\updefault}{\color[rgb]{0,0,0}$x$}%
}}}}
\put(4544,-2031){\makebox(0,0)[lb]{\smash{{\SetFigFont{12}{14.4}{\rmdefault}{\mddefault}{\updefault}{\color[rgb]{0,0,.69}$F_x$}%
}}}}
\end{picture}%

%% file: images/section.pdftex_t
\begin{picture}(0,0)%
\includegraphics{images/section}%
\end{picture}%
\setlength{\unitlength}{3947sp}%
\begingroup\makeatletter\ifx\SetFigFont\undefined%
\gdef\SetFigFont#1#2#3#4#5{%
  \reset@font\fontsize{#1}{#2pt}%
  \fontfamily{#3}\fontseries{#4}\fontshape{#5}%
  \selectfont}%
\fi\endgroup%
\begin{picture}(5185,3228)(-4596,-3280)
\put(-674,-3211){\makebox(0,0)[lb]{\smash{{\SetFigFont{12}{14.4}{\rmdefault}{\mddefault}{\updefault}{\color[rgb]{0,0,0}$B$}%
}}}}
\put(-2769,-2596){\makebox(0,0)[lb]{\smash{{\SetFigFont{12}{14.4}{\rmdefault}{\mddefault}{\updefault}{\color[rgb]{0,0,0}$x_1$}%
}}}}
\put(-1547,-2487){\makebox(0,0)[lb]{\smash{{\SetFigFont{12}{14.4}{\rmdefault}{\mddefault}{\updefault}{\color[rgb]{0,0,0}$x_2$}%
}}}}
\put(-2918,-1082){\makebox(0,0)[lb]{\smash{{\SetFigFont{12}{14.4}{\rmdefault}{\mddefault}{\updefault}{\color[rgb]{0,0,0}$0$}%
}}}}
\put(-1384,-974){\makebox(0,0)[lb]{\smash{{\SetFigFont{12}{14.4}{\rmdefault}{\mddefault}{\updefault}{\color[rgb]{0,0,0}$0$}%
}}}}
\put(-943,-1564){\makebox(0,0)[lb]{\smash{{\SetFigFont{12}{14.4}{\rmdefault}{\mddefault}{\updefault}{\color[rgb]{0,0,0}$F_{x_2}$}%
}}}}
\put(-3712,-1435){\makebox(0,0)[lb]{\smash{{\SetFigFont{12}{14.4}{\rmdefault}{\mddefault}{\updefault}{\color[rgb]{0,0,0}$F_{x_1}$}%
}}}}
\put(-4581,-757){\makebox(0,0)[lb]{\smash{{\SetFigFont{12}{14.4}{\rmdefault}{\mddefault}{\updefault}{\color[rgb]{0,0,1}section $s$}%
}}}}
\put(-3278,-682){\makebox(0,0)[lb]{\smash{{\SetFigFont{12}{14.4}{\rmdefault}{\mddefault}{\updefault}{\color[rgb]{0,0,1}$s(x_1)$}%
}}}}
\put(-1398,-635){\makebox(0,0)[lb]{\smash{{\SetFigFont{12}{14.4}{\rmdefault}{\mddefault}{\updefault}{\color[rgb]{0,0,1}$s(x_2)$}%
}}}}
\put(-692,-2935){\makebox(0,0)[lb]{\smash{{\SetFigFont{12}{14.4}{\rmdefault}{\mddefault}{\updefault}{\color[rgb]{0,0,0}Base space}%
}}}}
\put(-733,-1042){\makebox(0,0)[lb]{\smash{{\SetFigFont{12}{14.4}{\rmdefault}{\mddefault}{\updefault}{\color[rgb]{0,0,0}: Fiber space F}%
}}}}
\end{picture}%

%% file: images/moebius_section.pdftex_t
\begin{picture}(0,0)%
\includegraphics{images/moebius_section}%
\end{picture}%
\setlength{\unitlength}{3947sp}%
\begingroup\makeatletter\ifx\SetFigFont\undefined%
\gdef\SetFigFont#1#2#3#4#5{%
  \reset@font\fontsize{#1}{#2pt}%
  \fontfamily{#3}\fontseries{#4}\fontshape{#5}%
  \selectfont}%
\fi\endgroup%
\begin{picture}(4708,2068)(1336,-2684)
\end{picture}%

%% file: images/bundle_sphere_2.pdftex_t
\begin{picture}(0,0)%
\includegraphics{images/bundle_sphere_2}%
\end{picture}%
\setlength{\unitlength}{3947sp}%
\begingroup\makeatletter\ifx\SetFigFont\undefined%
\gdef\SetFigFont#1#2#3#4#5{%
  \reset@font\fontsize{#1}{#2pt}%
  \fontfamily{#3}\fontseries{#4}\fontshape{#5}%
  \selectfont}%
\fi\endgroup%
\begin{picture}(5480,3805)(3364,-4908)
\put(7201,-3661){\makebox(0,0)[lb]{\smash{{\SetFigFont{12}{14.4}{\rmdefault}{\mddefault}{\updefault}{\color[rgb]{0,0,0}$v_2$}%
}}}}
\put(7276,-2386){\makebox(0,0)[lb]{\smash{{\SetFigFont{12}{14.4}{\rmdefault}{\mddefault}{\updefault}{\color[rgb]{0,0,0}clutching function}%
}}}}
\put(7276,-2761){\makebox(0,0)[lb]{\smash{{\SetFigFont{12}{14.4}{\rmdefault}{\mddefault}{\updefault}{\color[rgb]{0,0,0}$e^{i\varphi(\theta)}$}%
}}}}
\put(5701,-2986){\makebox(0,0)[lb]{\smash{{\SetFigFont{12}{14.4}{\rmdefault}{\mddefault}{\updefault}{\color[rgb]{0,0,0}$\theta$}%
}}}}
\put(6526,-4186){\makebox(0,0)[lb]{\smash{{\SetFigFont{12}{14.4}{\rmdefault}{\mddefault}{\updefault}{\color[rgb]{0,0,0}$\varphi (2 \pi) = \varphi(0) + \mathcal{C} 2 \pi$}%
}}}}
\put(7051,-1936){\makebox(0,0)[lb]{\smash{{\SetFigFont{12}{14.4}{\rmdefault}{\mddefault}{\updefault}{\color[rgb]{0,0,0}$e^{i\varphi(\theta)}$}%
}}}}
\put(7342,-2176){\makebox(0,0)[lb]{\smash{{\SetFigFont{12}{14.4}{\rmdefault}{\mddefault}{\updefault}{\color[rgb]{0,0,0}$v_1$}%
}}}}
\put(6749,-1617){\makebox(0,0)[lb]{\smash{{\SetFigFont{12}{14.4}{\rmdefault}{\mddefault}{\updefault}{\color[rgb]{0,0,.69}$v_2$}%
}}}}
\put(4517,-2795){\makebox(0,0)[lb]{\smash{{\SetFigFont{12}{14.4}{\rmdefault}{\mddefault}{\updefault}{\color[rgb]{0,0,0}$H_1$}%
}}}}
\put(4499,-3843){\makebox(0,0)[lb]{\smash{{\SetFigFont{12}{14.4}{\rmdefault}{\mddefault}{\updefault}{\color[rgb]{0,0,0}$H_2$}%
}}}}
\put(6573,-4499){\makebox(0,0)[lb]{\smash{{\SetFigFont{12}{14.4}{\rmdefault}{\mddefault}{\updefault}{\color[rgb]{0,0,0}$\mathcal{C}$ : Chern index}%
}}}}
\end{picture}%

%% file: images/bundle_TS2.pdftex_t
\begin{picture}(0,0)%
\includegraphics{images/bundle_TS2}%
\end{picture}%
\setlength{\unitlength}{3947sp}%
\begingroup\makeatletter\ifx\SetFigFont\undefined%
\gdef\SetFigFont#1#2#3#4#5{%
  \reset@font\fontsize{#1}{#2pt}%
  \fontfamily{#3}\fontseries{#4}\fontshape{#5}%
  \selectfont}%
\fi\endgroup%
\begin{picture}(8616,3486)(1522,-3439)
\put(2776,-136){\makebox(0,0)[lb]{\smash{{\SetFigFont{12}{14.4}{\rmdefault}{\mddefault}{\updefault}{\color[rgb]{0,0,0}North Hemisphere $H_1$}%
}}}}
\put(2926,-3361){\makebox(0,0)[lb]{\smash{{\SetFigFont{12}{14.4}{\rmdefault}{\mddefault}{\updefault}{\color[rgb]{0,0,0}South Hemisphere $H_2$}%
}}}}
\put(8026,-1636){\makebox(0,0)[lb]{\smash{{\SetFigFont{12}{14.4}{\rmdefault}{\mddefault}{\updefault}{\color[rgb]{.69,0,0}$2\pi$}%
}}}}
\put(9601,-1636){\makebox(0,0)[lb]{\smash{{\SetFigFont{12}{14.4}{\rmdefault}{\mddefault}{\updefault}{\color[rgb]{.69,0,0}$4\pi= 2 \times 2\pi$}%
}}}}
\put(8851,-1636){\makebox(0,0)[lb]{\smash{{\SetFigFont{12}{14.4}{\rmdefault}{\mddefault}{\updefault}{\color[rgb]{.69,0,0}$3\pi$}%
}}}}
\put(6226,-1636){\makebox(0,0)[lb]{\smash{{\SetFigFont{12}{14.4}{\rmdefault}{\mddefault}{\updefault}{\color[rgb]{.69,0,0}$\varphi =0$}%
}}}}
\put(7201,-1636){\makebox(0,0)[lb]{\smash{{\SetFigFont{12}{14.4}{\rmdefault}{\mddefault}{\updefault}{\color[rgb]{.69,0,0}$\pi$}%
}}}}
\put(7951,-1336){\makebox(0,0)[lb]{\smash{{\SetFigFont{12}{14.4}{\rmdefault}{\mddefault}{\updefault}{\color[rgb]{0,0,0}$\theta$}%
}}}}
\put(4801,-436){\makebox(0,0)[lb]{\smash{{\SetFigFont{12}{14.4}{\rmdefault}{\mddefault}{\updefault}{\color[rgb]{0,0,0}$\theta$}%
}}}}
\put(5251,-1636){\makebox(0,0)[lb]{\smash{{\SetFigFont{12}{14.4}{\rmdefault}{\mddefault}{\updefault}{\color[rgb]{.69,0,0}$e^{i \varphi(\theta)}$}%
}}}}
\put(5776,-3361){\makebox(0,0)[lb]{\smash{{\SetFigFont{12}{14.4}{\rmdefault}{\mddefault}{\updefault}{\color[rgb]{.69,0,0}$\varphi(2\pi)=4\pi=2\times 2\pi$ hence  $\mathcal{C}=+2$}%
}}}}
\put(5026,-961){\makebox(0,0)[lb]{\smash{{\SetFigFont{12}{14.4}{\rmdefault}{\mddefault}{\updefault}{\color[rgb]{0,0,.56}$v_1$}%
}}}}
\put(5101,-2461){\makebox(0,0)[lb]{\smash{{\SetFigFont{12}{14.4}{\rmdefault}{\mddefault}{\updefault}{\color[rgb]{0,0,.56}$v_2$}%
}}}}
\end{picture}%

%% file: images/bundle_sphere_section2.pdftex_t
\begin{picture}(0,0)%
\includegraphics{images/bundle_sphere_section2}%
\end{picture}%
\setlength{\unitlength}{3947sp}%
\begingroup\makeatletter\ifx\SetFigFont\undefined%
\gdef\SetFigFont#1#2#3#4#5{%
  \reset@font\fontsize{#1}{#2pt}%
  \fontfamily{#3}\fontseries{#4}\fontshape{#5}%
  \selectfont}%
\fi\endgroup%
\begin{picture}(3738,3636)(2877,-4114)
\put(5018,-1487){\makebox(0,0)[lb]{\smash{{\SetFigFont{12}{14.4}{\rmdefault}{\mddefault}{\updefault}{\color[rgb]{0,0,0}$x$}%
}}}}
\put(3826,-3811){\makebox(0,0)[lb]{\smash{{\SetFigFont{12}{14.4}{\rmdefault}{\mddefault}{\updefault}{\color[rgb]{.56,0,0}$\mathcal{C} (TS^2)=1+1=2$}%
}}}}
\put(3226,-4036){\makebox(0,0)[lb]{\smash{{\SetFigFont{12}{14.4}{\rmdefault}{\mddefault}{\updefault}{\color[rgb]{0,0,0}Tangent bundle $TS^2$ }%
}}}}
\put(4726,-3436){\makebox(0,0)[lb]{\smash{{\SetFigFont{12}{14.4}{\rmdefault}{\mddefault}{\updefault}{\color[rgb]{.56,0,0}$\sigma =+1$}%
}}}}
\put(4351,-661){\makebox(0,0)[lb]{\smash{{\SetFigFont{12}{14.4}{\rmdefault}{\mddefault}{\updefault}{\color[rgb]{.56,0,0}$\sigma =+1$}%
}}}}
\put(5401,-2911){\makebox(0,0)[lb]{\smash{{\SetFigFont{12}{14.4}{\rmdefault}{\mddefault}{\updefault}{\color[rgb]{0,0,.56}global section}%
}}}}
\end{picture}%

%% file: images/bundle_sphere_3.pdftex_t
\begin{picture}(0,0)%
\includegraphics{images/bundle_sphere_3}%
\end{picture}%
\setlength{\unitlength}{3947sp}%
\begingroup\makeatletter\ifx\SetFigFont\undefined%
\gdef\SetFigFont#1#2#3#4#5{%
  \reset@font\fontsize{#1}{#2pt}%
  \fontfamily{#3}\fontseries{#4}\fontshape{#5}%
  \selectfont}%
\fi\endgroup%
\begin{picture}(5108,3540)(3961,-5014)
\put(7593,-2829){\makebox(0,0)[lb]{\smash{{\SetFigFont{11}{13.2}{\familydefault}{\mddefault}{\updefault}{\color[rgb]{0,0,.69}Clutching function }%
}}}}
\put(5251,-4936){\makebox(0,0)[lb]{\smash{{\SetFigFont{12}{14.4}{\rmdefault}{\mddefault}{\updefault}{\color[rgb]{0,0,0}$S^{2n}$}%
}}}}
\put(7576,-3061){\makebox(0,0)[lb]{\smash{{\SetFigFont{12}{14.4}{\rmdefault}{\mddefault}{\updefault}{\color[rgb]{0,0,.69}$g:S^{2n-1} \rightarrow U(r)$}%
}}}}
\put(7276,-2236){\makebox(0,0)[lb]{\smash{{\SetFigFont{12}{14.4}{\rmdefault}{\mddefault}{\updefault}{\color[rgb]{0,0,0}$\mathbb{C}^r$}%
}}}}
\put(7576,-3736){\makebox(0,0)[lb]{\smash{{\SetFigFont{12}{14.4}{\rmdefault}{\mddefault}{\updefault}{\color[rgb]{0,0,0}$\mathbb{C}^r$}%
}}}}
\put(3976,-3136){\makebox(0,0)[lb]{\smash{{\SetFigFont{12}{14.4}{\rmdefault}{\mddefault}{\updefault}{\color[rgb]{0,0,0}Equator}%
}}}}
\put(3976,-3391){\makebox(0,0)[lb]{\smash{{\SetFigFont{12}{14.4}{\rmdefault}{\mddefault}{\updefault}{\color[rgb]{0,0,0}$S^{2n-1}$}%
}}}}
\end{picture}%